\documentclass[journal,twoside,twocolumn]{IEEEtran}

\usepackage{cite}
\usepackage[T1]{fontenc}
\usepackage{graphicx}
\usepackage{amssymb}
\usepackage{amsmath}
\usepackage{paralist}
\usepackage{subfigure}
\usepackage{booktabs} 
\usepackage{algorithm}
\usepackage{algorithmic}
\usepackage{amsthm}
\usepackage{multirow}
\usepackage{microtype} 
\usepackage{tablefootnote}
\usepackage{balance}
\usepackage[dvipsnames]{xcolor}
\usepackage{soul}
\usepackage{array}
\usepackage{cuted}
\usepackage{varioref}
\usepackage{hyperref}

\usepackage{amssymb}
\usepackage{amsfonts}
\usepackage{mathrsfs}
\usepackage{xspace}
\usepackage{bm}
\usepackage{upgreek}

\newcommand{\safemath}[2]{\newcommand{#1}{\ensuremath{#2}\xspace}}

\safemath{\bma}{\mathbf{a}}
\safemath{\bmb}{\mathbf{b}}
\safemath{\bmc}{\mathbf{c}}
\safemath{\bmd}{\mathbf{d}}
\safemath{\bme}{\mathbf{e}}
\safemath{\bmf}{\mathbf{f}}
\safemath{\bmg}{\mathbf{g}}
\safemath{\bmh}{\mathbf{h}}
\safemath{\bmi}{\mathbf{i}}
\safemath{\bmj}{\mathbf{j}}
\safemath{\bmk}{\mathbf{k}}
\safemath{\bml}{\mathbf{l}}
\safemath{\bmm}{\mathbf{m}}
\safemath{\bmn}{\mathbf{n}}
\safemath{\bmo}{\mathbf{o}}
\safemath{\bmp}{\mathbf{p}}
\safemath{\bmq}{\mathbf{q}}
\safemath{\bmr}{\mathbf{r}}
\safemath{\bms}{\mathbf{s}}
\safemath{\bmt}{\mathbf{t}}
\safemath{\bmu}{\mathbf{u}}
\safemath{\bmv}{\mathbf{v}}
\safemath{\bmw}{\mathbf{w}}
\safemath{\bmx}{\mathbf{x}}
\safemath{\bmy}{\mathbf{y}}
\safemath{\bmz}{\mathbf{z}}
\safemath{\bmzero}{\mathbf{0}}
\safemath{\bmone}{\mathbf{1}}

\bmdefine{\biad}{a}
\bmdefine{\bibd}{b}
\bmdefine{\bicd}{c}
\bmdefine{\bidd}{d}
\bmdefine{\bied}{e}
\bmdefine{\bifd}{f}
\bmdefine{\bigd}{g}
\bmdefine{\bihd}{h}
\bmdefine{\biid}{i}
\bmdefine{\bijd}{j}
\bmdefine{\bikd}{k}
\bmdefine{\bild}{l}
\bmdefine{\bimd}{m}
\bmdefine{\bind}{n}
\bmdefine{\biod}{o}
\bmdefine{\bipd}{p}
\bmdefine{\biqd}{q}
\bmdefine{\bird}{r}
\bmdefine{\bisd}{s}
\bmdefine{\bitd}{t}
\bmdefine{\biud}{u}
\bmdefine{\bivd}{v}
\bmdefine{\biwd}{w}
\bmdefine{\bixd}{x}
\bmdefine{\biyd}{y}
\bmdefine{\bizd}{z}

\bmdefine{\bixid}{\xi}
\bmdefine{\bilambdad}{\lambda}
\bmdefine{\bimud}{\mu}
\bmdefine{\bithetad}{\theta}
\bmdefine{\biphid}{\phi}
\bmdefine{\bideltad}{\delta}

\safemath{\bmia}{\biad}
\safemath{\bmib}{\bibd}
\safemath{\bmic}{\bicd}
\safemath{\bmid}{\bidd}
\safemath{\bmie}{\bied}
\safemath{\bmif}{\bifd}
\safemath{\bmig}{\bigd}
\safemath{\bmih}{\bihd}
\safemath{\bmii}{\biid}
\safemath{\bmij}{\bijd}
\safemath{\bmik}{\bikd}
\safemath{\bmil}{\bild}
\safemath{\bmim}{\bimd}
\safemath{\bmin}{\bind}
\safemath{\bmio}{\biod}
\safemath{\bmip}{\bipd}
\safemath{\bmiq}{\biqd}
\safemath{\bmir}{\bird}
\safemath{\bmis}{\bisd}
\safemath{\bmit}{\bitd}
\safemath{\bmiu}{\biud}
\safemath{\bmiv}{\bivd}
\safemath{\bmiw}{\biwd}
\safemath{\bmix}{\bixd}
\safemath{\bmiy}{\biyd}
\safemath{\bmiz}{\bizd}

\safemath{\bmxi}{\bixid}
\safemath{\bmlambda}{\bilambdad}
\safemath{\bmmu}{\bimud}
\safemath{\bmtheta}{\bithetad}
\safemath{\bmphi}{\biphid}
\safemath{\bmdelta}{\bideltad}

\safemath{\bA}{\mathbf{A}}
\safemath{\bB}{\mathbf{B}}
\safemath{\bC}{\mathbf{C}}
\safemath{\bD}{\mathbf{D}}
\safemath{\bE}{\mathbf{E}}
\safemath{\bF}{\mathbf{F}}
\safemath{\bG}{\mathbf{G}}
\safemath{\bH}{\mathbf{H}}
\safemath{\bI}{\mathbf{I}}
\safemath{\bJ}{\mathbf{J}}
\safemath{\bK}{\mathbf{K}}
\safemath{\bL}{\mathbf{L}}
\safemath{\bM}{\mathbf{M}}
\safemath{\bN}{\mathbf{N}}
\safemath{\bO}{\mathbf{O}}
\safemath{\bP}{\mathbf{P}}
\safemath{\bQ}{\mathbf{Q}}
\safemath{\bR}{\mathbf{R}}
\safemath{\bS}{\mathbf{S}}
\safemath{\bT}{\mathbf{T}}
\safemath{\bU}{\mathbf{U}}
\safemath{\bV}{\mathbf{V}}
\safemath{\bW}{\mathbf{W}}
\safemath{\bX}{\mathbf{X}}
\safemath{\bY}{\mathbf{Y}}
\safemath{\bZ}{\mathbf{Z}}

\safemath{\bDelta}{\mathbf{\Delta}}
\safemath{\bLambda}{\mathbf{\UpLambda}}
\safemath{\bPhi}{\mathbf{\Upphi}}
\safemath{\bSigma}{\mathbf{\Upsigma}}
\safemath{\bOmega}{\mathbf{\Upomega}}
\safemath{\bTheta}{\mathbf{\Uptheta}}

\bmdefine{\biAd}{A}
\bmdefine{\biBd}{B}
\bmdefine{\biCd}{C}
\bmdefine{\biDd}{D}
\bmdefine{\biEd}{E}
\bmdefine{\biFd}{F}
\bmdefine{\biGd}{G}
\bmdefine{\biHd}{H}
\bmdefine{\biId}{I}
\bmdefine{\biJd}{J}
\bmdefine{\biKd}{K}
\bmdefine{\biLd}{L}
\bmdefine{\biMd}{M}
\bmdefine{\biOd}{N}
\bmdefine{\biPd}{O}
\bmdefine{\biQd}{P}
\bmdefine{\biRd}{R}
\bmdefine{\biSd}{S}
\bmdefine{\biTd}{T}
\bmdefine{\biUd}{U}
\bmdefine{\biVd}{V}
\bmdefine{\biWd}{W}
\bmdefine{\biXd}{X}
\bmdefine{\biYd}{Y}
\bmdefine{\biZd}{Z}

\bmdefine{\biDelta}{\Delta}
\bmdefine{\biLambda}{\Lambda}
\bmdefine{\biPhi}{\Phi}
\bmdefine{\biSigma}{\Sigma}
\bmdefine{\biOmega}{\Omega}
\bmdefine{\biTheta}{\Theta}

\safemath{\bimA}{\biAd}
\safemath{\bimB}{\biBd}
\safemath{\bimC}{\biCd}
\safemath{\bimD}{\biDd}
\safemath{\bimE}{\biEd}
\safemath{\bimF}{\biFd}
\safemath{\bimG}{\biGd}
\safemath{\bimH}{\biHd}
\safemath{\bimI}{\biId}
\safemath{\bimJ}{\biJd}
\safemath{\bimK}{\biKd}
\safemath{\bimL}{\biLd}
\safemath{\bimM}{\biMd}
\safemath{\bimN}{\biNd}
\safemath{\bimO}{\biOd}
\safemath{\bimP}{\biPd}
\safemath{\bimQ}{\biQd}
\safemath{\bimR}{\biRd}
\safemath{\bimS}{\biSd}
\safemath{\bimT}{\biTd}
\safemath{\bimU}{\biUd}
\safemath{\bimV}{\biVd}
\safemath{\bimW}{\biWd}
\safemath{\bimX}{\biXd}
\safemath{\bimY}{\biYd}
\safemath{\bimZ}{\biZd}

\safemath{\bimDelta}{\biDelta}
\safemath{\bimLambda}{\biLambda}
\safemath{\bimPhi}{\biPhi}
\safemath{\bimSigma}{\biSigma}
\safemath{\bimOmega}{\biOmega}
\safemath{\bimTheta}{\biTheta}

\safemath{\setA}{\mathcal{A}}
\safemath{\setB}{\mathcal{B}}
\safemath{\setC}{\mathcal{C}}
\safemath{\setD}{\mathcal{D}}
\safemath{\setE}{\mathcal{E}}
\safemath{\setF}{\mathcal{F}}
\safemath{\setG}{\mathcal{G}}
\safemath{\setH}{\mathcal{H}}
\safemath{\setI}{\mathcal{I}}
\safemath{\setJ}{\mathcal{J}}
\safemath{\setK}{\mathcal{K}}
\safemath{\setL}{\mathcal{L}}
\safemath{\setM}{\mathcal{M}}
\safemath{\setN}{\mathcal{N}}
\safemath{\setO}{\mathcal{O}}
\safemath{\setP}{\mathcal{P}}
\safemath{\setQ}{\mathcal{Q}}
\safemath{\setR}{\mathcal{R}}
\safemath{\setS}{\mathcal{S}}
\safemath{\setT}{\mathcal{T}}
\safemath{\setU}{\mathcal{U}}
\safemath{\setV}{\mathcal{V}}
\safemath{\setW}{\mathcal{W}}
\safemath{\setX}{\mathcal{X}}
\safemath{\setY}{\mathcal{Y}}
\safemath{\setZ}{\mathcal{Z}}
\safemath{\emptySet}{\varnothing}

\safemath{\colA}{\mathscr{A}}
\safemath{\colB}{\mathscr{B}}
\safemath{\colC}{\mathscr{C}}
\safemath{\colD}{\mathscr{D}}
\safemath{\colE}{\mathscr{E}}
\safemath{\colF}{\mathscr{F}}
\safemath{\colG}{\mathscr{G}}
\safemath{\colH}{\mathscr{H}}
\safemath{\colI}{\mathscr{I}}
\safemath{\colJ}{\mathscr{J}}
\safemath{\colK}{\mathscr{K}}
\safemath{\colL}{\mathscr{L}}
\safemath{\colM}{\mathscr{M}}
\safemath{\colN}{\mathscr{N}}
\safemath{\colO}{\mathscr{O}}
\safemath{\colP}{\mathscr{P}}
\safemath{\colQ}{\mathscr{Q}}
\safemath{\colR}{\mathscr{R}}
\safemath{\colS}{\mathscr{S}}
\safemath{\colT}{\mathscr{T}}
\safemath{\colU}{\mathscr{U}}
\safemath{\colV}{\mathscr{V}}
\safemath{\colW}{\mathscr{W}}
\safemath{\colX}{\mathscr{X}}
\safemath{\colY}{\mathscr{Y}}
\safemath{\colZ}{\mathscr{Z}}

\safemath{\opA}{\mathbb{A}}
\safemath{\opB}{\mathbb{B}}
\safemath{\opC}{\mathbb{C}}
\safemath{\opD}{\mathbb{D}}
\safemath{\opE}{\mathbb{E}}
\safemath{\opF}{\mathbb{F}}
\safemath{\opG}{\mathbb{G}}
\safemath{\opH}{\mathbb{H}}
\safemath{\opI}{\mathbb{I}}
\safemath{\opJ}{\mathbb{J}}
\safemath{\opK}{\mathbb{K}}
\safemath{\opL}{\mathbb{L}}
\safemath{\opM}{\mathbb{M}}
\safemath{\opN}{\mathbb{N}}
\safemath{\opO}{\mathbb{O}}
\safemath{\opP}{\mathbb{P}}
\safemath{\opQ}{\mathbb{Q}}
\safemath{\opR}{\mathbb{R}}
\safemath{\opS}{\mathbb{S}}
\safemath{\opT}{\mathbb{T}}
\safemath{\opU}{\mathbb{U}}
\safemath{\opV}{\mathbb{V}}
\safemath{\opW}{\mathbb{W}}
\safemath{\opX}{\mathbb{X}}
\safemath{\opY}{\mathbb{Y}}
\safemath{\opZ}{\mathbb{Z}}
\safemath{\opZero}{\mathbb{O}}
\safemath{\identityop}{\opI}

\safemath{\veca}{\bma}
\safemath{\vecb}{\bmb}
\safemath{\vecc}{\bmc}
\safemath{\vecd}{\bmd}
\safemath{\vece}{\bme}
\safemath{\vecf}{\bmf}
\safemath{\vecg}{\bmg}
\safemath{\vech}{\bmh}
\safemath{\veci}{\bmi}
\safemath{\vecj}{\bmj}
\safemath{\veck}{\bmk}
\safemath{\vecl}{\bml}
\safemath{\vecm}{\bmm}
\safemath{\vecn}{\bmn}
\safemath{\veco}{\bmo}
\safemath{\vecp}{\bmp}
\safemath{\vecq}{\bmq}
\safemath{\vecr}{\bmr}
\safemath{\vecs}{\bms}
\safemath{\vect}{\bmt}
\safemath{\vecu}{\bmu}
\safemath{\vecv}{\bmv}
\safemath{\vecw}{\bmw}
\safemath{\vecx}{\bmx}
\safemath{\vecy}{\bmy}
\safemath{\vecz}{\bmz}

\safemath{\veczero}{\bmzero}
\safemath{\vecone}{\bmone}
\safemath{\vecxi}{\bmxi}
\safemath{\veclambda}{\bmlambda}
\safemath{\vecmu}{\bmmu}
\safemath{\vectheta}{\bmtheta}
\safemath{\vecphi}{\bmphi}
\safemath{\vecdelta}{\bmdelta}

\safemath{\matA}{\bA}
\safemath{\matB}{\bB}
\safemath{\matC}{\bC}
\safemath{\matD}{\bD}
\safemath{\matE}{\bE}
\safemath{\matF}{\bF}
\safemath{\matG}{\bG}
\safemath{\matH}{\bH}
\safemath{\matI}{\bI}
\safemath{\matJ}{\bJ}
\safemath{\matK}{\bK}
\safemath{\matL}{\bL}
\safemath{\matM}{\bM}
\safemath{\matN}{\bN}
\safemath{\matO}{\bO}
\safemath{\matP}{\bP}
\safemath{\matQ}{\bQ}
\safemath{\matR}{\bR}
\safemath{\matS}{\bS}
\safemath{\matT}{\bT}
\safemath{\matU}{\bU}
\safemath{\matV}{\bV}
\safemath{\matW}{\bW}
\safemath{\matX}{\bX}
\safemath{\matY}{\bY}
\safemath{\matZ}{\bZ}
\safemath{\matzero}{\bmzero}

\safemath{\matDelta}{\bDelta}
\safemath{\matLambda}{\bLambda}
\safemath{\matPhi}{\bPhi}
\safemath{\matSigma}{\bSigma}
\safemath{\matOmega}{\bOmega}
\safemath{\matTheta}{\bTheta}

\safemath{\matidentity}{\matI}
\safemath{\matone}{\matO}

\safemath{\rnda}{A}
\safemath{\rndb}{B}
\safemath{\rndc}{C}
\safemath{\rndd}{D}
\safemath{\rnde}{E}
\safemath{\rndf}{F}
\safemath{\rndg}{G}
\safemath{\rndh}{H}
\safemath{\rndi}{I}
\safemath{\rndj}{J}
\safemath{\rndk}{K}
\safemath{\rndl}{L}
\safemath{\rndm}{M}
\safemath{\rndn}{N}
\safemath{\rndo}{O}
\safemath{\rndp}{P}
\safemath{\rndq}{Q}
\safemath{\rndr}{R}
\safemath{\rnds}{S}
\safemath{\rndt}{T}
\safemath{\rndu}{U}
\safemath{\rndv}{V}
\safemath{\rndw}{W}
\safemath{\rndx}{X}
\safemath{\rndy}{Y}
\safemath{\rndz}{Z}

\safemath{\rveca}{\bimA}
\safemath{\rvecb}{\bimB}
\safemath{\rvecc}{\bimC}
\safemath{\rvecd}{\bimD}
\safemath{\rvece}{\bimE}
\safemath{\rvecf}{\bimF}
\safemath{\rvecg}{\bimG}
\safemath{\rvech}{\bimH}
\safemath{\rveci}{\bimI}
\safemath{\rvecj}{\bimJ}
\safemath{\rveck}{\bimK}
\safemath{\rvecl}{\bimL}
\safemath{\rvecm}{\bimM}
\safemath{\rvecn}{\bimN}
\safemath{\rveco}{\bomO}
\safemath{\rvecp}{\bimP}
\safemath{\rvecq}{\bimQ}
\safemath{\rvecr}{\bimR}
\safemath{\rvecs}{\bimS}
\safemath{\rvect}{\bimT}
\safemath{\rvecu}{\bimU}
\safemath{\rvecv}{\bimV}
\safemath{\rvecw}{\bimW}
\safemath{\rvecx}{\bimX}
\safemath{\rvecy}{\bimY}
\safemath{\rvecz}{\bimZ}

\safemath{\rvecxi}{\bmxi}
\safemath{\rveclambda}{\bmlambda}
\safemath{\rvecmu}{\bmmu}
\safemath{\rvectheta}{\bmtheta}
\safemath{\rvecphi}{\bmphi}

\safemath{\rmatA}{\bimA}
\safemath{\rmatB}{\bimB}
\safemath{\rmatC}{\bimC}
\safemath{\rmatD}{\bimD}
\safemath{\rmatE}{\bimE}
\safemath{\rmatF}{\bimF}
\safemath{\rmatG}{\bimG}
\safemath{\rmatH}{\bimH}
\safemath{\rmatI}{\bimI}
\safemath{\rmatJ}{\bimJ}
\safemath{\rmatK}{\bimK}
\safemath{\rmatL}{\bimL}
\safemath{\rmatM}{\bimM}
\safemath{\rmatN}{\bimN}
\safemath{\rmatO}{\bimO}
\safemath{\rmatP}{\bimP}
\safemath{\rmatQ}{\bimQ}
\safemath{\rmatR}{\bimR}
\safemath{\rmatS}{\bimS}
\safemath{\rmatT}{\bimT}
\safemath{\rmatU}{\bimU}
\safemath{\rmatV}{\bimV}
\safemath{\rmatW}{\bimW}
\safemath{\rmatX}{\bimX}
\safemath{\rmatY}{\bimY}
\safemath{\rmatZ}{\bimZ}

\safemath{\rmatDelta}{\bimDelta}
\safemath{\rmatLambda}{\bimLambda}
\safemath{\rmatPhi}{\bimPhi}
\safemath{\rmatSigma}{\bimSigma}
\safemath{\rmatOmega}{\bimOmega}
\safemath{\rmatTheta}{\bimTheta}

\usepackage{amssymb}
\usepackage{amsfonts}
\usepackage{mathrsfs}
\usepackage{xspace}
\usepackage{bm}
\usepackage{fancyref}
\usepackage{textcomp}

\usepackage{multirow}
\usepackage{stmaryrd}

\newenvironment{textbmatrix}{	\setlength{\arraycolsep}{2.5pt}%
								\big[\begin{matrix}}{\end{matrix}\big]%
								\raisebox{0.08ex}{\vphantom{M}}}

\def\be{\begin{equation}}
\def\ee{\end{equation}}
\def\een{\nonumber \end{equation}}
\def\mat{\begin{bmatrix}}
\def\emat{\end{bmatrix}}
\def\btm{\begin{textbmatrix}}
\def\etm{\end{textbmatrix}}

\def\ba#1\ea{\begin{align}#1\end{align}}
\def\bas#1\eas{\begin{align*}#1\end{align*}}
\def\bs#1\es{\begin{split}#1\end{split}}
\def\bg#1\eg{\begin{gather}#1\end{gather}}
\def\bml#1\eml{\begin{multline}#1\end{multline}}
\def\bi#1\ei{\begin{itemize}#1\end{itemize}}

\newcommand{\lefto}{\mathopen{}\left}

\DeclareMathOperator{\Tr}{\opT r}			
\DeclareMathOperator*{\argmin}{arg\;min}		
\DeclareMathOperator*{\argmax}{arg\;max}		
\DeclareMathOperator{\kron}{\otimes}			
\DeclareMathOperator{\Exop}{\opE}			

\newcommand{\Ex}[2]{\ensuremath{\Exop_{#1}\lefto[#2\right]}} 	
\newcommand{\abs}[1]{\lefto\lvert#1\right\rvert}		

\newcommand{\vecnorm}[1]{\lefto\lVert#1\right\rVert}		

\safemath{\dirac}{\delta}					
\safemath{\krond}{\dirac}					

\safemath{\upto}{\uparrow}
\safemath{\downto}{\downarrow}
\safemath{\iu}{j}							
\safemath{\ev}{\lambda}						
\safemath{\hilseqspace}{l^{2}}				
\newcommand{\banachfunspace}[1]{\setL^{#1}}	
\safemath{\hilfunspace}{\banachfunspace{2}}	

\safemath{\SNR}{\textit{SNR}} 				
\safemath{\PAR}{\textit{PAR}} 				
\safemath{\No}{N_0}							
\safemath{\Es}{E_s}							
\safemath{\Eb}{E_b}							
\safemath{\EbNo}{\frac{\Eb}{\No}}
\safemath{\EsNo}{\frac{\Es}{\No}}

\DeclareMathOperator{\CHop}{\ensuremath{\opH}} 
\safemath{\tvir}{\rndh_{\CHop}}				
\safemath{\tvtf}{\rndl_{\CHop}}				
\safemath{\spf}{\rnds_{\CHop}}				
\safemath{\bff}{H_{\CHop}}					

\safemath{\ircf}{r_{h}}						
\safemath{\tftvcf}{r_{s}}					
\safemath{\tfcf}{r_{l}}						
\safemath{\bfcf}{r_{H}}						

\safemath{\tcorr}{c_h}						
\safemath{\scf}{c_{s}}						
\safemath{\tfcorr}{c_{l}}					
\safemath{\fcorr}{c_{H}}						

\safemath{\mi}{I}							
\safemath{\capacity}{C}						

\safemath{\normal}{\mathcal{N}}			
\safemath{\jpg}{\mathcal{CN}}			
\safemath{\mchain}{\leftrightarrow}		

\safemath{\dB}{\,\mathrm{dB}}
\safemath{\dBm}{\,\mathrm{dBm}}
\safemath{\Hz}{\,\mathrm{Hz}}
\safemath{\kHz}{\,\mathrm{kHz}}
\safemath{\MHz}{\,\mathrm{MHz}}
\safemath{\GHz}{\,\mathrm{GHz}}
\safemath{\s}{\,\mathrm{s}}
\safemath{\ms}{\,\mathrm{ms}}
\safemath{\mus}{\,\mathrm{\text{\textmu}s}}
\safemath{\ns}{\,\mathrm{ns}}
\safemath{\ps}{\,\mathrm{ps}}
\safemath{\meter}{\,\mathrm{m}}
\safemath{\mm}{\,\mathrm{mm}}
\safemath{\cm}{\,\mathrm{cm}}
\safemath{\m}{\,\mathrm{m}}
\safemath{\W}{\,\mathrm{W}}
\safemath{\mW}{\, \mathrm{mW}}
\safemath{\J}{\,\mathrm{J}}
\safemath{\K}{\,\mathrm{K}}
\safemath{\bit}{\,\mathrm{bit}}
\safemath{\nat}{\,\mathrm{nat}}

\safemath{\define}{\triangleq}			

\safemath{\equivalent}{\sim}
\safemath{\distas}{\sim}					
\safemath{\sdiff}{\Delta}				

\safemath{\reals}{\mathbb{R}}
\safemath{\positivereals}{\reals_{+}}
\safemath{\integers}{\mathbb{Z}}
\safemath{\posint}{\integers_{+}}
\safemath{\naturals}{\mathbb{N}}
\safemath{\posnaturals}{\naturals_{+}}
\safemath{\complexset}{\mathbb{C}}
\safemath{\rationals}{\mathbb{Q}}

\newcommand*{\fancyrefapplabelprefix}{app}		
\newcommand*{\fancyrefthmlabelprefix}{thm}		
\newcommand*{\fancyreflemlabelprefix}{lem}		
\newcommand*{\fancyrefcorlabelprefix}{cor}		
\newcommand*{\fancyrefdeflabelprefix}{def}		
\newcommand*{\fancyrefproplabelprefix}{prop}		
\newcommand*{\fancyrefexmpllabelprefix}{exmpl}
\newcommand*{\fancyrefalglabelprefix}{alg}		
\newcommand*{\fancyreftbllabelprefix}{tbl}		

\frefformat{vario}{\fancyrefseclabelprefix}{Section~#1}
\frefformat{vario}{\fancyrefthmlabelprefix}{Theorem~#1}
\frefformat{vario}{\fancyreftbllabelprefix}{Table~#1}
\frefformat{vario}{\fancyreflemlabelprefix}{Lemma~#1}
\frefformat{vario}{\fancyrefcorlabelprefix}{Corollary~#1}
\frefformat{vario}{\fancyrefdeflabelprefix}{Definition~#1}
\frefformat{vario}{\fancyreffiglabelprefix}{Fig.~#1}
\frefformat{vario}{\fancyrefapplabelprefix}{Appendix~#1}
\frefformat{vario}{\fancyrefeqlabelprefix}{(#1)}
\frefformat{vario}{\fancyrefproplabelprefix}{Proposition~#1}
\frefformat{vario}{\fancyrefexmpllabelprefix}{Example~#1}
\frefformat{vario}{\fancyrefalglabelprefix}{Algorithm~#1}

 \newtheorem{thm}{Theorem}

\safemath{\dictab}{[\,\dicta\,\,\dictb\,]}

\safemath{\ysig}{\bmy}
\safemath{\ysighat}{\hat{\ysig}}
\safemath{\ysigdim}{M}
\safemath{\xsig}{\bmx}
\safemath{\xsigdim}{N}
\safemath{\nx}{n_x}
\safemath{\zsig}{\bmz}
\safemath{\zsigdim}{\ysigdim}
\safemath{\rsig}{\bmr}
\safemath{\Adict}{\bA}
\safemath{\Adicttilde}{\widetilde{\Adict}}
\safemath{\Adictdim}{\outputdim\times\xsigdim}
\safemath{\avec}{\bma}
\safemath{\avectilde}{\tilde{\avec}}
\safemath{\Bdict}{\bB}
\safemath{\Bdicttilde}{\widetilde{\Bdict}}
\safemath{\Cdict}{\bC}
\safemath{\cvec}{\bmc}
\safemath{\Ddict}{\bD}
\safemath{\Ddictdim}{\ysigdim\times\xsigdim}
\safemath{\dvec}{\bmd}
\safemath{\Ddicttilde}{\widetilde{\bD}}
\safemath{\Bonb}{\bB}
\safemath{\bvec}{\bmb}
\safemath{\Bonbdim}{\ysigdim\times\ysigdim}
\safemath{\noise}{\bmn}
\safemath{\noisedim}{\ysigim}
\safemath{\err}{\bme}
\safemath{\errdim}{\ysigdim}
\safemath{\errset}{\setE}
\safemath{\nerr}{n_e}
\safemath{\delop}{\bP_\errset}
\safemath{\delopc}{\bP_{{\errset}^c}}

\safemath{\cplxi}{\imath}
\safemath{\cplxj}{\jmath}

\safemath{\dict}{\matD}
\safemath{\inputdim}{N}		
\safemath{\outputdim}{M}		
\safemath{\sparsity}{S}	
\safemath{\inputdimA}{{N_a}}	
\safemath{\inputdimB}{{N_b}}	
\safemath{\elemA}{{n_a}}	
\safemath{\elemB}{{n_b}}	
\safemath{\resA}{\matR_a}	
\safemath{\resB}{\matR_b}	
\safemath{\subD}{\matS} 
\safemath{\subA}{\matS_a} 
\safemath{\subB}{\matS_b} 
\safemath{\dicta}{\matA} 	
\safemath{\dictb}{\matB} 	
\safemath{\hollowS}{H}
\safemath{\hollowA}{H_a}
\safemath{\hollowB}{H_b}
\safemath{\cross}{Z}
\safemath{\coh}{\mu_d}			
\safemath{\coha}{\mu_a}			
\safemath{\cohb}{\mu_b}			
\safemath{\mubs}{\nu}	
\safemath{\cohm}{\mu_m} 
\safemath{\dictset}{\setD}	
\safemath{\dictsetp}{\dictset(\coh,\coha,\cohb)}	
\safemath{\dictsetgen}{\dictset_\text{gen}}
\safemath{\dictsetgenp}{\dictsetgen(\coh)}
\safemath{\dictsetonb}{\dictset_\text{onb}}
\safemath{\dictsetonbp}{\dictsetonb(\coh)}

\safemath{\leftside}{U}
\safemath{\rightsideA}{R_a}
\safemath{\rightsideB}{R_b}

\safemath{\indexS}{\setI_S} 

\safemath{\na}{n_a}			
\safemath{\nb}{n_b}			
\safemath{\coeffa}{p_i}	
\safemath{\coeffb}{q_j}	
\safemath{\seta}{\setP}		
\safemath{\setb}{\setQ}     
\safemath{\setw}{\setW}	
\safemath{\setz}{\setZ}	
\safemath{\cola}{\veca}		
\safemath{\colb}{\vecb}		
\safemath{\cold}{\vecd}		
\safemath{\inputvec}{\vecx} 	
\safemath{\error}{\vece}	
\safemath{\noiseout}{\vecz} 	
\safemath{\inputvecel}{x}
\safemath{\inputveca}{\vecx_a}
\safemath{\inputvecb}{\vecx_b}
\safemath{\outputvec}{\vecy}	
\safemath{\lambdamin}{\lambda_{\mathrm{min}}}

\newcommand{\normtwo}[1]{\vecnorm{#1}_2}
\newcommand{\normone}[1]{\vecnorm{#1}_1}

\newcommand{\norminftilde}[1]{\vecnorm{#1}_{\widetilde\infty}}
\newcommand{\normfro}[1]{\vecnorm{#1}_F}

\safemath{\elltwo}{\ell_2}
\safemath{\ellone}{\ell_1}
\safemath{\ellzero}{\ell_0}
\safemath{\ellinf}{\ell_\infty}
\safemath{\ellinftilde}{\ell_{\widetilde\infty}}
\safemath{\licard}{Z(\coh,\coha,\cohb)}
\safemath{\xsol}{\hat{x}}
\safemath{\xbord}{x_b}		
\safemath{\xstat}{x_s}		
\safemath{\xstatLone}{\tilde{x}_s}
\safemath{\order}{\mathcal{O}} 
\safemath{\scales}{\Theta} 
\safemath{\ones}{\mathbf{1}} 
\safemath{\zeroes}{\mathbf{0}} 
\safemath{\thlone}{\kappa(\coh,\cohb)} 
\safemath{\constoneA}{\delta} 
\safemath{\constoneB}{\epsilon} 
\safemath{\nlarge}{L}				   
\safemath{\sumlarge}{S_\nlarge}
\safemath{\maxlarger}{P_\nlarge}	   
\safemath{\Pzero}{\textrm{P0}}	
\safemath{\Pone}{\textrm{P1}}
\safemath{\vecfir}{\vecw}			 
\safemath{\vecsec}{\vecz}
\safemath{\elvecfir}{w}              
\safemath{\elvecsec}{z}				 
\safemath{\nlargefir}{n}
\safemath{\normout}{\gamma}
\safemath{\auxfun}{h}
\safemath{\supp}{\textrm{supp}}

\safemath{\indexa}{\ell}
\safemath{\indexb}{r}
\safemath{\indexc}{i}
\safemath{\indexd}{j}

\safemath{\project}{P}

\newcommand{\PAP}{\ensuremath{\bP_\text{AP}}}
\newcommand{\PUE}{\ensuremath{\bP_\text{UE}}}
\newcommand{\sdj}{[\bS_D]_{u}}
\newcommand{\bsdj}{[\bar\bS_D]_{u}}
\newcommand{\bOne}[2]{\ones_{#1\times #2}}
\newcommand{\bZero}[2]{\mathbf{0}_{#1\times #2}}
\newcommand{\innR}[1]{\langle #1\rangle_\mathfrak{R}}
\newcommand{\nbfZt}{\nabla f( \bZ^{* (t) } )}

\newcommand{\aln}[1]{\begin{align}#1\end{align}}

\newcommand{\matb}{\left( \begin{matrix*}[r] }
\newcommand{\mate}{\end{matrix*}\right)}
\newcommand{\entj}[1]{\text{H} \left(  #1 \right) }

\newcommand{\half}{\frac{1}{2}}

\newcommand{\kp}{^{(t+1)}}

\newcommand{\ko}{^{(t)}}

\newcommand{\taut}{\tau^{(t)}}

\renewcommand\Re{\operatorname{\mathfrak{Re}}}
\renewcommand\Im{\operatorname{Im}}

\newtheorem{lemma}{Lemma}
\theoremstyle{definition}

\newtheorem{remark}{Remark}

\newcommand{\revision}[1]{\textcolor{blue}{#1}}

\begin{document}

\title{Joint Channel Estimation and Data Detection \\  in Cell-Free Massive MU-MIMO Systems}

\author{
\IEEEauthorblockN{Haochuan Song, Tom Goldstein, Xiaohu You,  Chuan Zhang,  Olav Tirkkonen, and Christoph Studer }\\
\thanks{A short version of this paper has been presented at IEEE SPAWC 2020~\cite{songMinimizing2020}.}
\thanks{H.~Song, X. You, and C. Zhang are with the LEADS, Southeast University, Nanjing, China, also with the National Mobile Communications Research Laboratory Southeast University,
Nanjing, China, and also with the Purple Mountain Laboratories,
Nanjing, China; email: hcsong, xhyu, chzhang@seu.edu.cn}
\thanks{T.~Goldstein is with the Department of Computer Science at University of Maryland, College Park, MD; email: tomg@cs.umd.edu}
\thanks{O. Tirkkonen is with the Department of Communications and Networking, Aalto University, Espoo, Finland;  email: olav.tirkkonen@aalto.fi}
\thanks{C. Studer is with the Department of Information Technology and Electrical Engineering at ETH Zurich, Zurich, Switzerland; email: studer@ethz.ch}
}
\maketitle
\vspace{-1.8cm}
\begin{abstract}
We propose a joint channel estimation and data detection (JED) algorithm for densely-populated cell-free massive multiuser (MU) multiple-input multiple-output (MIMO) systems, which reduces the channel training overhead caused by the presence of hundreds of simultaneously transmitting user equipments (UEs).
Our algorithm iteratively solves a relaxed version of a maximum a-posteriori JED problem and simultaneously exploits the sparsity of cell-free massive MU-MIMO channels as well as the boundedness of QAM constellations.
In order to improve the performance and convergence of the algorithm, we propose methods that permute the access point and UE indices to form so-called virtual cells, which leads to better initial solutions.
We assess the performance of our algorithm in terms of root-mean-squared-symbol error, bit error rate, and mutual information, and we demonstrate that JED significantly reduces the pilot overhead compared to orthogonal training, which enables reliable communication with short packets to a large number of UEs.
\end{abstract}

\vspace{-0.5cm}

\section{Introduction}
\IEEEPARstart{C}{ell-free} massive multi-user (MU) multiple-input multiple-output (MIMO) wireless systems promise significant enhancements in spectral efficiency compared to traditional cellular systems\cite{ngoCellfree2017,buzzi2017cell,hoang2018cell,zhang2020Prospective}.
The distributed nature of such systems assures that every user equipment (UE) is able to communicate with multiple nearby access points (APs)~\cite{basharUplink2019,basharMax2019}.
Cell-free massive MU-MIMO systems are envisioned to operate in time-division duplex (TDD) mode.
The ideal case for channel estimation in the uplink would be to use orthogonal pilot sequences.
However, densely-populated cell-free massive MU-MIMO systems, in which hundreds or even thousands of UEs communicate in the same time-frequency resource, prevent the use of orthogonal training sequences as it would reduce the achievable data rates.  
While nonorthogonal pilots can certainly mitigate this issue, without taking special precautions, the accuracy of the extracted channel estimates will be severely compromised, resulting in poor spectral efficiency. 

In order to address this issue, recent research has mainly focused on pilot reuse, and maximizing the signal-to-interference-plus-noise ratio (SINR) with linear estimation and equalization methods while taking pilot contamination into account~\cite{maiPilot2018,sabbaghPilot2018,doanPerformance2017,parkOptimizing2018,liu2020Graph}.
While such approaches have relatively low complexity, 
they allow for high spectral efficiency only in
scenarios in which a large number of AP antennas serve a far smaller number of UEs~\cite{attarifarModified2019,atzeni2021Distributed,basharMax2019,basharEnhanced2018,maryopiUplink2019}.
In other words, as the number of UEs approaches or even exceeds the number of AP antennas, the performance of such densely-populated cell-free massive MU-MIMO systems degrades considerably, especially when relying on linear channel estimation and data detection methods~\cite{attarifar2018random}.

\subsection{Contributions}
We propose a novel joint channel estimation and data detection (JED) algorithm tailored to densely populated cell-free massive MU-MIMO systems in which the number of UEs is close to or larger than the number of AP antennas.
The distributed placement of UEs and APs results in sparse channel matrices, as every UE is  only nearby to a small number of APs.
The proposed JED algorithm simultaneously exploits the sparsity of cell-free massive MU-MIMO channels and the boundedness of constellation sets in order to minimize the pilot overhead while providing high spectral efficiency.
Our algorithm approximately solves a relaxed version of the maximum a-posteriori (MAP) JED problem using forward backward splitting (FBS).
To improve the performance and reduce the complexity of our JED algorithm, we combine  nonorthogonal pilot sequences with novel permutation strategies of AP and UE indices, which enable us to find better initializers.
We present simulation results that demonstrate the advantages of JED compared to traditional methods that separate channel estimation from data detection in terms of the root-mean-squared-symbol error (RMSSE), bit error-rate (BER), mutual information (MI), and channel-estimation mean-square error~(MSE).

\subsection{Relevant Prior Art}
\subsubsection{Channel Estimation and Data Detection}
The majority of research on uplink transmission in cell-free massive MU-MIMO systems has focused on linear methods that separate channel estimation from data detection, such as maximum ratio combining (MRC), zero forcing (ZF), and linear minimum mean-square error (L-MMSE)  equalization \cite{basharCellFree2018,basharEnhanced2018,basharMax2019,basharPerformance2018,interdonato2019Scalability,bjornsonScalable2020,bjornsonMaking2020}. 
The optimization targets of such linear methods are typically the MSE of channel estimation and/or signal estimation, or maximizing post-equalization SINR.
Albeit computationally efficient and easy to analyze, linear methods do not perform well 
in systems (i) that use nonorthogonal pilot sequences or (ii) where the number of UE antennas approaches the number of AP antennas~\cite{attarifar2018random}. 
The situation is further aggravated in overloaded systems, where the number of UE antennas exceeds the number of AP antennas. 
In contrast, our nonlinear JED algorithm enables reliable transmission in densely-populated systems with (often significantly) fewer pilots than UEs. 
In addition, our JED algorithm will not cause an increase in fronthaul data rates compared to the centralized data detectors put forward in \cite{bjornsonMaking2020}.
Concretely, given a cell-free massive MU-MIMO system with $B$ APs, each equipped with $N$ antennas (we assume $N=1$), and $U$ single antenna UEs transmitting pilots and payload for $K$ time slots within one coherence block, the total amount of fronthaul signaling is $BNK$ complex scalars. This is the same as that of the level $4$ centralized method in~\cite[Tbl.~I]{bjornsonMaking2020} (where   $\tau_c=K$ and $L=B$).
Furthermore, our JED algorithm does not require knowledge of second-order statistics on the UEs channel vectors, which further reduces the pilot and fronthaul overheads compared to the methods in \cite{bjornsonMaking2020}.
As a drawback, the complexity of our JED algorithm is substantially higher than that of linear methods.
Nonetheless, as we will show in \fref{sec:numerical results}, linear methods that separate channel estimation from data detection, even when performed in a centralized manner, perform only poorly in densely-populated scenarios.
Moreover, decentralized linear data detectors as proposed in \cite{bjornsonScalable2020,bjornsonMaking2020}, which excel in complexity and scalability, perform even worse than their centralized counterparts and are not suitable for densely-populated scenarios.

\subsubsection{Joint Channel Estimation and Data Detection}
JED has been studied in the small-scale MIMO literature~\cite{alshamary2015optimal,pham2009joint,prasad2015joint,kofidis2017joint,wen15b}.
While the complexity of such methods does not scale well to large systems, 
an efficient JED algorithm has been proposed in~\cite{castaneda2017vlsi} for massive single-input multiple-output (SIMO) systems. 
For massive MU-MIMO systems, JED algorithms have been proposed only recently in~\cite{jiangJoint2020,Yilmaz19a,feng2017noncoherent,zhangBlind2018,liuSuperresolution2019,dingSparsity2019,xueBlind2020}.  
To the best of our knowledge, none of these methods exploit the specifics of cell-free massive MU-MIMO systems. For example,
reference~\cite{xueBlind2020} maximizes the $ \ell_3 $-norm to exploit beamspace sparsity of millimeter-wave (mmWave) MIMO systems.
Message passing (MP) algorithms have also been used for JED in \cite{jiangJoint2020,zhangBlind2018}. 
In contrast, our method exploits the sparse nature of cell-free massive MU-MIMO channels in combination with the boundedness of QAM constellations. 
Furthermore, our UE and AP permutation methods discussed in \fref{sec: initialization} could also improve the performance of MP-based algorithms.

\subsubsection{Sparsity in Cell-free Massive MU-MIMO systems}
Sparsity of the channel matrices in cell-free massive MU-MIMO systems has, up to now, not been exploited extensively. 
In~\cite{binbindaiSparse2013,hanSparse2019,shiGroup2014,vanchien2020Joint}, the authors exploit the sparsity of beamforming vectors during downlink transmission by only serving a small portion of UEs.
Exploiting sparsity to identify active UEs was proposed in~\cite{guoDistributed2019}. 
Reference~\cite{shiLargescale2015} formulates channel estimation as a convex optimization problem using a sparsity-inducing $ \ell_1 $-norm penalty. 
In \cite{jinChannel2019a,mirfarshbafanBeamspace2019}, channel sparsity in beamspace domain has been exploited for cellular mmWave communication systems.
In contrast, we exploit sparsity for the JED algorithm in cell-free massive MU-MIMO systems, which comes from the distributed placement of APs and UEs, and the fact that the path loss between UEs and APs causes only a small number of strong links to be present.

\subsubsection{Pilot Design and Reuse}\label{sec:intro pilots}
In densely-populated cell-free massive MU-MIMO systems, the shortage of pilots has been identified as a major concern.
A straightforward approach is to assume that the number of UEs is smaller than the number of available pilot sequences, which enables the use of orthogonal training~\cite{interdonato2019Downlink,hanSparse2019,huCellfree2019}.
If the number of UEs exceeds the number of available pilots, either pilot reuse or nonorthogonal training is necessary. 
Reference \cite{doanPerformance2017,riera-palouClustered2018} propose to divide the UEs into fixed groups, each assigned with one pilot, whereas reference~\cite{sabbaghPilot2018} proposes a dynamic pilot assignment strategy.
Nonorthogonal pilot sequence design has been studied in~\cite{maiPilot2018,parkOptimizing2018} aiming at minimizing channel estimation MSE.
Nonorthogonal pilot reuse strategies have been proposed in~\cite{attarifar2018random,liuTabuSearchBased2020}.
JED algorithm natively enables the use of nonorthogonal pilot sequences as the data symbols are also used to estimate the channel matrix.

\subsubsection{Clustering for UEs and APs}
UE-centric and AP-centric clustering schemes that aim at reducing the backhaul data transfer have been studied in network MIMO \cite{binbindaiSparse2013} and rediscovered in cell-free massive MU-MIMO~\cite{buzzi2017cell}.
UE-centric clustering enables the UEs to communicate with only a few nearby APs, which has been studied in~\cite{buzziUserCentric2018,hanSparse2019,interdonatoUbiquitous2019,vanchien2020Joint,dandrea2020User,bjornsonScalable2020}.
AP-centric clustering only serves a few nearby UEs and has been studied in \cite{basharCellFree2018,basharMax2019,basharUplink2019}.
As mentioned above, clustering strategies for pilot reuse have been discussed in  \cite{doanPerformance2017,riera-palouClustered2018}. Clustering to facilitate channel estimation has been proposed in \cite{huangEfficient2020}.
In contrast, our approach only clusters AP and UE indices and dynamically constructs virtual cells in which we perform orthogonal channel training as interference among virtual cells is minimized---to minimize intra-virtual-cell interference, we propose to use mutually unbiased bases  \cite{durt2010mutually} as pilot sequences.

\subsection{Notation}
Lower case and upper case boldface letters denote matrices and vectors, respectively. We use $ A_{b,u} $,  $\bma_u$, and $ a_k $ to represent the entry in the $ b $th row and $ u $th column of the matrix $ \bA $, the $u$th column of the matrix $\bA$, and the $ k $th element of the vector~$ \bma $, respectively. 
We use $ \bI_M $, $ \bOne{L}{M} $, and $ \bZero{L}{M} $ for the $ M\times M $ identity, $ L\times M $ all-ones, and $ L\times M $ all-zeros matrix, respectively.
The superscripts~$^* $, $^T $, and $^H$ refer to the complex conjugate, transpose, and Hermitian transpose, respectively. 
For the matrices \bA and \bB, we define the real-valued inner product as $\innR{\bA,\bB} = \Re \{ \Tr (\bA^H\bB) \}$, where $\Re\{x\}$ extracts the real part of $x\in\complexset$ and $\Tr$ is the matrix trace.
For the vectors~$\bma$ and \bmb, we define $ \innR{\bma,\bmb}  = \Re \{ \bma^H\bmb \}$. Consequently, we have $ \innR{\bA,\bB} = \innR{\text{vec}(\bA),\text{vec}(\bB) } $, where $\text{vec}(\cdot)$ is the vectorization operator. 
The matrix operators $ \kron $ and $ \circ $ denote the Kronecker product and Hadamard product, respectively. 
The operator $\overset{e}{\geq}$ denotes element-wise larger-equal-than.
For a matrix $\bA$, we will use the following entrywise norms: $ \normfro{\bA} = \sqrt{\innR{\bA,\bA} }$,  $ \normone{\bA} = \sum_{b,u}\abs{A_{b,u}}$, and ${ \norminftilde \bA = \max_i  \abs{\tilde{a}_i} }$ with $ \tilde{\bma}  = [ \Re\left\lbrace \text{vec}(\bA)\right\rbrace^T , \Im\left\lbrace  \text{vec}(\bA)\right\rbrace^T  ]^T$.

\subsection{Paper Outline}
The rest of the paper is organized as follows. 
\fref{sec:sys model} introduces the system model. \fref{sec:JED} formulates the JED problem and details our FBS algorithm.
\fref{sec: initialization} proposes principled initialization schemes for the  nonconvex JED problem.
\fref{sec:perm} discusses UE and AP permutation based on CSI and on physical locations.
\fref{sec:numerical results} analyzes the computational complexity  and demonstrates the efficacy of our method via simulation results. We conclude in \fref{sec:conclusion}.
%
%
\section{Prerequisites}\label{sec:sys model}
We now introduce the cell-free massive MU-MIMO system and summarize the channel model.

\begin{figure}[t]
	\centering
	\includegraphics[width=0.8\linewidth]{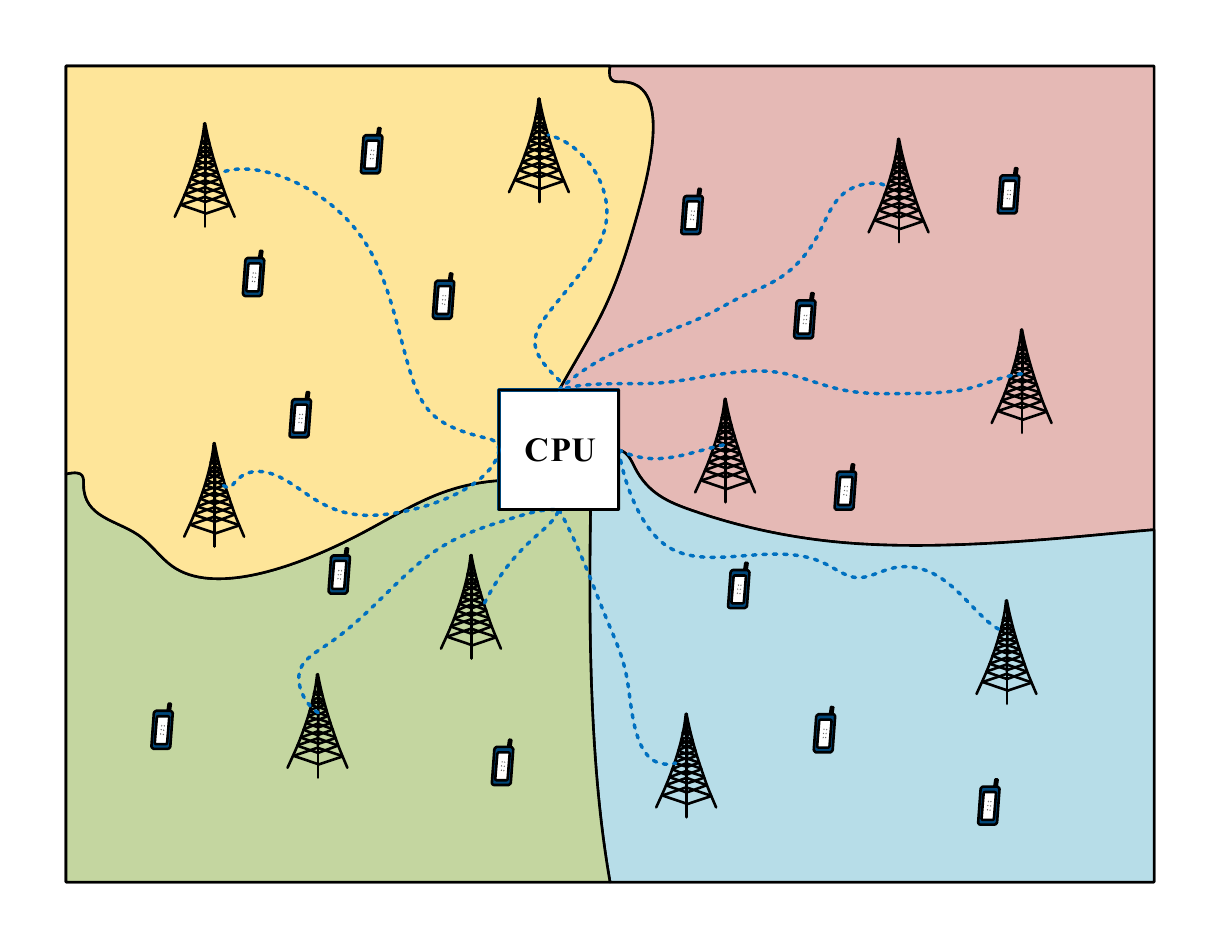}
	\caption{A cell-free massive MU-MIMO system where we dynamically construct so-called virtual cells with minimal inter-virtual-cell interference.}
	\label{fig:cfree mimo}
\end{figure}

\subsection{System Model}\label{sec: sys def}
We focus on the uplink in a cell-free massive MU-MIMO system with $ B $ distributed single-antenna APs and $U $ single-antenna UEs. 
As shown in \fref{fig:cfree mimo}, all the APs are connected to a central processing unit (CPU) via a backhaul network. Due to the distributed nature of cell-free massive MU-MIMO systems and the channel's sparsity, the area can be divided into virtual cells shown with different colors in \fref{fig:cfree mimo}. 
Each virtual cell will be constructed dynamically (i.e., dependent on the channel matrix or the physical UE/AP locations) to minimize inter-cell interference---this approach will be detailed in \fref{sec:perm}. 
We assume a block-fading scenario with TDD and a coherence time of $ K =T+D$ time slots, where $ T $ time slots are reserved for pilot-based channel training and~$ D $ time slots for payload data. 
The input-output relation of the considered frequency-flat\footnote{For frequency-selective channels, we can use orthogonal frequency-division multiplexing (OFDM) to obtain an equivalent system model per subcarrier. } cell-free massive MU-MIMO system is given by~\cite{gesbert2003theory}
\begin{align}\label{eq:mimo model}
\bY = \bH\bS + \bN,
\end{align}
where $ \bY\in \complexset^{B\times K} $ is the receive-signal matrix, $ \bH\in\complexset^{B\times U} $ is the MIMO channel matrix, $ \bS $ contains two parts and will be introduced below, and $ \bN \in \complexset^{B\times K}$ models noise, with entries assumed to be i.i.d.\ circularly-symmetric complex Gaussian with variance~$ N_0 $ per complex entry. 
To simplify notation, we separate training from payload by rewriting \fref{eq:mimo model} as follows:
\begin{align}\label{key}
[\bY_T,\bY_D ]= \bH \, [\bS_T,\bS_D] + \bN.
\end{align}
Here, the matrices $ \bS_T\in \complexset^{U\times T} $ and $ \bS_D\in \setQ^{U\times {D}}$ contain training pilots and data symbols, respectively; 
the pilot sequences $\bS_T$ are designed as tight frames (see \fref{sec:tight frames} for the details) and the entries of $ \bS_D$ are chosen from the constellation~$ \setQ $;
the matrices $ \bY_T \in \complexset^{B\times T}$ and $ \bY_D \in \complexset^{B\times D}$ contain the received pilot and data symbols, respectively. 
Our goal is to jointly estimate the channel matrix $\bH$ and detect the entries in $\bS_D$  from the received signals in~$ \bY$ and the known training-pilot matrix~$ \bS_T$.

\begin{figure*}[tp]
	\centering
	\subfigure[Arbitrary order (no permutation).\label{fig: chnl_none}]{\includegraphics[width=0.3\textwidth]{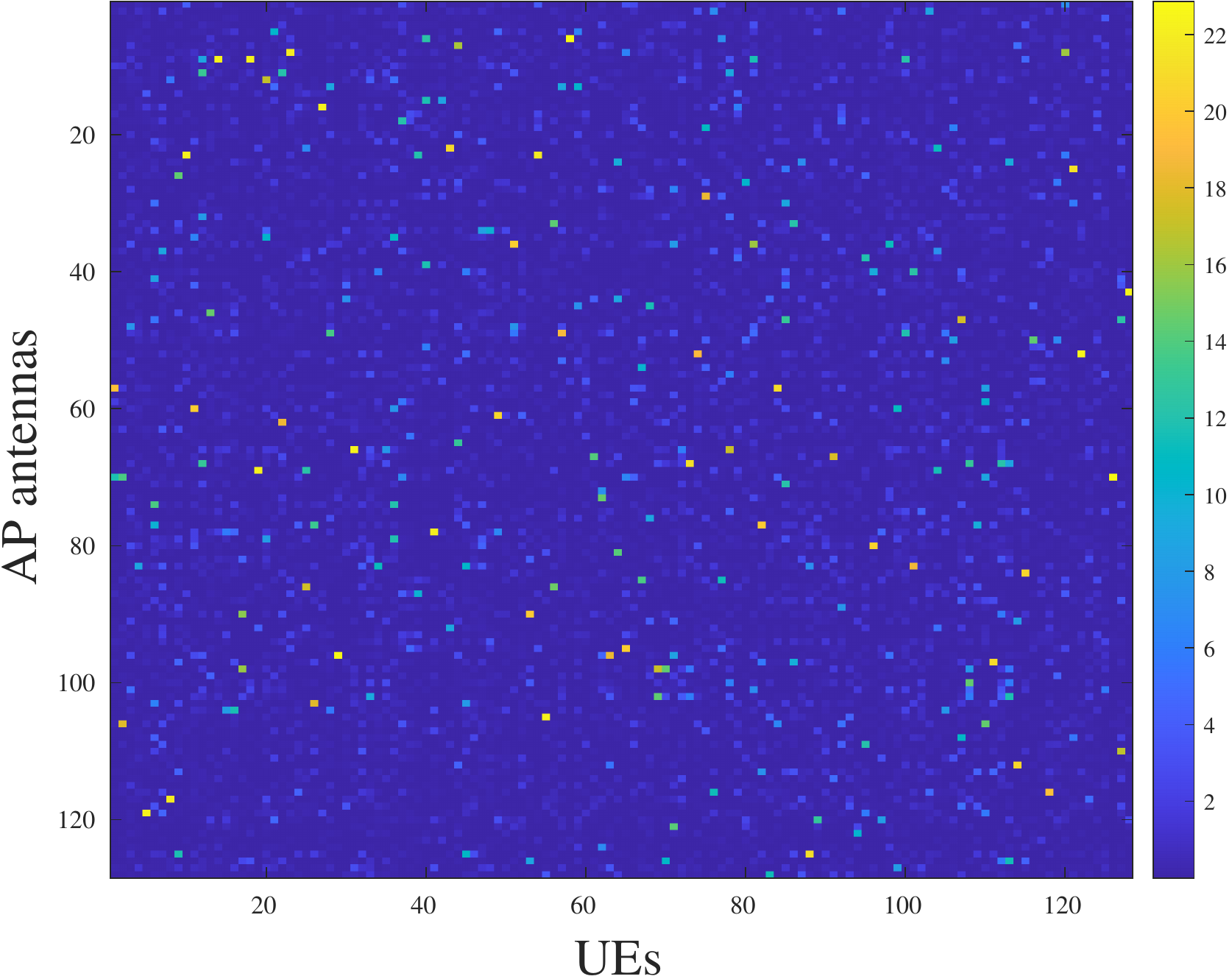}}
	\hspace{0.25cm}
	\subfigure[CSI-based permutation.\label{fig: chnl_csi}]{\includegraphics[width=0.3\textwidth]{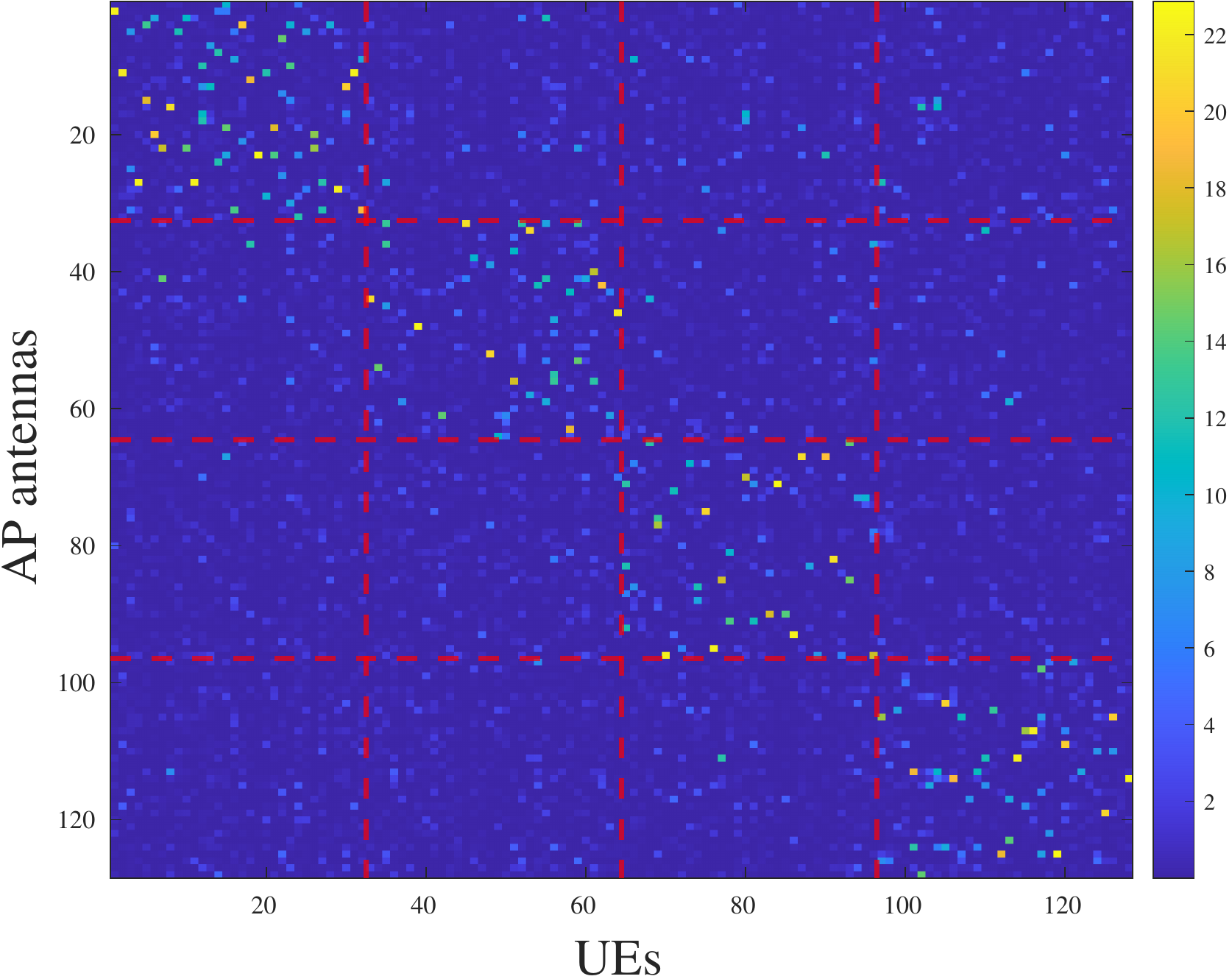}}
	\hspace{0.25cm}
	\subfigure[Location-based permutation.\label{fig: chnl_phy}]{\includegraphics[width=0.3\textwidth]{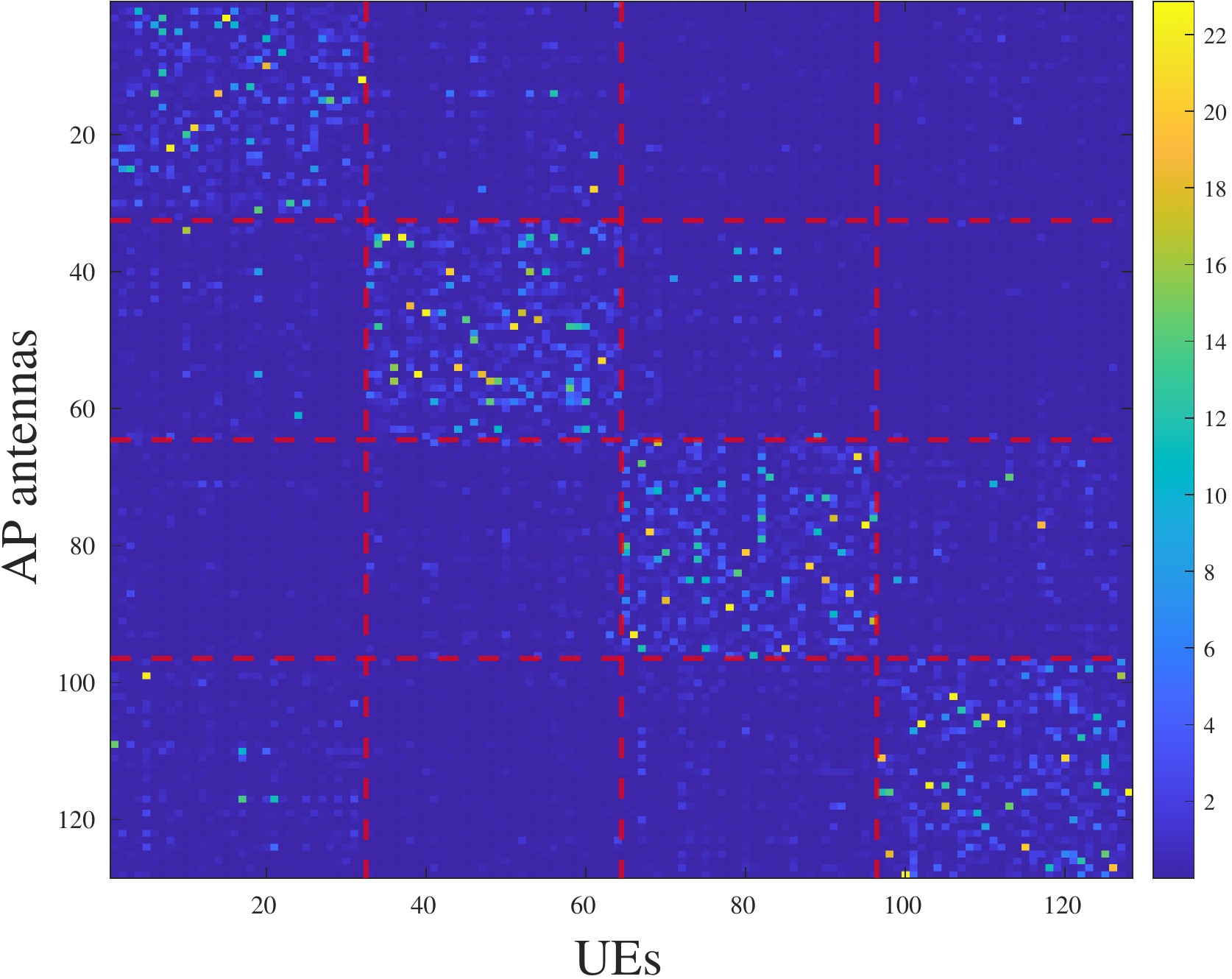}}\\
	\vspace{-0.1cm}
	\caption{Entry-wise visualization  of  $ \abs{H_{b,u}}  $  from a  cell-free massive MU-MIMO channel matrix: (a) The original (unpermuted) channel matrix; (b) a permuted channel matrix based on CSI; and (c) a permuted channel matrix based on physical locations. By reindexing APs and UEs at the CPU one can construct ``virtual cells," for which  inter-virtual-cell interference is minimized in order to simplify nonorthogonal channel estimation and improves spectral efficiency.}
	\label{fig:channel plots}
	\vspace{-0.2cm}
\end{figure*}

\subsection{Cell-free Massive MU-MIMO Channels}\label{sec:cell free model}
To develop a JED algorithm for cell-free massive MU-MIMO communication, we use the channel model put forward in~\cite{ngoCellfree2017} and consider single-antenna APs (see Remark~\ref{rem:multiantennaAPs} for a possible generalization to multi-antenna APs).
For this model, the channel matrix in~\fref{eq:mimo model} is decomposed as
\aln{
\bH=\sqrt{\rho_u}\bG\mathbf{\Lambda}.  \label{eq: channel matrix}
}
where $ \rho_u $ denotes the normalized uplink transmit signal-to-noise ratio (SNR), $ \bG \in\complexset^{B\times U}$ is the cell-free channel matrix, and $ \boldsymbol\Lambda \in  \reals^{U\times U}$ is a diagonal power control matrix.
The average power of each transmit symbol in $\bS$ is normalized so that $ \Exop \left[ |S_{u,k}|^2 \right] = 1$, and the entries of $ \bN $ are normalized so that $\No=1$.
Following the model in~\cite{ngoCellfree2017}, the entries of $\bG$ are modeled as $ G_{b,u}=\sqrt{\beta_{b,u}}\theta_{b,u} $ where~$ \beta_{b,u} $ and $ \theta_{b,u} $ characterize large-scale and small-scale fading between the $ b $th receive antenna and the $ u $th UE, respectively. 
We assume $\theta_{b,u}\sim  \mathcal{CN}(0,1) $ and $\beta_{b,u}$ is detailed in \fref{sec:comp under cfree}.
The power control matrix~$ \boldsymbol\Lambda$ is used to attenuate the transmit symbols in~$\bS$ and we absorb its effect in the channel matrix $ \bH $.

Since  in cell-free massive MU-MIMO systems with random placement of UEs and APs, the UEs are only close to a few APs, most of the entries in \bG will be small---a central property which we will discuss further in \fref{sec:channel sparsity}. 
Since $ \Exop \left[ |S_{u,k}|^2 \right] = 1$, the total received power for the $ u $th UE is~$ \normtwo{\bmh_u}^2 $, which may vary substantially among UEs.
To this end, we use a per-UE power control scheme that limits the maximum receive power by restricting the transmission power of some UEs based on their channel condition.
Concretely, we set an upper limit on the received power 
such that UEs whose received power would exceed the limit have to transmit with lower power---weak users continue transmitting at their nominal power.
To achieve this goal, we define the entries of the diagonal power-control matrix $ \boldsymbol\Lambda=\mathrm{diag}(\lambda_1,\ldots,\lambda_U)$ as follows:
\begin{align}\label{eq:power control}
\lambda_u^2 &= \textstyle \min\!\left\lbrace 
\normtwo{\bmg_u}^2,  10^{\frac{P}{10}}\min_{u'=1,\ldots,U} \|\bmg_{u'}\|^2_2
\right\rbrace \!/  \normtwo{\bmg_u}^{2}.
\end{align}
Here, $ P $ defines the maximum dynamic range between the weakest and strongest UE received power in decibels. 
This power control scheme relies on the magnitudes of the channel matrix entries, which change only significantly at the timescale of large-scale fading. 
Hence, it is possible that APs transmit the power level to the UEs in the downlink phase and the UEs could back-off accordingly. 
While this power control scheme is merely to confine the dynamic range of the received signals, it does not fundamentally alter the channel's sparsity property discussed next.

\subsection{Channel Sparsity of Cell-free Massive MU-MIMO Systems}\label{sec:channel sparsity}
Due to the distributed and random placement of APs and UEs, each UE is likely to be close to only a few APs---this property causes most links to be weak and the channel matrix in  \fref{eq: channel matrix} to be sparse.
 \fref{fig: chnl_none} illustrates this key property, where we show the absolute values of a channel matrix for $ 128 $ APs and $ 128 $ UEs placed randomly in a $ 1 $\,km$ ^2 $ square area. 

Since the enumeration of APs and UEs is arbitrary, and only a few entries of the channel matrix contain most of the energy, it is key to realize that one can permute the rows (APs) and columns (UEs) of the channel matrix to approximate a block-diagonal structure, by merely re-indexing the UEs and APs from the CPU's viewpoint.
As an example,  \fref{fig: chnl_csi} and \fref{fig: chnl_phy} show block-diagonal structures that can be obtained by leveraging either channel-state information (CSI) or physical UE location, respectively. 
Interestingly, the UEs within each diagonal block will experience strong inter-UE interference, whereas UEs in different blocks will experience only little interference. Effectively, such clustering strategies create virtual cells, which can be used to perform orthogonal training within each virtual cell and nonorthogonal training among virtual cells where interference is minimized. See  \fref{sec:perm} for the details. 
\section{ Joint Channel Estimation and Data Detection}\label{sec:JED}
We now formulate the JED problem and then relax it so that it can be solved  approximately using FBS~\cite{goldsteinField2016}.  

\subsection{The MAP-JED Problem}\label{sec:objective setting}
Using Bayes' theorem and the assumption made in \fref{sec:sys model}, the channel matrix $ \bH $ and data matrix $ \bS_D $ can be recovered jointly by maximizing the posterior probability density function (PDF) as follows:
\begin{align}\label{eq:MAP original}
\big\{\widehat \bH,\widehat\bS_D\big\} 
& = \!  \argmax_{\substack{\bH\in\complexset^{B\times U}\\\bS_D\in\setQ^{U \times D}} }  p(\bY|\bH,\bS_D)  p(\bH).	
\end{align}
Here, we assume that $ \bH $ and $ \bS_D $ are independent, and the entries of $ \bS_D $ are i.i.d. taken from the constellation set~$\setQ$. 
Since the entries  of $ \bN $ are assumed to be i.i.d. circularly-symmetric complex normal, the conditional PDF in \fref{eq:MAP original} is given by
\begin{align}
p(\bY|\bH,\bS_D) 
 = \dfrac{1}{\pi^{BK}\No}\exp\!\bigg(\! & - \frac{\normfro{\bY_T-\bH\bS_T}^2}{\No}   \notag \\
&  - \frac{\normfro{\bY_D-\bH\bS_D}^2}{\No} \bigg) . \label{eq:MAP pdf1}
\end{align}
Due to channel sparsity, we assume that the channel coefficients in \bH follow a sparsity-inducing complex-valued Laplace prior.
With the definition in \cite[Eq.~14]{pedersenSparse2015a} and the assumption that  the entries in \bH are i.i.d.,  the joint PDF is
\aln{
p\left( \bH  \right) = \left(\frac{\tilde\mu^2}{ 2\pi }\right)^{BU} \exp \left( -\tilde\mu\normone{\bH} \right), \quad \tilde\mu \in\reals_+. \label{eq:Laplace}
}
By inserting \fref{eq:MAP pdf1} and \fref{eq:Laplace} into \fref{eq:MAP original}, we obtain the following equivalent MAP-JED problem:
\begin{align}\label{eq:opt1}
\big\{\widehat \bH,\widehat\bS_D\big\} = \!  \argmin_{\substack{\bH\in\complexset^{B\times U}\\\bS_D\in\setQ^{U \times D}} } \textstyle \normfro{\bY-\bH[\bS_T,\bS_D]}^2+\mu\normone{\bH}\!.
\end{align}
Here, the parameter $\mu=\tilde\mu\No$ controls the channel's sparsity, where larger values promote sparser channel matrices.  
\begin{remark} \label{rem:multiantennaAPs}
For multi-antenna APs, we can generalize our problem formulation by leveraging block-sparsity~\cite{eldar2010Blocksparse}. This requires us to replace the Laplace prior in \fref{eq:Laplace} by $p(\bH) \propto \exp \left( \sum_{b=1}^{B}\sum_{u=1}^{U} \!-\tilde\mu \normtwo{\bmh_{b,u}}\right)$, where $\tilde\mu\in\reals_+$, $b\in\{1,2,\ldots,B\}$, and  $u\in\{1,2,\ldots,U\}$. Here, $\bmh_{b,u}\in \complexset^{N\times1}$ is the channel vector of UE $u$ to AP $b$, where $N$ is the number of antennas for each AP.
The permutation technique and the JED solver can also be adapted to this block-sparsity prior. 
While the multi-antenna AP case might be more practical, we stick to the single-antenna AP case for simplicity of exposition. 
\end{remark}

\begin{remark}
For cell-free massive MU-MIMO systems that have access to second-order statistics for each UE channel vector, as, e.g., in \cite{bjornsonScalable2020,bjornsonMaking2020}, one can adapt our JED problem formulation with a suitable Gaussian prior for $p(\bH)$ instead of the Laplace prior in  \fref{eq:Laplace}.
While our JED formulation in \fref{eq:opt1} requires only one hyperparameter (namely $\mu$), a MAP-JED method that exploits such second-order statistics would require additional pilot resources \cite{bjornsonScalable2020}. 
For the sake of brevity, a detailed comparison between the two approaches is left for future work.
\end{remark}
We note that JED in cell-free massive MU-MIMO systems is different from that in cellular massive MIMO systems for two reasons.
First, channel sparsity naturally arises in cell-free massive MU-MIMO channels as a result of APs and UEs placement in space. 
The collocated antennas at BSs in cellular massive MU-MIMO system results in approximately the same path loss, which eliminates channel sparsity.
The second reason is that channel sparsity in cell-free massive MU-MIMO systems enables us to group APs and UEs in such a way that the sparse channel matrices are approximately block diagonal (cf.~\fref{fig:channel plots}).
This structure enables us to deploy fewer pilots than UEs, while the UEs within each virtual cell (corresponding to a block in the block-diagonal matrix) can still be furnished with orthogonal pilots and UEs among virtual cells with near-orthogonal pilots. We will further detail this idea in \fref{sec: initialization}.

\subsection{Biconvex Relaxation of the JED Problem}\label{sec:biconvex relx}
We now provide means that enable us to approximately solve   the MAP-JED problem in  \fref{eq:opt1} with manageable complexity. 
We start by relaxing the discrete constellation set $\setQ$ to its convex hull, which is defined as \cite{castaneda2017vlsi} 
\begin{align}
\setC = \textstyle \left\{\sum_{i=1}^{|\setQ|}\delta_i q_i \mid (\delta_i\in\reals_+,\forall i) \wedge \sum_{i=1}^{|\setQ|}\delta_i =1 )\right\},
\end{align}
where $ q_i $ is the $ i $th symbol in $ \setQ $. 
Note that for QPSK with $\{\pm\sqrt{\frac12}\pm\sqrt{\frac{1}{2}} j\}$, the convex hull $\setC$ is a box around the four constellation points. 
This relaxation enables us to find solutions in a continuous region $\complexset^{U\times D}$ instead of a discrete set and has been  used recently for massive MIMO data detection which separates channel estimation from data detection~\cite{jeon16a,abbasi2019performance,shahabuddin2017admm}.

To improve the performance of the relaxed problem, we additionally use a strategy put forward in \cite{shah2016biconvex}. 
Intuitively, for QPSK, we are favoring solutions near the four corner points.
Thus, we add a concave regularizer $-\gamma\|\bS_D\|_F^2$ with parameter $\gamma\in\reals_+$ to the objective of the relaxed problem which pushes the solution towards the corners of the convex hull:
\begin{align}
\big\{\widehat \bH,\widehat\bS_D\big\} = \!  \argmin_{\substack{\bH\in\complexset^{B\times U}\\\bS_D\in\setC^{U \times D}} }   \textstyle \normfro{\bY\!-\!\bH[\bS_T,\bS_D]}^2  
+\mu\normone{\bH} \!- \textstyle \gamma \normfro{\bS_D}^2\!. \label{eq:opt3}
\end{align}
While the problem \fref{eq:opt3} remains nonconvex, the following lemma establishes conditions for which the problem is biconvex in $\bH$ and $\bS_D$. A short proof is given in \fref{app:biconvexity}.  
\begin{lemma} \label{lem:biconvexity}
The problem in \fref{eq:opt3} is biconvex in \bH and $ \bS_D $ if $ \lambda_\text{min}  \geq \gamma$, where $\lambda_\text{min}$ is the smallest eigenvalue of $\bH^H\bH $.
\end{lemma}

\subsection{Uniqueness of the JED Solution}\label{sec:ambiguity}
Since our goal is to simultaneously recover the channel matrix~$\bH$ and the data symbols in $\bS_D$, certain nonuniqueness issues of the solution may arise. We now show how such ambiguities can be avoided with suitable pilot matrices $\bS_T$. 
Define a diagonal phase-shift matrix $ \bD =  \mathrm{diag}\left( e^{j\phi_1}, e^{j\phi_2},  \cdots, e^{j\phi_U} \right)  $ with $ \phi_u\in[0,2\pi] $, $u=1,\ldots,U$,  and a permutation matrix $ \bP =  \left[ \bme_{\pi_{(1)}}, \bme_{\pi_{(2)}}, \cdots, \bme_{\pi_{(U)}} \right] $ where $ \bme_{\pi_{(u)}} $ is a standard basis vector, which is one in the $ \pi(u) $th entry and zero otherwise.
Let $ \big\{\widehat\bH,[ \bS_T,\widehat\bS_D]\big\} $ be a solution to \fref{eq:opt3}.
Then, the alternative tuple $\big\{\widehat\bH\bD^H\bP^T,\bP\bD [\bS_T, \widehat\bS_D] \big\}$  can also be a solution as it has exactly the same cost in \fref{eq:opt3} and satisfies the constraints, as long as the phase shifts $\phi_u$ satisfy  $e^{j\phi_u}s_T=s_T'$, where $s_T,s_T'\in\setQ$ are entries of the training matrix $ \bS_T $.
Such nonuniqueness issues have been studied in  \cite{dingSparsity2019,liuSuperresolution2019,ngoEVDbased2012,xueBlind2020,zhangBlind2018} and can be resolved in various ways.
Pilot-based systems with orthogonal pilots avoid such issues entirely. 
Since our goal is to undertrain channels with nonorthogonal pilots, uniqueness of a solution to \fref{eq:opt3} is no longer guaranteed. 
We now provide a simple condition for which no phase-permutation ambiguity can arise. A short proof is given in \fref{app:phase-permutation-ambiguity}. 
\begin{lemma} \label{lem:phasepermutationambiguity}
Fix a pilot matrix $ \bS_T = \sqrt{U}\bF^H $, where \bF has normalized columns so that ${\normtwo{\bmf_b}^2 = \nu},\, b =1,\ldots,B $.
Let 
\aln{
\kappa = \max_{b\neq b'} \, \abs{ \bmf_{b}^H\bmf_{b'} } \label{eq:coherence}
} 
be the \emph{coherence} of the matrix \bF. If $ \kappa <\nu $, then no phase-permutation ambiguity can exist. 
\end{lemma}
In \fref{sec:tight frames}, we provide pilot matrices that avoid the phase-permutation ambiguity and enable accurate channel estimation even in heavily undertrained systems.

\subsection{JED via Forward-Backward Splitting}\label{sec:FASTA}
We now show an FBS-based approach to approximately solve the problem in \fref{eq:opt3} at low complexity.
Due to the nonconvex nature of \fref{eq:opt3}, FBS is not guaranteed to find an optimal solution. 
Nevertheless, we show in \fref{sec:convergence} that FBS is guaranteed to converge to a stationary point with a proper stepsize. 
Furthermore, we show in \fref{sec:numerical results} that our algorithm performs well for various performance metrics. 

FBS is an efficient numerical method to iteratively solve convex optimization problems of the following form \cite{goldsteinField2016}: 
\begin{align} \label{eq:fbsproblem}
\hat\bmx = \argmin_{\bmx} f(\bmx)+g(\bmx).
\end{align}
Here, the function $f$ is differentiable and convex, and $ g $ is a more general (not necessarily smooth or bounded) convex function. 
Given a non-analytic function $ f:\complexset^{M} \rightarrow\reals$, we use the Wirtinger derivatives \cite{hjorungnes11a} to define the gradient.
To this end, after initializing the algorithm with $\bmx^{(1)}$, FBS solves the problem in~\fref{eq:fbsproblem} via iterations $  t=1,2,\ldots $ by computing 
\begin{align}\label{eq:fbs alg}
\bmx^{(t+1)} = \text{prox}_g\!\left(\bmx^{ (t) }- \tau^{(t)} \nabla f( \bmx^{* (t) } ); \tau^{(t)} \right)\!,
\end{align}
where $ \nabla f( \bmx^{* (t) }) $ is the gradient of $ f $ with respect to $ \bmx^{*(t)} $, and $ \tau^{(t)}$ is a per-iteration stepsize.
We use the adaptive stepsize selection proposed in~\cite[Sec. 4.1]{goldsteinField2016} to accelerate the convergence.
The proximal operator for~$ g(\bmx) $ is defined as
\begin{align}\label{eq:proximal def}
\text{prox}_g(\bmz;\tau) = \argmin_\bmx \, \tau g(\bmx) + \textstyle \frac{1}{2}\normtwo{\bmx-\bmz}^2.
\end{align}

Instead of performing alternating optimization in~$\bH$ and~$\bS_D$, we use FBS to solve for both matrices simultaneously. 
We group the two matrices together by defining $ \bZ =  [ \bH^H\; \bS] ^H$, where $\bS=[\bS_T,\bS_D]$ and $\bS_T$ is known and fixed throughout the iterations; the matrices $\bH$ and $\bS_D$ contain the optimization variables. 
We define the functions $f$ and $g$ in \fref{eq:fbsproblem} as 
\begin{align}
f(\bZ) &= f(\bH,\bS_D) = \textstyle \frac{1}{2}\normfro{\bY-\bH\bS}^2, \label{eq:fterm}\\
g(\bZ) &= g(\bH,\bS_D)= \textstyle \mu\normone{\bH}-\frac{\gamma}{2}\normfro{\bS_D}^2+\chi_\setC \left( \bS_D\right)\!, \label{eq:gterm}
\end{align}
where we define the indicator function as
\begin{align}\label{eq:indicator}
\chi_\setC \left( \bS_D\right) =\left\{
\begin{array}{ll}
0 \qquad &\bS_D \in \setC^{U\times D} \\
\infty \qquad &\bS_D \notin \setC^{U\times D}.
\end{array}
\right.
\end{align}

With the above definitions, the objective function consisting of a sum of \fref{eq:fterm} and \fref{eq:gterm} is not analytic. 
That said, the objective function is not only dependent on the complex matrix \bZ, but also implicitly on $ \bZ^* $ and hence the quantities $ \frac{\partial f}{\partial \bZ} $ and $ \frac{\partial f}{\partial\bZ^*} $ are both gradients of $ f $ with respect to \bZ and~$ \bZ^* $~\cite{hjorungnes11a}. 
According to \cite[Eq.~4.49]{hjorungnes11a}, the steepest descent direction is simply~$ \frac{\partial f}{\partial\bZ^*}  $ and thus the complex-valued gradient of $f$ is given by 
\begin{align}\label{eq:gradient}
\nabla f(\bZ^*) = \frac{\partial f}{\partial\bZ^*} 
=\left[\begin{array}{c} \frac{\partial f}{\partial \bH^*}\\[0.1cm]  \frac{\partial f}{\partial\bS^T}\end{array}\right]
=\left[\begin{array}{c}  (\bH\bS-\bY)\bS^H \\[0.1cm]  (\bH\bS-\bY)^H\bH \end{array}\right]\!.
\end{align}

The proximal operator for $ \bH $ is given by
\aln{
\text{prox}_{g}(\bH;\tau^{(t)}) 
	= &  \argmin_{\bX} \Big\{ \textstyle  \tau^{(t)} \mu\normone{\bX} \notag  \\
	& \textstyle + \half \normfro{\bX-\bH}^2 \Big\}
	= \eta(\bH;\mu\tau^{(t)}),
}
where $ \eta(\bH;\mu\tau^{(t)}) $ is the shrinkage operator \cite{goldsteinField2016} defined as 
\begin{align}
\eta(H_{b,u};\mu\tau^{(t)}) = \frac{H_{b,u}}{|H_{b,u}|} \max \left\{ \abs{H_{b,u}}-\mu\tau^{(t)}, 0 \right\}\!,
\end{align}
where division and multiplication are interpreted entry-wise.
Here, $ \mu $ is the sparsity parameter, $\tau^{(t)}$ is the per-iteration stepsize, and we define $x/|x|=0$ for $x=0$. 

The proximal operator for  $ \bS_D $ can be derived from \fref{eq:proximal def} as 
\begin{align}
\text{prox}_{g}(\bS_D;\tau^{(t)}) 
= \argmin_{\bX\in\setC^{U\times D}} \textstyle -\frac{\tau^{(t)} \gamma}{2}\normfro{\bX}^2+\frac{1}{2}\normfro{\bX-\bS_D}^2, \label{eq: NC proximal S}
\end{align}
where we moved the indicator function in $g(\bZ)$ back to the constraint. 
By completing the square, the solution to \fref{eq: NC proximal S} is 
\begin{align}\label{eq:proximal equality}
	\text{prox}_{g}(\bS_D;\tau^{(t)}) = \textstyle \text{proj} _\setC\!\left(\frac{1}{1-\rho^{(t)}}\bS_D;\tau^{(t)}\right)\!,
\end{align}
where $ \rho^{(t)} = \tau^{(t)}\gamma$. Note, for stability we must choose $\tau^{(t)}$ to be small enough that  $ \rho^{(t)} \in [0,1)$.
The right-hand side proximal operator is the projection onto the convex hull $ \setC $. For complex-valued QAM constellations, each element of the projection is applied independently to real and imaginary parts as 
\begin{align}\label{eq:inf proximal}
\text{proj} _\setC(\Re\{X_{u,d}\}) &= \min\{\max\{\abs{\Re\{X_{u,d}\}},-\alpha\},\alpha\}\\
\text{proj} _\setC(\Im\{X_{u,d}\}) &= \min\{\max\{\abs{\Im\{X_{u,d}\}},-\alpha\},\alpha\},
\end{align}
where $d\in \left\{ 1,2,\ldots,D \right\}$ and $ \alpha $ defines the radius of the box around $\setQ$.

\subsection{Convergence of FBS for the JED Problem in \fref{eq:opt3}}\label{sec:convergence}
Since the problem in \fref{eq:opt3} is nonconvex, we now analyze the convergence properties of FBS which depend on the initial choice of the initialization variable $\bZ^{(1)}$.
For this reason, rather than identifying a specific stepsize $\taut$ to use, it is simpler to guarantee convergence when a simple backtracking line search is used~\cite{goldsteinField2016}. The following result is proven in~\fref{app:fbsproof}.
\begin{thm} \label{thm:monotone}
Let $ h (\bZ) = f(\bZ)+g(\bZ)$ be the objective function given by \fref{eq:fterm} and  \fref{eq:gterm}.
Suppose that the stepsizes $\taut$ of FBS are bounded away from zero, and selected small enough to satisfy the following backtracking line search condition:
\aln{
	f(\bZ\kp) \leq \, & 
		f(\bZ\ko) +  \innR{\bZ\kp - \bZ\ko, \nbfZt}  \notag \\
		& + \frac{1}{2\taut}\normfro{\bZ\kp - \bZ\ko}^2, \label{eq:linesearch}
}
where $ \nbfZt $ is the gradient for $ f $ at $ \bZ^{*(t)} $.
Then, the objective $h$ decreases monotonically, i.e.,  we have
	\aln{h(\bZ\kp) < h(\bZ\ko).}
In addition, if $\taut$ is further restricted to satisfy $\taut < 1/\gamma$, then the sequence of iterates converges.
\end{thm} 

Note that while FBS for solving \fref{eq:opt3} is guaranteed to converge if the stepsizes are chosen appropriately, it is not guaranteed to converge to an optimal solution.
We reiterate that our FBS solver only \emph{approximately} solves the formulated MAP-JED problem in \fref{eq:opt3} but our simulation results in \fref{sec:numerical results} demonstrate that it converges to excellent stationary points. Establishing stronger optimality guarantees is extremely challenging and left for future work.


\section{Initializing FBS-JED}\label{sec: initialization}
We now show methods to initialize our FBS-JED algorithm that improve performance and reduce complexity. 

\subsection{Pilot Sequence Design}\label{sec:tight frames}
One key aspect for JED is designing suitable pilot sequences, which is particularly important as we focus on undertraining the channel matrix with nonorthogonal pilots. 
Concretely, we will use pilot matrices with low coherence as defined in \fref{eq:coherence}.

As in \fref{lem:phasepermutationambiguity}, let the pilot matrix be $ \bS_T = \sqrt{U}\bF^H $, where~$\bF$ has unit-norm rows and normalized columns $  {\normtwo{\bmf_b}^2 = \nu = \frac{T}{U}},\, b =1,\ldots,B $.
A prominent instance of matrices with near-orthogonal columns are equiangular tight frames (ETFs)  \cite{troppDesigning2005}  for which all pairs of inner products, i.e., $\kappa=|\bmf^H_b\bmf_{b'}|$ for $b
\neq b'$ achieve the same coherence $\kappa= \frac{T}{U}\sqrt{\frac{U-T}{T(U-1)}} $, given by the Welch lower bound~\cite{welchLower1974}. 
Furthermore, ETFs have orthonormal rows, i.e., $ \bF\bF^H = \bI_T $. 
Another useful class of matrices with low-coherence are mutually unbiased bases (MUBs). MUBs have the following block structure $\bF = [\bF_1,\ldots,\bF_N]$, where $\bF_n\in\complexset^{T\times T}$ with $n=1,\ldots,N$ blocks. 
Besides having orthonormal rows, MUBs also have orthogonal blocks, i.e., $\bF_n^H\bF_n=\frac{T}{U}\bI_{\frac{U}{N}}$, $n=1,\ldots,N$, and the coherence between any two columns of two different blocks is $\kappa=\frac{T}{U}\sqrt{\frac{1}{T}}$.

From the perspective of sparse signal recovery, nonorthogonal pilots $ \bS_T\in\complexset^{U\times T} $ where $ U>T $ and noise may cause ambiguity and render estimation difficult. 
ETFs and MUBs have low coherence and thus enable stable recovery of sparse signals using $ \ell_1 $-norm penalty as shown in~\cite{troppJust2006}.
Even though MUBs have higher coherence than ETFs, the orthogonal structure within each block matrix $\bF_n$ is particularly helpful for block-permuted  channel matrices that form virtual cells (cf.~\fref{fig:channel plots}).
We refer the interested readers to \cite{troppDesigning2005,durt2010mutually} for detailed properties and construction steps of MUBs and ETFs.
Note that we use ETFs for the $\ell_1$-norm-based channel estimator~\fref{eq:l1 chEst} for unpermuted channels, and we use MUBs to estimate each block on the diagonal separately for permuted channels; see \fref{sec: chEst JS} for the details.

\subsection{Permutation of APs and UEs}\label{sec:permutation}
The distributed  nature of APs and UEs in cell-free massive MU-MIMO systems promotes sparsity in the channel matrix.
Given this property together with the fact that enumeration of APs and UEs is arbitrary, we can permute the rows and columns of the channel matrix to attain approximately block-diagonal structure (cf.~\fref{fig:channel plots}). 
Mathematically, we introduce two permutation matrices $\PAP$ and $\PUE$ to reformulate input-output relation as:
\aln{
	\PAP\bY = (\PAP\bH\PUE)( \PUE^T\bS) + \PAP\bN.
}
By defining $ \widetilde{\bY} = \PAP\bY$, $\widetilde{\bH} = \PAP\bH\PUE $, and $ [\widetilde{\bS}_T,\widetilde{\bS}_D] = \PUE^T\bS $, we see that all of the assumptions for $\bN,\bH,\bS_D$ in \fref{sec:sys model} and also the cost function in \fref{eq:opt3} are invariant to such permutation.
Note that all of the assumptions on $\bN$, $\bH$, and $\bS_D$ in \fref{sec:sys model} are invariant to such permutations, i.e., the problem \fref{eq:opt3} can be posed equivalently as 
\begin{align}
	\big\{\widehat \bH,\widehat\bS_D\big\} = \,  \argmin_{\substack{\widetilde{\bH}\in\complexset^{B\times U}\\ \widetilde{\bS}_D\in\setC^{U \times D}} } \Big\{
	 & \textstyle \normfro{\widetilde{\bY}-\widetilde{\bH}[\widetilde{\bS}_T,\widetilde{\bS}_D]}^2  \notag \\
	  & \textstyle+\mu\normone{\widetilde{\bH}} - \textstyle \gamma \normfro{\widetilde{\bS}_D}^2 \Big\}, \label{eq:opt4}
\end{align}
where $ \widetilde{\bY} = \PAP\bY$, $\widetilde{\bH} = \PAP\bH\PUE $, and $ [\widetilde{\bS}_T,\widetilde{\bS}_D] = \PUE^T\bS $. 
For the assumptions on $ \bH $ and~$ \bN $, the JED problem remains unaffected by such AP and UE permutations, and the i.i.d.\ Laplace distribution of each entry in $ \widetilde{ \bH} $ and the i.i.d.\ circular symmetry of the Gaussian noise also remains. 
In fact, the new variables $\widetilde{\bH}$ and $\widetilde{\bS}_D$ are merely two new matrices in $ \complexset^{B\times U} $ and $ \setC^{U \times D} $, respectively.
Nonetheless, the goal behind such  AP and UE permutations are to (i) assign suitable pilot sequences to the UEs and (ii) find better initializers for both $ \widetilde\bH $ and $ \widetilde\bS_D $, which matters as we   are solving a nonconvex problem using FBS.
By permuting the channel matrix to obtain approximately block-diagonal structures  as illustrated in \fref{fig: chnl_csi} and \fref{fig: chnl_phy}, we can separately perform channel estimation and data detection within these blocks, which we call ``virtual cells" in \fref{fig:channel plots}, with MUBs that approximately decouple the interference among both blocks.

We note that clustering methods in papers on UE-centric cell-free systems, e.g.,~\cite{bjornsonScalable2020}, typically form overlapping cells, which is cruicial for  the design of decentralized and scalable data detection methods. However, we do not perform our permutation approach in that way for two reasons.
First, our virtual cells only serve the purpose of assigning pilots to UEs and simplifying channel estimation. Overlapping cells would not help us in accomplishing this goal.
Second, the vectors in each sub-block of MUBs are orthogonal, while the vectors from different sub-blocks are correlated.
If we were to form overlapping cells, then the block-wise orthogonality of MUBs can no longer be exploited.

\subsection{Initialization of the Channel Matrix}\label{sec: chEst JS}
In our conference paper \cite{songMinimizing2020}, we used a least-square (LS) channel estimator to initialize \bH.
Here, we show that we can improve upon this approach using the permutation idea introduced above. 
Consider an example with two virtual cells in~$\widetilde{ \bH}$, where the input-output relation during the training phase is given by the following block structure:
\aln{
	\begin{bmatrix}
		\widetilde{\bY}_{T_1} \\ \widetilde{\bY}_{T_2} 
	\end{bmatrix} = 
	\begin{bmatrix}
		 \widetilde{\bH}_{11} & \widetilde{\bH}_{12} \\
		\widetilde{\bH}_{21} & \widetilde{\bH}_{22}
	\end{bmatrix} 
	\begin{bmatrix}
		\bT_1 \\ \bT_2
	\end{bmatrix}+
	\begin{bmatrix}
		\widetilde{\bN}_{T_1} \\ \widetilde{\bN}_{T_2}
	\end{bmatrix}\!.
\label{eq:perm IO}
}
Here, we use the permutation approach to create two virtual cells $\widetilde{ \bH}_{11} $ and $\widetilde{ \bH}_{22} $ whose entries are much stronger than those in the block-off-diagonal matrices $\widetilde{ \bH}_{12} $ and $\widetilde{ \bH}_{21} $. See \fref{fig: chnl_csi}  and \ref{fig: chnl_phy} for an illustration of four virtual cells.
Assume that the pilot matrix is constructed by an MUB with {$N=2$, where $ \bT_1\bT_1^H = \bT_2\bT_2^H = T\bI_{\frac{U}{N}} $ and the inter-cell correlation is $  \bT_1\bT_2^H =  \bT_2\bT_1^H =  \sqrt{T}\bOne{ \frac{U}{N} }{ \frac{U}{N} }  $.} 
Due to the facts that (i) the entries in the block-off-diagonal matrices are much weaker than the block-diagonal matrices and (ii) the coherence between~$ \bT_1 $ and~$ \bT_2 $ is low,  we can perform independent training within the two virtual cells, assuming the off-diagonal blocks are zero.

We now illustrate this approach for estimating $ \widetilde{ \bH}_{11} $; the case for $ \widetilde{ \bH}_{22} $ is analogous.
Since $ \bT_1 $ is orthogonal and the entries in $ \widetilde{\bH}_{12} $ are close to zero, we can perform least-squares (LS) channel estimation\footnote{We utilize LS channel estimation instead of the linear mean-square error (L-MMSE) estimator as we lack the necessary statistical knowledge of the interference caused by UEs from an adjacent virtual cell, i.e., the distribution of $\widetilde\bH_{12}\bT_{2}$ is unknown.}  
\aln{\label{eq:chest block}
	\widehat{\bH}_{11,\text{LS}} = \widetilde{\bY}_{T_1}\bT_1^{-1} =  \widetilde{\bH}_{11} + \bar \bN,
}
where $ \bar{\bN}  = \widetilde{\bN}_{T_1}\bT_1^{-1} + \widetilde{\bH}_{12}\bT_2\bT_1^{-1} $.
The property of MUBs helps to mitigate the noise $\widetilde{\bN}_{T_1}\bT_1^{-1}$ and the interference $\widetilde{\bH}_{12}\bT_2\bT_1^{-1}$ present in $ \bar{\bN} $.
To see this, recall that the entries in $ \widetilde{\bN}_{T1}  $ are i.i.d.\ circularly-symmetric complex Gaussian with variance $ N_0 =1 $.
Furthermore, we have $ \bT_1^{-1} = \frac{1}{T}\bT_1^H $.
Hence, the covariance and the interference of the noise after LS channel estimation 
are $\frac{1}{T}\bI_{\frac{B}{N}}$ and $\frac{1}{\sqrt{T}}\widetilde{ \bH}_{12} \, \bOne{\frac{U}{N}}{\frac{U}{N}}$, respectively.
The interference is small as long as the entries in $\widetilde{ \bH}_{12}$ are small. The presented permutation strategy is designed to ensure this property, i.e., inter-virtual-cell interference is minimized.  
Consequently, the use of MUBs for initial channel training is sensible for the following reasons: (i) Orthogonal pilots are used within each virtual cell and (ii) noise and inter-virtual-cell interference are further suppressed due to the incoherence between MUB blocks.
We reiterate that our clustering approach only improves initialization of our FBS algorithm---\fref{sec: complexity} shows that this approach results in low MSE and can significantly reduce our algorithm's complexity. The CPU still solves the JED problem in~\fref{eq:opt4} as a whole.
\subsection{James-Stein Estimator and Median Absolute Deviation}\label{sec:JS est}
To further reduce the MSE of initial channel estimation, we propose to use the James-Stein (JS) estimator \cite{jamesEstimation1992a}.
By treating~$ \bar\bN $ as a matrix consisting of circularly-symmetric complex Gaussian random entries with variance~$ \bar{N}_0 $ per complex  entry, we can improve the initial channel estimate as follows.  
Let $\hat{\bmh}_{\text{LS}} = \tilde{\bmh} +  \bar\bmn$ represent one column of $\widehat{\bH}_{11,\text{LS}} $ in \fref{eq:chest block}. Assuming that $ \bar\bmn $ is circularly-symmetric complex Gaussian with variance  $ \bar N_0 $, the complex-valued version of the JS estimator is given by 
\aln{
\hat \bmh_{\text{JS}} = \left(  1 - \frac{(B-1)\bar{N}_0}{\big\|\hat\bmh_{\text{LS}}\big\|^2}  \right)\! \hat \bmh_{\text{LS}},
	\label{eq: JS est}
}
which results in lower channel estimation MSE compared to the traditional LS estimator if $B>1$~\cite{jamesEstimation1992a}. 

The remaining piece of the puzzle is to identify the unknown variance~$\bar{N}_0$ of the noise and interference term $ \bar \bmn $.
Fortunately, reference \cite{gallyas-sanhueza2021Blind} recently provided a computationally efficient way to estimate the noise variance in systems where sparse signals are observed in complex Gaussian noise.
By exploiting the sparsity of $ \tilde{\bmh} $, we can estimate the noise variance $ \bar N_0 = \mathrm{median}\big( |\hat \bmh_{\text{LS}} |^2\big)/\log (2)$  \cite[Eq.~4]{gallyas-sanhueza2021Blind} where $ \mathrm{median}(\cdot)$ refers to the sample median and  $ |\hat \bmh_{\text{LS}} |^2$ refers to the entry-wise absolute value square of the vector $\hat \bmh_{\text{LS}}$.

\subsection{Initialization of the Data Matrix}\label{sec: init Sd}
We now show how to initialize the data matrix $ \bS_D $.
We first {define a vector~$ \bar \bmh $ that is the vectorized $ \widehat \bH $ and then compute
 $ \widehat N_{\bar \bmh} =  {\mathrm{median}(| \bar \bmh |^2)}/{\log (2)}$}
followed by L-MMSE estimation from the received payload data matrix~$ \bY_D $ as	$\widehat\bS_D = \left( \widehat\bH^H\widehat\bH+ \widehat N_{\widehat \bH} \bI_U  \right)^{-1} \widehat \bH^H \bY_D $
since $ \Exop \left[ |S_{u,k}|^2 \right] =1$.
We reiterate that the proposed methods to initialize the channel and data matrices can also improve the performance and complexity of other JED algorithms. 


\section{AP and UE Permutation}\label{sec:perm}
We now propose two algorithms that perform AP and UE permutation with the goal of constructing virtual cells. 

\subsection{CSI-based Channel Matrix Permutation}\label{sec:CSI perm}
We start by focusing on a CSI-based permutation approach. 
Our goal is to cluster the entries in the channel matrix $\bH$ into an approximately block-diagonal structure as shown in \fref{fig: chnl_csi}. 
To this end, we define the auxiliary matrix $ \bA \in \reals^{B\times U}$ where $ A_{b,u} \triangleq \abs{H_{b,u}}^2 $ for all $b,u$. 
Given $\bA$, we can permute its rows and columns by $\widetilde{\bA} = \PAP\bA\PUE$ where $\PAP$ and $\PUE$ are
{permutation matrices for the rows and columns of \bA.}  
Permuting $\bA$ into an approximately block-diagonal matrix can be formulated as a nonconvex optimization problem:
\begin{align}\label{eq:combinatorial problem}
	\underset{\substack{\PAP\in \Pi_B\\\PUE\in\Pi_U}}{\mathrm{maximize}} \,\,
	 \textstyle \bOne{1}{B} \big[ \bM_N \circ(\PAP\bA\PUE) \big] \bOne{U}{1}.
\end{align}
Here, $ \Pi_M $ is the set of all possible $ M\times M $ permutation matrices, {$ \bM_N \in \reals^{B\times U}$} is a mask which determines the structure of the permuted matrix $ \widetilde{\bA} = \bM_N \circ(\PAP\bA\PUE) $, and $ N $ indicates there are $ N $ virtual cells to be constructed. 
To arrive at an approximately block-diagonal structure with $ N $ virtual cells on the diagonal, we set $\bM$ to be block-diagonal with $N$ diagonal blocks  $\bOne{\frac{B}{N}}{\frac{U}{N}}$. 
In our simulations, we set $ N\in\{2,4\} $ for different modulation schemes which are divisible by $ B $ and $ U $.

Since the complexity of enumerating all possible pair of permutation matrices in \fref{eq:combinatorial problem} is prohibitive, we use a convexification method put forward in \cite{fogelConvex2015}.
Specifically, we relax the set of permutation matrices $ \Pi_M $ to the set of doubly-stochastic matrices:
\begin{align}
\setD_M = &\big\lbrace  \bX\in\reals^{M\times M}:\bX\overset{e}{\geq}0, \notag  \\
  & \quad \bX\bOne{M}{1}=\bOne{M}{1}, \bX^T \bOne{M}{1} = \bOne{M}1 \big\rbrace\!.\label{eq:doubly set}
\end{align}
While a solution in \fref{eq:doubly set} is not necessarily a pair of permutation matrices (as the entries may lie in the set $ [0,1] $), we use the technique from \cite{shah2016biconvex} to gently push the results to the corners of $\setD_B, \setD_U $. 
The resulting problem to solve therefore becomes 
\begin{align}
\underset{\substack{\PAP\in \setD_B\\\PUE\in\setD_U}}{\mathrm{minimize}} \, \Big\{\! &
-\textstyle \bOne{1}{B} \big[ \bM_N \circ(\PAP\bA\PUE) \big] \bOne{U}{1} \notag  \\
&  - \frac{\varrho}{2}\normfro{\PAP}^2 - \frac{\varrho}{2}\normfro{\PUE}^2 \Big\},	 \label{eq:penalty 1}
\end{align}
with the parameter $ \varrho\in\reals_{+} $.
Such a problem can be solved with FBS as well.
The gradient can be calculated by exploiting the property of Hadamard product.
The proximal can be solved with Douglas-Rachford splitting (DRS).
We omit the details of such a solver due to the lack of space.

\subsection{Physical Location-Based Channel Matrix Permutation}\label{sec:PHY perm}
The block-diagonal structure can also be attained by permuting the channel matrix using information on physical locations as shown in \fref{fig: chnl_phy}.
Given Euclidean distances, we group fixed APs into $N$ balanced-sized clusters where we assign all of the UEs accordingly.

To obtain balanced-sized clusters, we employ the algorithm put forward in \cite{tangOptimizing2019}, which consists of an assignment step and an update step for cluster grouping and centroid updating, respectively.
For AP clustering, we iterate the two steps until convergence, whereas we only run the assignment step once for UE clustering since the centroids are already obtained in the AP clustering.
In what follows, we only demonstrate how to formulate the AP assignment problem in the form that is solvable with FBS and DRS since the update step is obvious.

In order to minimize the overall Euclidean distance between APs and their corresponding centroids, the relaxed version of AP assignment problem can be formulated as
\aln{\label{eq: phy obj}
\underset{\bQ\in\tilde\setD_{BN}}{\mathrm{minimize}}\,\, \textstyle 
	\ones_{1\times B} \left( \bQ \circ \bD \right)\ones_{N\times 1} - \frac{\omega}{2} \normfro{\bQ}^2,
}
where  
\aln{
	\tilde\setD_{BN} = \big \lbrace & \bX\in\reals^{B\times N} : \bX\overset{e}{\geq}0, \notag \\ 
& \textstyle 
	\bX^T { \bOne{B}{1}}=\frac{B}{N}\bOne{N}{1}, 
	\bX { \bOne{N}{1}}=\bOne{B}{1} \big\rbrace.
}
Here, \bQ is the partition matrix where $ Q_{b,n} = 1 $ indicates the $ b $th AP belongs to the $ n $th cluster, 
$ \bD\in\reals^{B\times N} $ is the distance matrix consisting of the Euclidean distances between all the APs and their corresponding centroids,
and the second term with a parameter $ \omega \in \reals_{+} $ indicates that we are favoring solutions close to the corners of $ \tilde\setD_{BN} $.
Consequently, \fref{eq: phy obj} is basically a simplified version of \fref{eq:penalty 1}.

\section{Numerical Results}
\label{sec:numerical results}
We now demonstrate the efficacy of our JED algorithm. 
\subsection{Simulation Setup }\label{sec:comp under cfree}
We evaluate our algorithm with the cell-free channel model detailed in \fref{sec:cell free model} and consider a square area of $1$\,km$^2$ with $U=128$ randomly positioned UEs. 
%
As in our previous study~\cite{songMinimizing2020}, we assess the performance of BPSK, QPSK, 16-QAM with $ B=64 $, $ B=128 $, and $ B=256 $ randomly positioned APs, respectively.
The maximum UE transmission power is $100$\,mW and we use the \mbox{per-UE} power control with $P=12$\,dB discussed in \fref{sec:cell free model}. The carrier frequency is $1.9$\,GHz and the bandwidth $20$\,MHz.
The receive and UE antennas are at a height of $15$\,m and $1.65$\,m, respectively.
We use the three-slope path-loss model defined in \cite{tang2001mobile}. The small-scale fading and large-scale fading parameters between the $ b $th antenna and the $u$th UE are  $ \theta_{u,b}\sim \mathcal{CN}(0,1) $ and $ \beta_{u,b} =\text{PL}_{u,b}10^{\frac{\sigma_\text{sh}z_{u,b}}{10}}$ where $\text{PL}_{u,b}$ 
is the path loss, $\sigma_{\text{sh}}$ is $8$\,dB, and $ z_{u,b}\sim \mathcal{N}(0,1) $ is shadow fading with variance $\sigma_{\text{sh}}^2$.
We permute the channel matrices using CSI-based and physical locations-based methods shown in \fref{sec:perm}.
Pilots are tailored for different channels. 
We design $ \bS_T $ with ETFs and MUBs for unpermuted and permuted channels, respectively.
We perform the $\ell_1$-norm-based channel estimator (cf. \fref{eq:l1 chEst}) for unpermuted channels.
For permuted channels, we estimate channels in the way discussed in \fref{sec: initialization}.
\subsection{Performance Metrics and {Baseline Algorithms}}
In a cell-free massive MU-MIMO system, the UEs are experiencing different SNRs which prevents us from generating conventional BER vs. SNR plots.
Thus, we characterize the per-UE cumulative density function (CDF) for the RMSSE, BER, and channel estimation MSE to examine our algorithm's efficacy from different aspects.
Also, instead of providing a spectral efficiency (SE) analysis, which is difficult due to the nonlinearity of our JED algorithm and would require Gaussian codebooks instead of discrete transmit constellations (which is what our JED algorithm exploits), 
we numerically calculate the mutual information (MI) between the discrete transmit signals  and   soft-symbol estimates  generated by the data detector for each UE individually, and we show the resulting distribution. Our performance metrics are as follows.

\subsubsection{Per-UE BER}
We define the BER for the $ u $th UE as  $\textit{BER}_u = \frac{\varepsilon_u}{n_{q}\cdot D}$, where $\varepsilon_u$ is the total number of bit errors for UE $u$ over~$D$ payload data slots and $n_q$ is the number of bits per symbol. 
\subsubsection{Per-UE RMSSE}
We define the RMMSE for UE $ u $ over $D$ payload data slots  as 
\begin{align}
\textit{RMSSE}_u = \textstyle \sqrt{\frac{\sum_{k=1}^D \abs{[\widehat{\bS}_D]_{uk} -[\bS_D]_{uk}}^2 }{\sum_{k=1}^D \abs{[\bS_D]_{uk}}^2}},
\end{align}
where $[\widehat{\bS}_D]_{uk}$ and $[\bS_D]_{uk}$ denote the estimated and transmitted data symbols of the $u$th UE at time slot $k$, respectively. 
\subsubsection{Per-UE MI}
In the interest of a SE analysis, we numerically simulate the MI for each UE given the discrete input constellation \cite{shannon48a}.
The MI for the $ u $th UE is defined as
\aln{
\textit{MI}_u = \frac{K-T}{K}\left(\entj{\sdj} - \entj{\sdj| \bsdj}\right)\!.
}
Here, $ T $ and $ K $ are the pilot time slots and total time slots, respectively. The prefactor $(K-T)/K$ takes the pilot overhead into account and decreases the per-UE MI by the fraction of used pilots (as they do not carry any payload data). The quantities $ \sdj $ and $ \bsdj $ denote the transmitted and quantized estimated data symbols of the $ u $th UE over $ D $ data slots, 
$  \entj{\sdj} $ is the empirical source entropy of the  $ u $th UE over $ D $ data slots, 
and $ \entj{\sdj| \bsdj} $ is the empirical conditional entropy of the $ u $th UE over $ D $ data slots. 
Since the output of JED is continuous, we quantize the output of JED and numerically compute the empirical entropies. 
\subsubsection{Per-UE MSE}
We define the channel estimation MSE of the $u$th UE as $\textit{MSE}_u = \frac{1}{B} \Ex{}{\|\hat\bmh_u-\bmh_u\|_2^2}$,
where $ \hat\bmh_u $ and $ \bmh_u\ $ are the estimated and true channel vectors, respectively.
\subsubsection{CDFs}
By treating all of the above performance quantities as random variables, we use Monte-Carlo simulations to characterize their CDFs over multiple UE and antenna placements, noise realizations, and data transmissions. 
The fraction of Monte-Carlo trials for which the per-UE RMSSE was below $x$ is defined as ${\Pr[\textit{RMSSE}<x]}$; the quantities ${\Pr[\textit{BER}<x]}$ and  ${\Pr[\textit{MSE}<x]}$ are defined analogously. 
To ensure consistency among all performance metrics (i.e., good performance is indicated by a curve in the upper-left of the respective plot), we define the per-UE MI as $ {\Pr[\textit{MI}>x]} $ (which is technically a complementary CDF) on the y-axis and show the largest  per-UE MI value on the left-hand-side of the x-axis---this is in stark contrast to classical CDF plots for the Gaussian SE in the literature (see, e.g.,~\cite{ngoCellfree2017}).
We note that all of the above performance metrics come with their own shortcomings. We thus demonstrate the efficacy of our JED problem with all four metrics.

\subsubsection{Baseline Algorithms}
To characterize the performance of our JED algorithm, we introduce two baseline algorithms for comparison. The first one is the L-MMSE symbol detector defined in \fref{sec: init Sd}. 
To obtain a channel estimate for this detector, we employ the $\ell_1$-norm-based channel estimator to exploit the sparsity in cell-free massive MU-MIMO systems:
\begin{align}\label{eq:l1 chEst}
	\widehat{\bH}= \!  \argmin_{\substack{\bH\in\complexset^{B\times U}} } \textstyle \frac{1}{2}\normfro{\bY_T-\bH\bS_T}^2+\mu_1\normone{\bH}\!,
\end{align}
where we tune the sparsity parameter $\mu_1$ for each scenario.
The other benchmark is the 
single-input multiple-output (SIMO) lower bound,
which perfectly cancels MU interference in a genie-aided fashion~\cite{zhang2006non}.
Both baselines are simulated with permuted channels.
\begin{figure*}[tp]
	\centering
	\subfigure[RMSSE]{\includegraphics[width=0.22\textwidth]{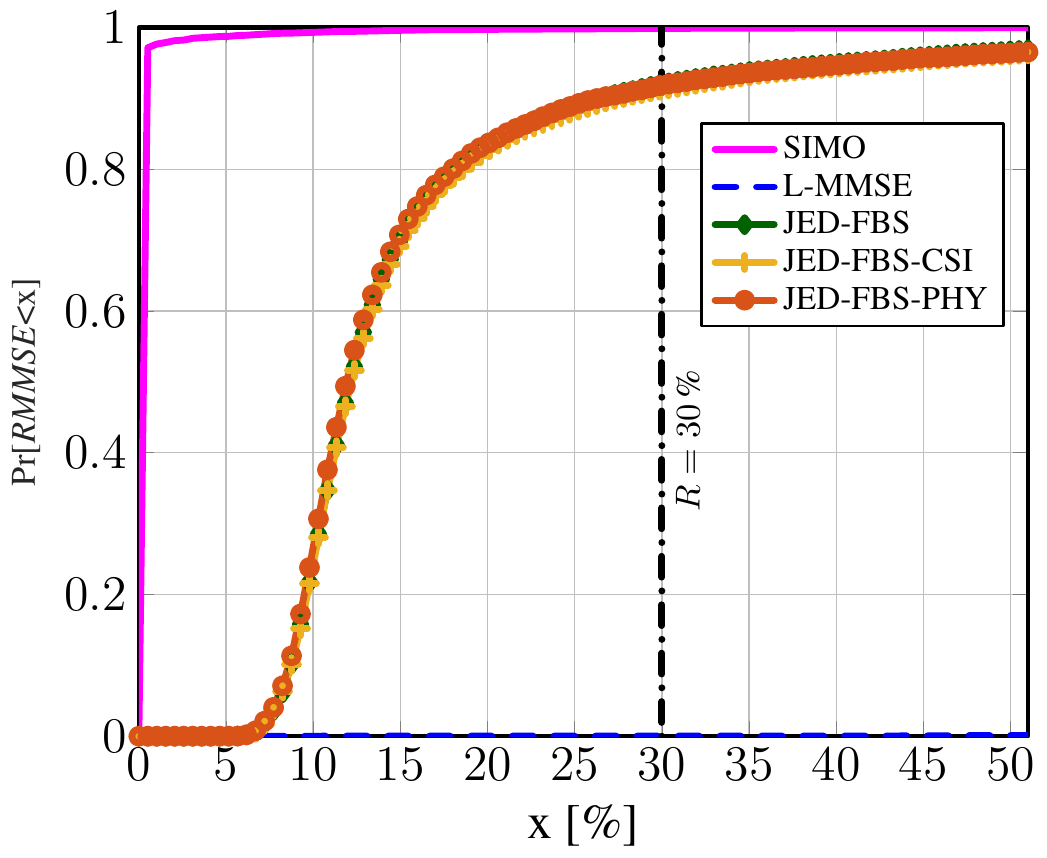}}
	\hspace{0.2cm}
	\subfigure[BER]{\includegraphics[width=0.22\textwidth]{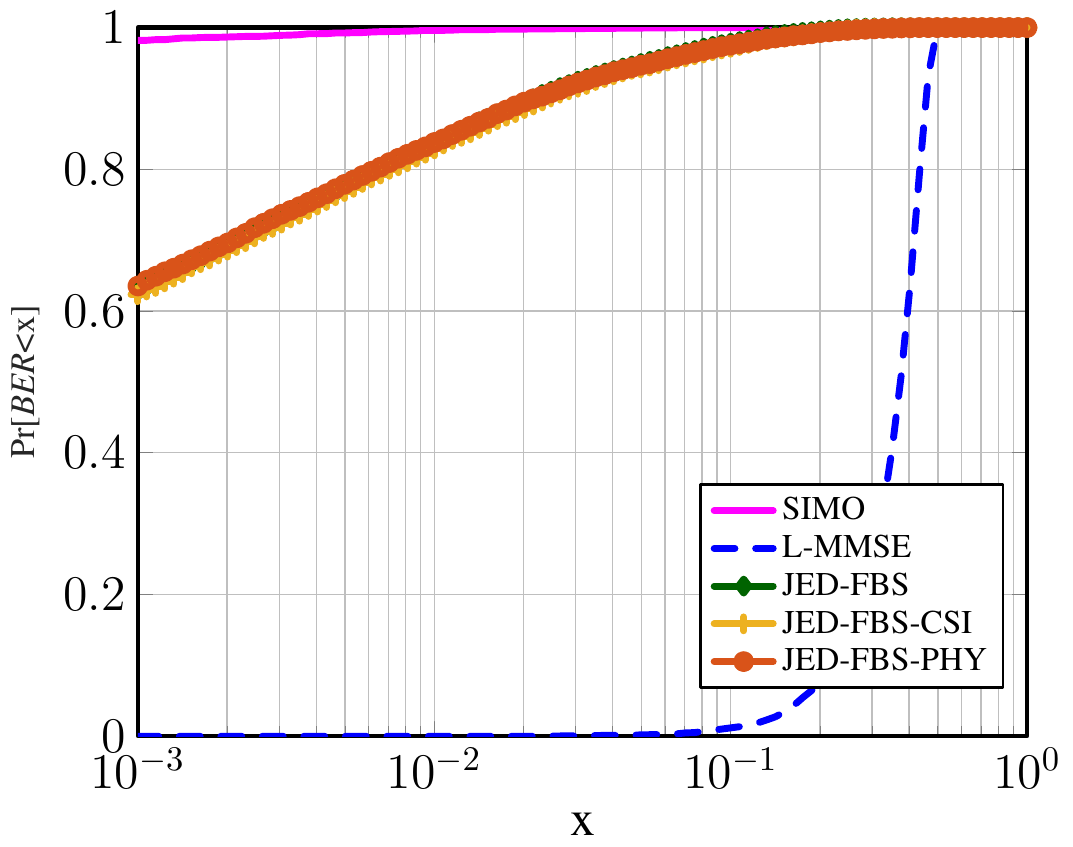}}
	\hspace{0.2cm}
	\subfigure[MI]{\includegraphics[width=0.22\textwidth]{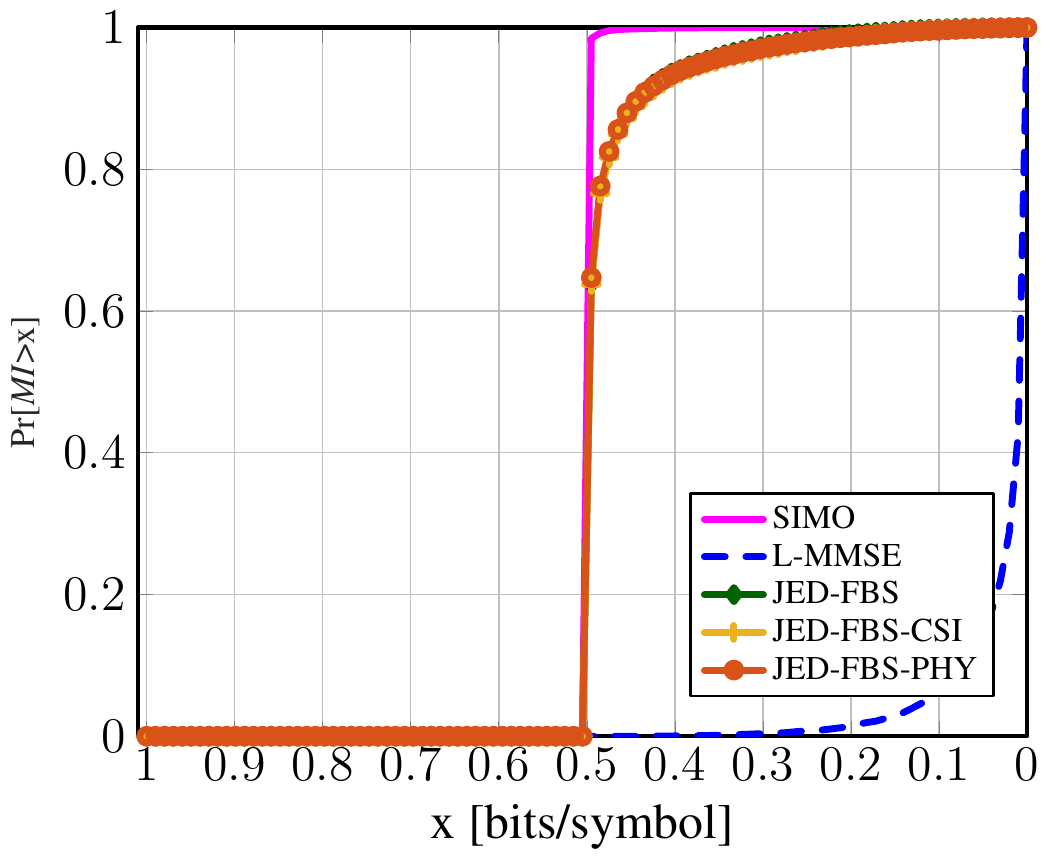}}
	\hspace{0.2cm}
	\subfigure[MSE]{\includegraphics[width=0.22\textwidth]{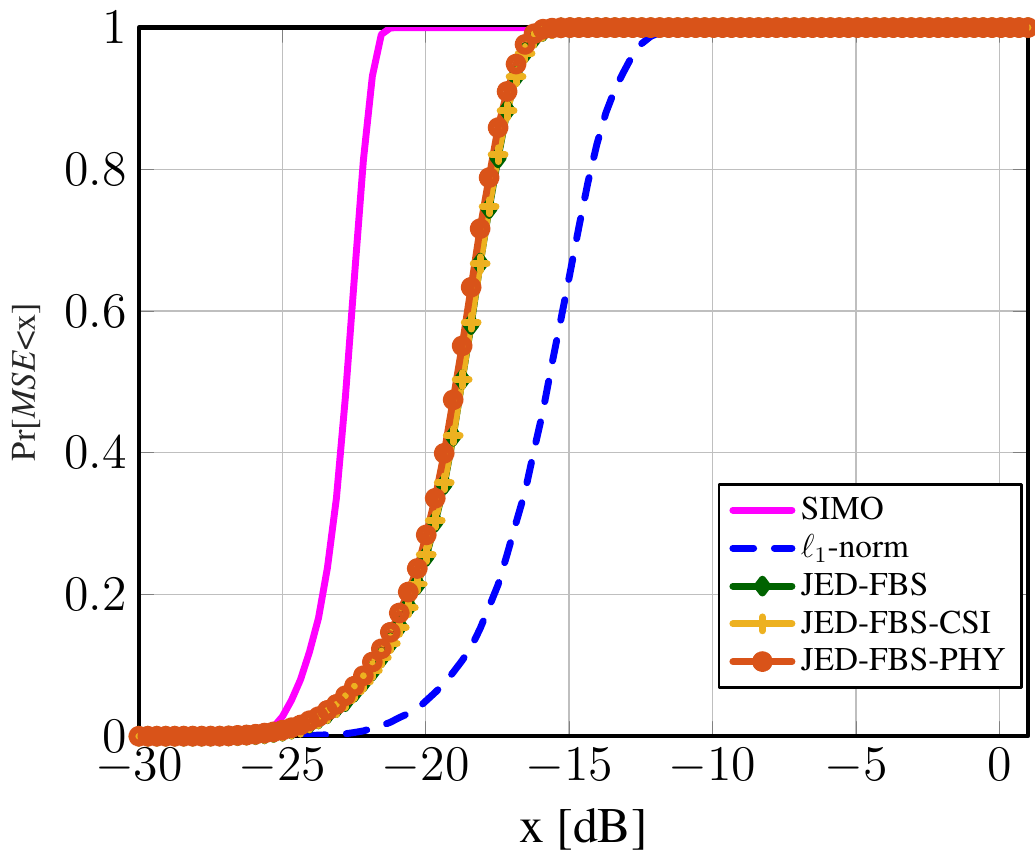}}\\
	\vspace{-0.1cm} 
	\caption{RMSSE (a), BER (b), MI (c), and MSE (d) performance for a cell-free massive MU-MIMO system with $B=64$ receive antennas, $U=128$ UEs transmitting BPSK, $K=128$ time slots, and $T=64$ ($ 50 $\%) nonorthogonal training symbols. {The permuted channels have two virtual cells which form two $ 32\times64 $ blocks on the diagonal.}
	{In such an overloaded system,} the proposed JED algorithm supports over $ 90 $\% of the UEs with an RMSSE of $30$\% {and over $ 60 $\% of the UEs with an uncoded BER of $ 10^{-3} $}; and enables $ 90 $\% of the UEs to achieve a per-UE MI of $ 0.43 $\,bits/symbol; $\ell_1$-norm training-based L-MMSE data detection fails completely.}
	\label{fig:BPSK_64x128x128}
	\vspace{-0.2cm}
\end{figure*}

\begin{figure*}[tp]
	\centering
	\subfigure[RMSSE]{\includegraphics[width=0.22\textwidth]{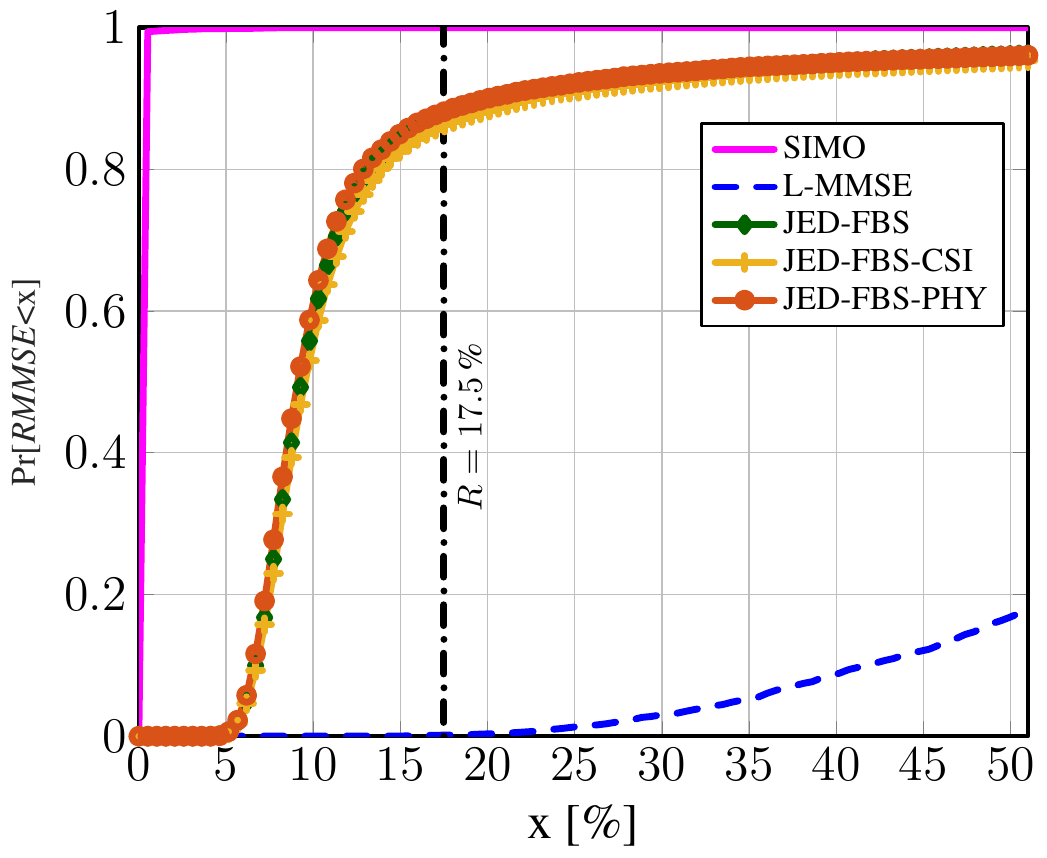}}
	\hspace{0.2cm}
	\subfigure[BER]{\includegraphics[width=0.22\textwidth]{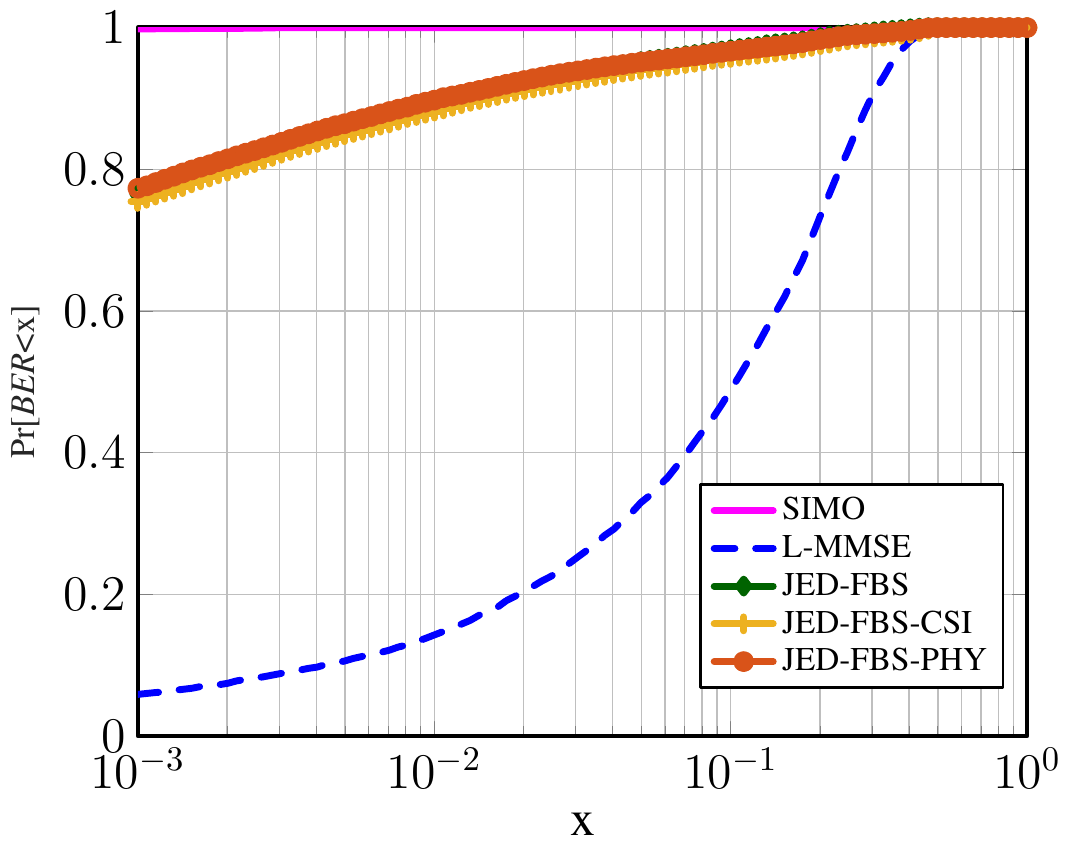}}
	\hspace{0.2cm}
	\subfigure[MI]{\includegraphics[width=0.22\textwidth]{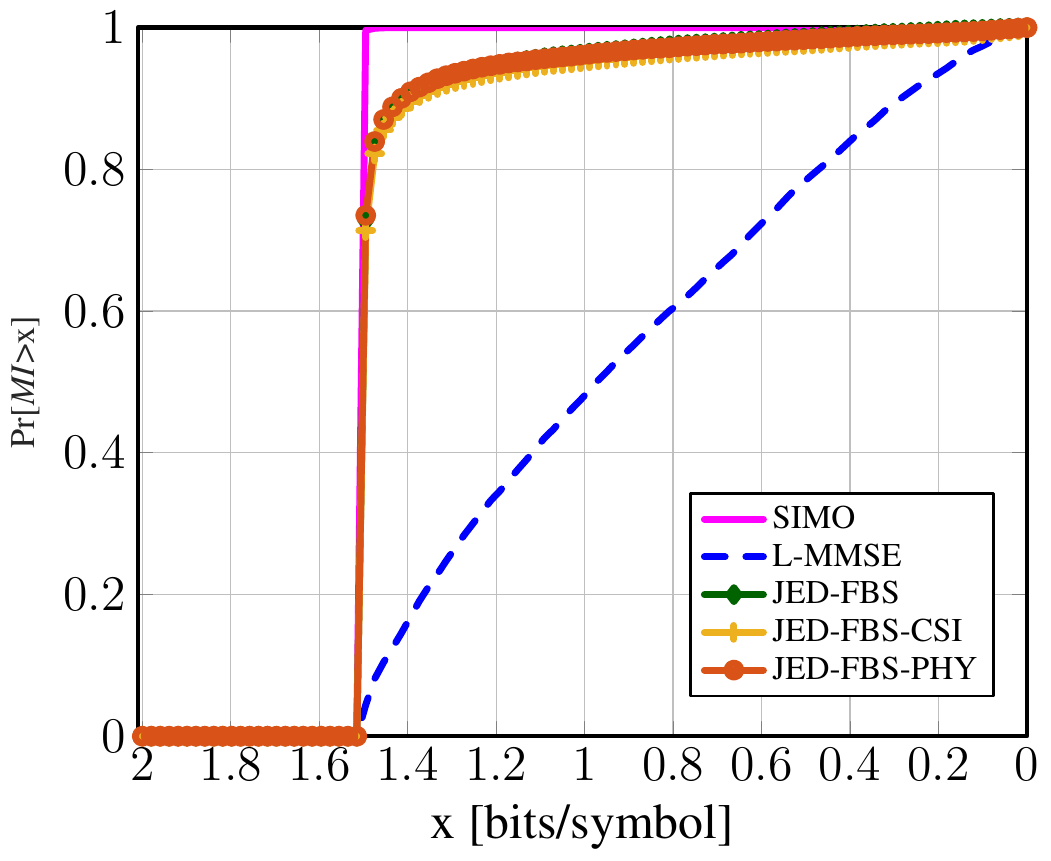}}
	\hspace{0.2cm}
	\subfigure[MSE]{\includegraphics[width=0.22\textwidth]{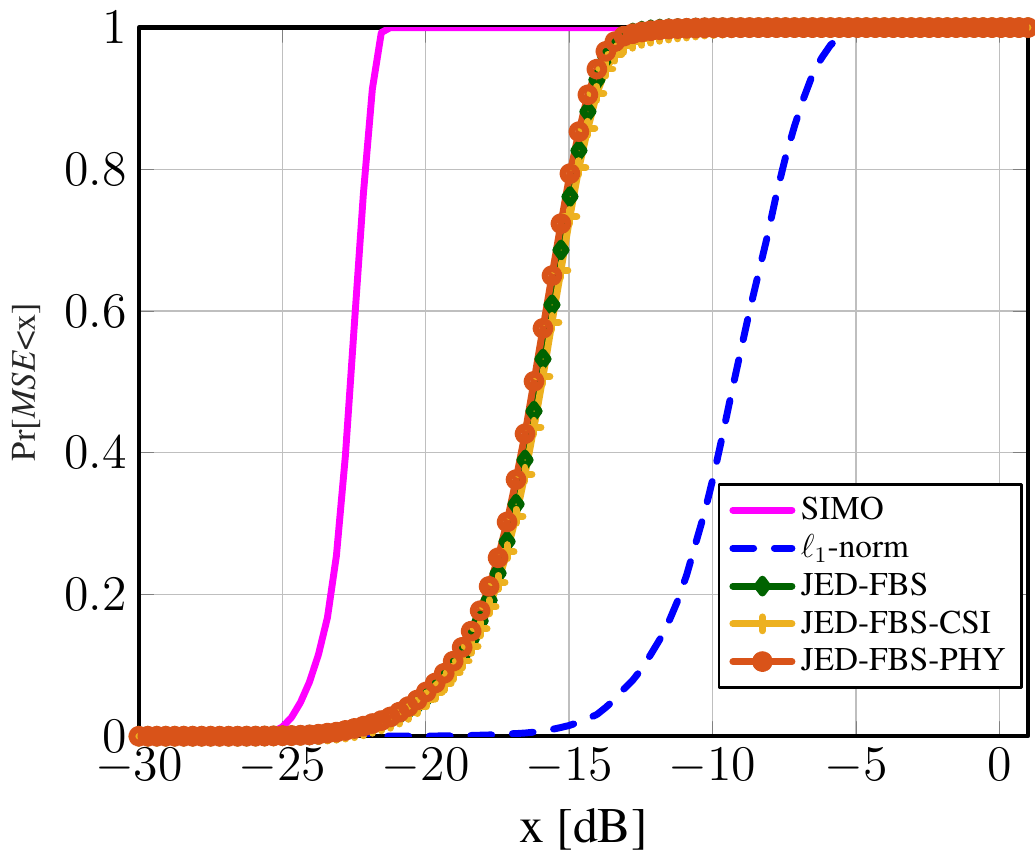}}\\
	\vspace{-0.1cm}
	\caption{RMSSE (a), BER (b), MI (c), and MSE (d) performance for a cell-free massive MU-MIMO system with $B=128$ receive antennas, $U=128$ UEs transmitting QPSK, $K=128$ time slots, and $T=32$ ($ 25 $\%) nonorthogonal training symbols.{The permuted channels have four virtual cells which form four $ 32\times 32 $ blocks on the diagonal.}
	{In such a fully-loaded system,} the proposed JED algorithm supports over {$ 88 $\%} of the UEs with an RMSSE of $17.5$\% {and nearly $ 80 $\% of the UEs with an uncoded BER of $ 10^{-3}$}; and enables $ 90 $\% of the UEs to achieve a per-UE MI of {$ 1.4 $}\,bits/symbol; $\ell_1$-norm training-based L-MMSE data detection is unable to achieve acceptable performance.}
	\label{fig:QPSK_128x128x128}
	\vspace{-0.2cm}
\end{figure*}

\begin{figure*}[tp]
	\centering
	\subfigure[RMSSE]{\includegraphics[width=0.22\textwidth]{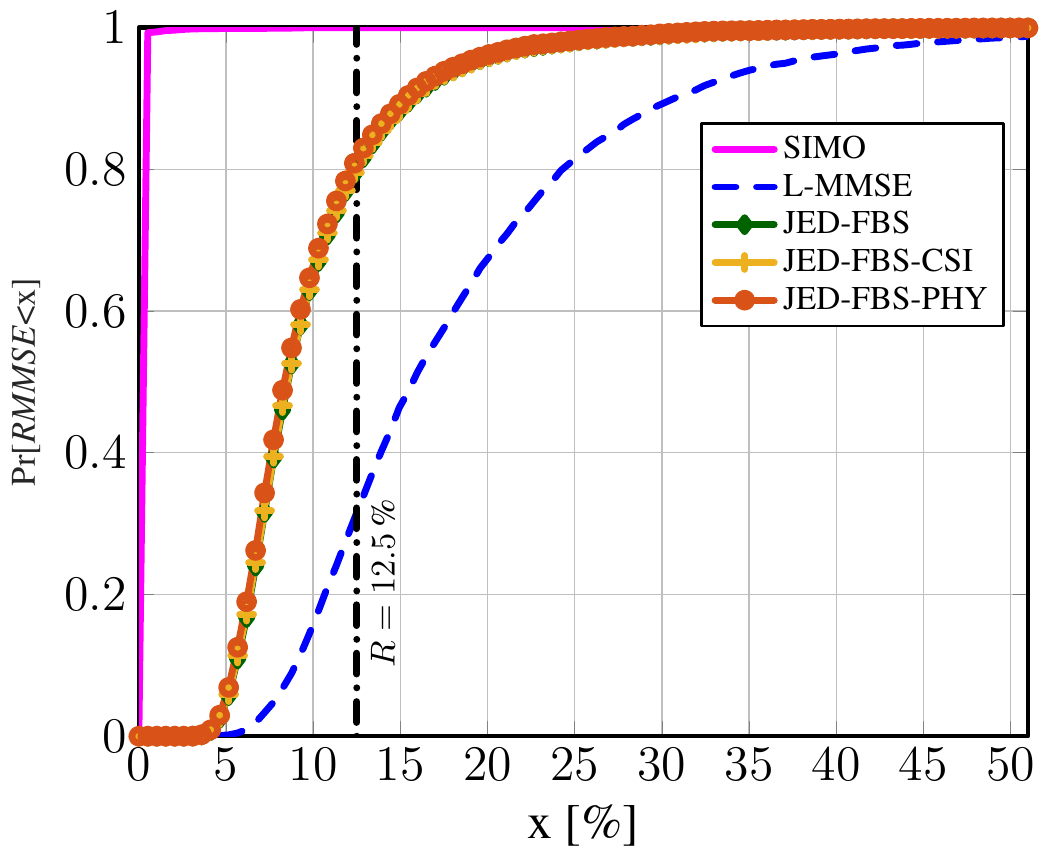}}
	\hspace{0.2cm}
	\subfigure[BER]{\includegraphics[width=0.22\textwidth]{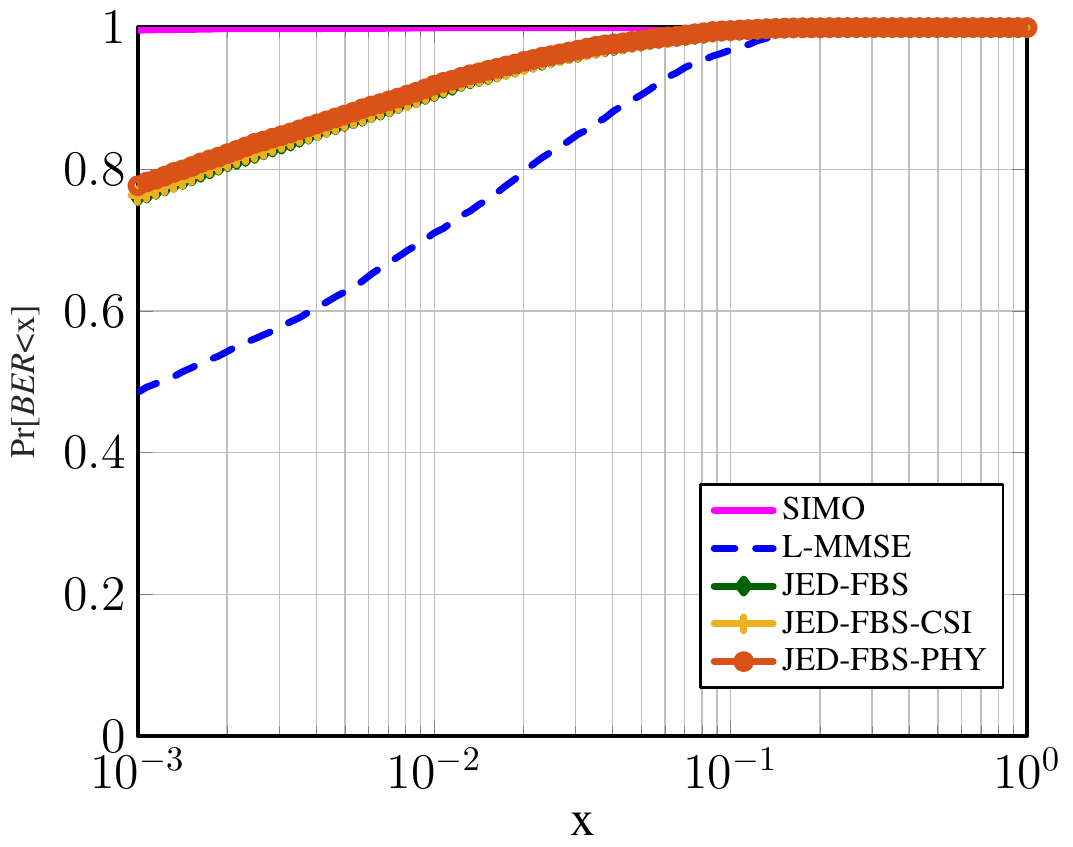}}
	\hspace{0.2cm}
	\subfigure[MI]{\includegraphics[width=0.22\textwidth]{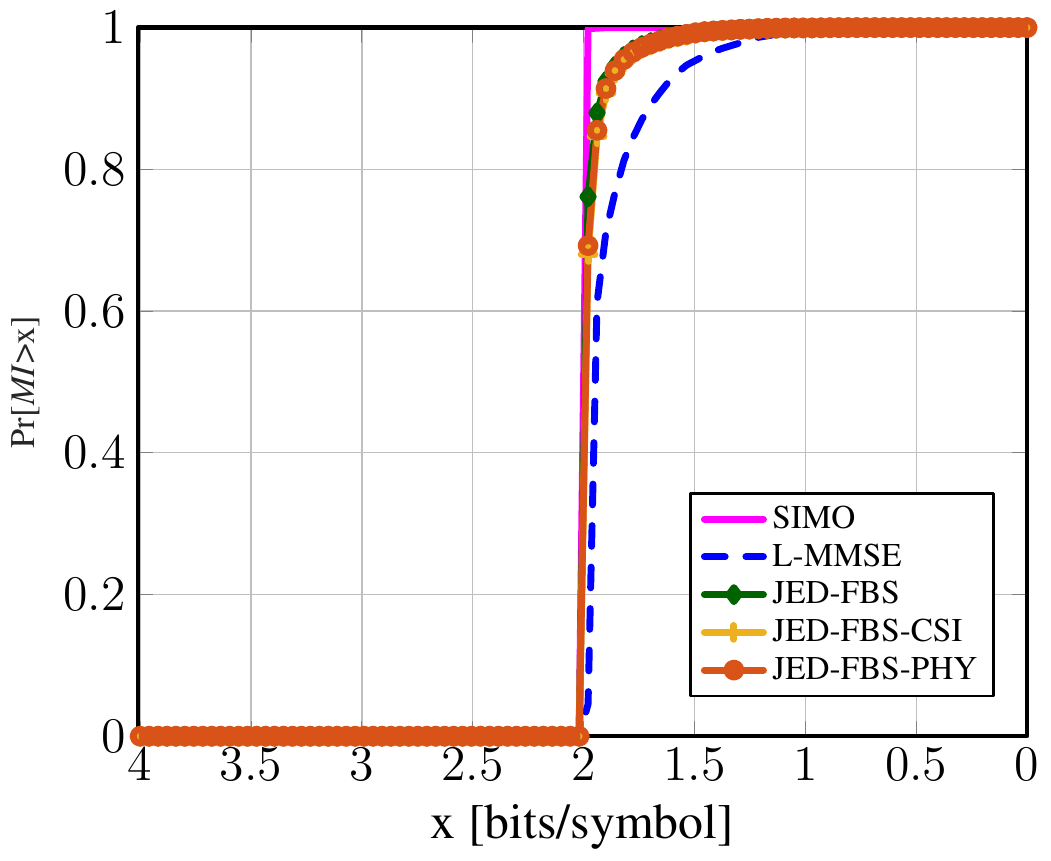}}
	\hspace{0.2cm}
	\subfigure[MSE]{\includegraphics[width=0.22\textwidth]{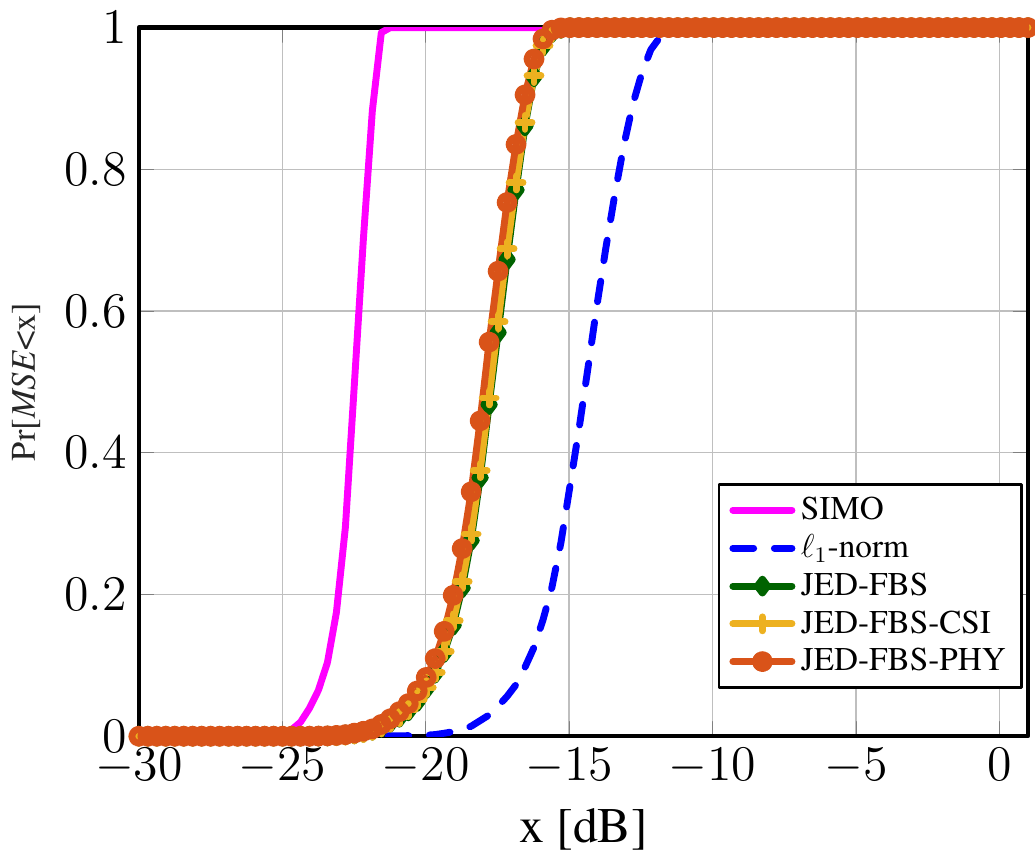}}\\
	\vspace{-0.1cm}
	\caption{RMSSE (a), BER (b), MI (c), and MSE (d) performance for a cell-free massive MU-MIMO system with $B=256$ receive antennas, $U=128$ UEs transmitting $ 16 $-QAM, $K=128$ time slots, and $T=64$ ($50$\%) nonorthogonal training symbols. {The permuted channels have two virtual cells which form two $ 128\times64 $ blocks on the diagonal.}
	{In a conventional massive MU-MIMO system,} the proposed JED algorithm supports over 80\% of the UEs with an RMSSE of $12.5$\% {and nearly $ 80 $\% of the UEs with an uncoded BER of $ 10^{-3}$}; and enables $ 90 $\% of the UEs to achieve a per-UE MI of $ 1.9 $\,bits/symbol; $\ell_1$-norm training-based L-MMSE data detection does not perform as well as JED.}
	\label{fig:16QAM_256x128x128}
	\vspace{-0.4cm}
\end{figure*}
\begin{figure*}[!ht]
	\minipage[t]{0.32\textwidth}
	\includegraphics[width=\linewidth]{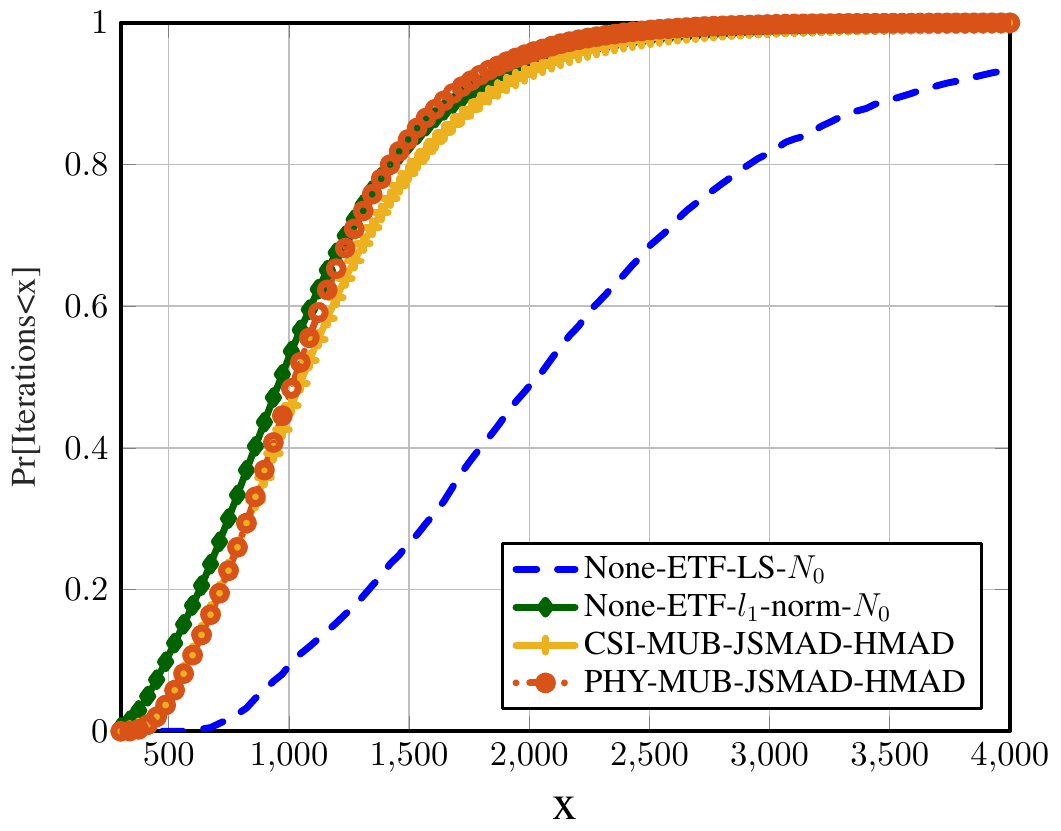}
	\caption{Numbers of iterations required for convergence of FBS with different initialization methods for a system same as \fref{fig:QPSK_128x128x128}. 
	In permuted channels, the proposed initialization techniques support $ 90 $\% of the UEs to converge at the same speed as the $ \ell_1 $-norm based benchmark.
	The FBS with naive LS channel estimation and L-MMSE data detection has the lowest convergence speed among all the cases, which requires $ 2\times $ more iterations.}
	\label{fig:QPSK_iterations}
	\endminipage\hfill
	\minipage[t]{0.32\textwidth}
	\includegraphics[width=\linewidth]{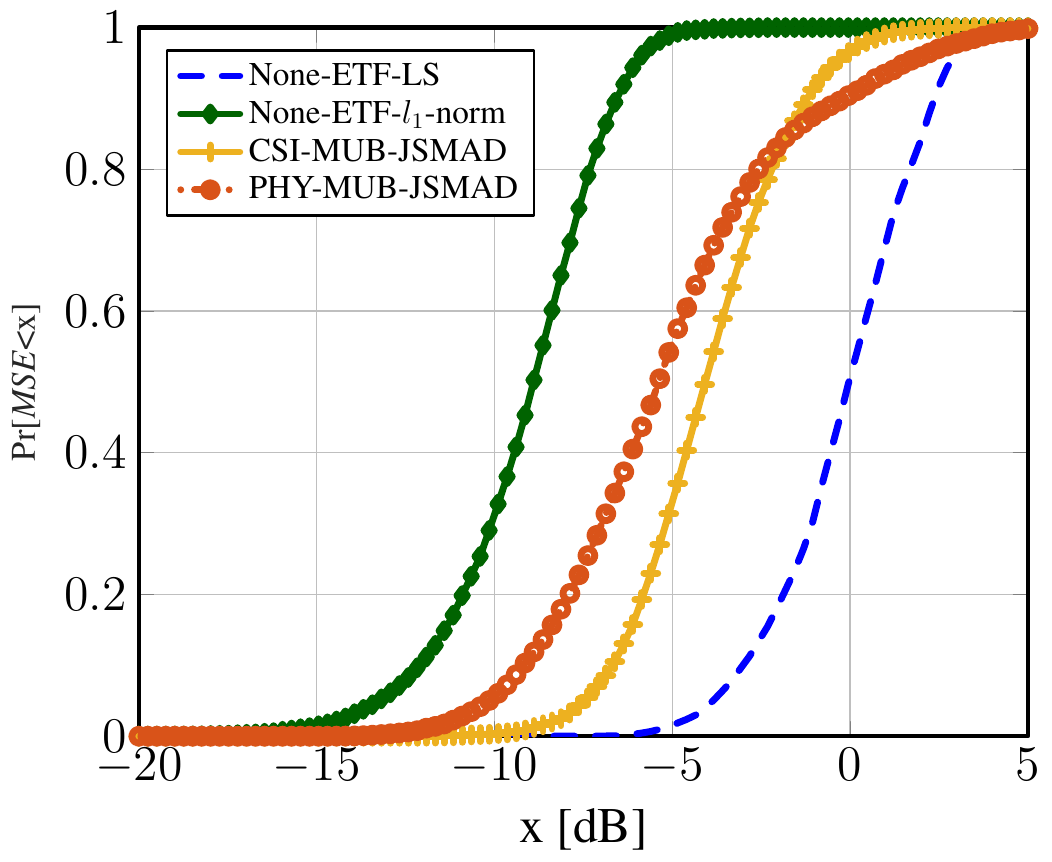}
	\caption{MSE of different initialization methods for a system same as \revision{in}~\fref{fig:QPSK_128x128x128}. 
	In unpermuted channels, the $ \ell_1 $-norm-based channel estimation method serves at the benchmark; LS channel estimation only achieves $ 2.4 $\,dB MSE for $ 90 $\% of the UEs.
	In permuted channels, our proposed initialization methods and the mutually unbiased bases (MUBs) together can support $ 90 $\% of the UEs to achieve $ -1.7 $\,dB and $ -2.47 $\,dB MSE, respectively.
	}
	\label{fig:complexity_reduction_CHEST_MSE}
	\endminipage\hfill
	\minipage[t]{0.32\textwidth}%
	\includegraphics[width=\linewidth]{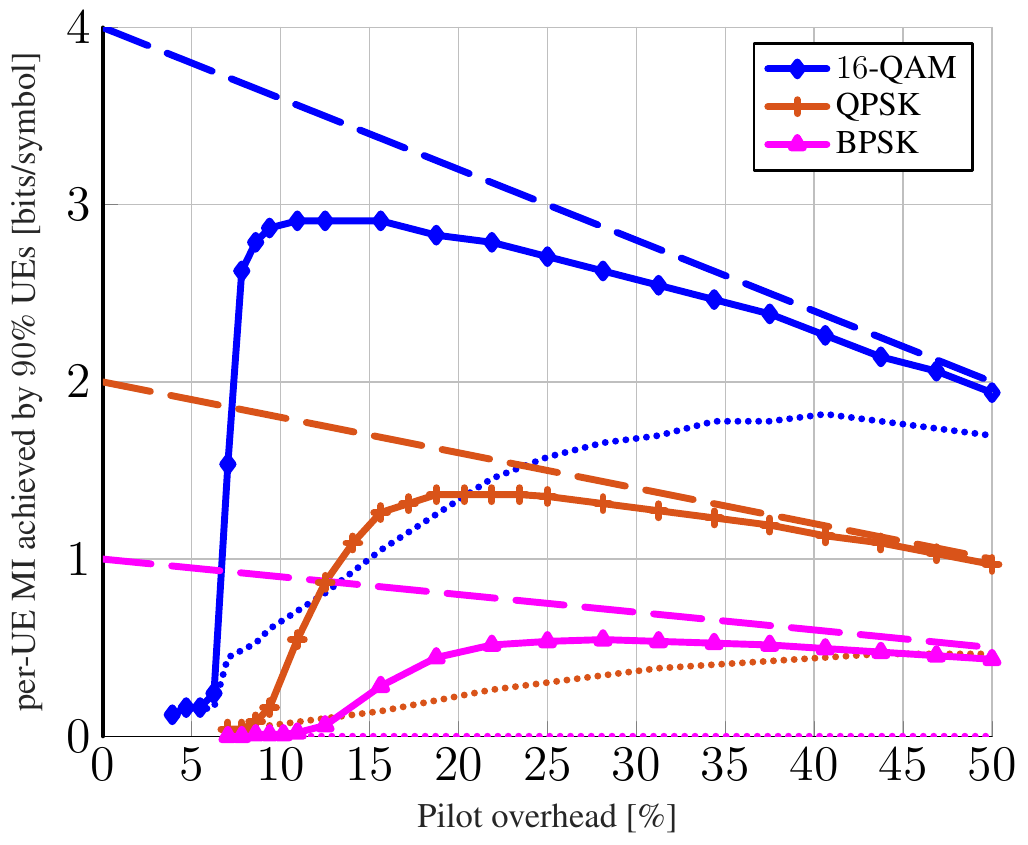}
	\caption{
	Trade-off between pilot overhead and the MI achieved by $ 90\% $ of UEs for three $ U=128 $ UE cell-free systems with $K=128$ time slots: (diamond) $B=256$ antennas with 16-QAM; (plus sign) $ B = 128 $ antennas with QPSK; (triangle) $ B=64 $ antennas with BPSK. Dotted lines show the trade-off of L-MMSE.
	We see that less training is required for massive MIMO systems whereas overloaded systems require more training.
	}
	\label{fig:CFJED_Training_TradOff}
	\endminipage
\end{figure*}

\subsection{RMSSE, BER,  MI, and MSE  Results}\label{sec:MI results}

We now {compare the performance of JED with unpermuted and permuted channel matrices to our  baseline algorithms.} 
Figure~\ref{fig:BPSK_64x128x128} shows simulation results for a $ B = 64 $ antenna system with $ U=128 $ UEs transmitting pilots and BPSK payload data over $ K=128 $ time slots, where $ T=64 $ ($ 50 $\%) pilots are used. The permuted channels have two virtual cells. 
Figure~\ref{fig:QPSK_128x128x128} shows simulation results for a $B=128$ antenna system with $U=128$ UEs transmitting pilots and QPSK payload data over $K=128$ time slots, where $T={ 32}$ ($ 25 $\%) pilots are used. The permuted channels have four virtual cells.
Figure~\ref{fig:16QAM_256x128x128} shows simulation results for a $B=256$ antenna system with $U=128$ UEs transmitting pilots and $ 16 $-QAM payload data over $K=128$ time slots, where $T= 64$ ($ 50 $\%) pilots are used. The permuted channels have two virtual cells.
Note that the differences among the three setups are the number of APs, modulation scheme, and the number of virtual cells.
We do not further investigate the case where the available time slots~$K$ is smaller than the number of UEs $U$; the interested readers are encouraged to simulate such cases using our code that will be made available on GitHub after possible acceptance of the paper.
To understand the performance in terms of RMSSE, it is instructive to compare the resulting RMSSEs to the {error vector magnitudes} (EVMs) allowed in UE implementation. The EVM characterizes the distortion caused by transmitter hardware on the digital constellation. The 3GPP 5G NR technical specification~\cite[Tbl.~6.4.2.1-1]{3gpp138-101} allows 
UEs to distort the BPSK, QPSK, and $16$-QAM constellations by an EVM of $30, 17.5$, and $12.5$\,\%, respectively.  
In \fref{fig:BPSK_64x128x128}(a), \fref{fig:QPSK_128x128x128}(a) and \fref{fig:16QAM_256x128x128}(a), we see that JED enables more than {$ 90 $\%, $ 88 $\% and $ 80 $\%} of the UEs to have an RMSSE at the receiver that is {\it smaller} than the EVM allowed to UE hardware. 
In \fref{fig:BPSK_64x128x128}(b), \fref{fig:QPSK_128x128x128}(b) and \fref{fig:16QAM_256x128x128}(b), we see that JED enables an $ 10^{-3} $ uncoded BER for $ 60 $\%, $78$\%, and $ 76 $\% of the UEs for BPSK, QPSK, and 16-QAM, respectively. 
In~\fref{fig:BPSK_64x128x128}(c), \fref{fig:QPSK_128x128x128}(c) and \fref{fig:16QAM_256x128x128}(c), we see that the JED supports $ 90$\% of the UEs to achieve a transmission rate at $ 0.43 $\,bits/symbol, $ 1.4 $\,bits/symbol, and $ 1.9 $\,bits/symbol for BPSK, QPSK, and $16$-QAM, respectively.
Particularly, the per-UE MI results suggest that if cell-free massive MU-MIMO systems are equipped with strong error correction codes, over $ 90 $\% of the UEs could transmit at high data rates, without the common assumption \cite{attarifarModified2019,atzeni2021Distributed,basharMax2019,basharEnhanced2018,maryopiUplink2019} that the number of UEs should be far lower than the number of receive antennas.
In \fref{fig:BPSK_64x128x128}(d), \fref{fig:QPSK_128x128x128}(d), and \fref{fig:16QAM_256x128x128}(d), we see that JED provides $ 3 $\,dB,  $5$\,dB, and $ 3 $\,dB lower channel estimation MSE than the {$\ell_1$-norm} channel estimator in \fref{eq:l1 chEst} for BPSK, QPSK, and 16QAM, respectively. 
Notably, our JED algorithm achieves satisfactory performance in overloaded ($ B<U $) and fully-loaded ($ B=U $) MU-MIMO systems.
More specifically, we observe that the L-MMSE detector, even when performed in a \emph{centralized} manner with an $\ell_1$-norm-based channel estimator, still performs significantly worse than our JED algorithm.
Decentralized data detectors, although attractive due to their low complexity and scalability \cite{bjornsonScalable2020,bjornsonMaking2020},  perform even worse.
In fact, this observation is particularly valid in densely-populated systems as shown in \fref{fig:BPSK_64x128x128} and \ref{fig:QPSK_128x128x128} where centralized L-MMSE data detection completely fails in the overloaded case ($B<U$) and is almost $6\times$ worse than our JED algorithm in the fully-loaded case ($B=U$).

We also note that in user-centric cell-free massive MU-MIMO systems~\cite{bjornsonScalable2020}, the number of APs that serve each UE is does not necessarily depend on the ratio ${B}/{U}$.
However, we observe from the above simulations that increasing the ratio ${B}/{U}$ from $0.5$ to $2$ significantly improves the efficacy of linear methods.
This property applies for both centralized and distributed data detectors---corresponding simulations are omitted due to the page limit.
\subsection{Computational Complexity Analysis }\label{sec: complexity}
We now analyze the complexity of our JED algorithm.
We start by measuring the complexity of the L-MMSE equalizer and our FBS solver by counting the number of real-valued multiplications (and ignore the complexity of additions, square roots, reciprocals, etc.).
We assume that one complex-valued multiplication requires four real-valued multiplications.
In what follows, the numbers in parentheses refer to the complexity.

The L-MMSE equalizer corresponds to computing ${ \widehat{\bS}_D = (\rho\bI_U + \bH^H\bH)^{-1} \bH^H \bY_D}$. 
From \cite{castanedaFiniteAlphabet2020}, we have that the total complexity is $ 2U^3+6BU^2-2BU+4BUD-2U+1 $.
In each iteration of FBS, we first compute $   \bH\bS-\bY \, (4BUK)  $.
Then, we multiply it with $ \bS^H \, (4BUK) $ and $ \bH \, (4BUK) $.
The next step is to scale $ \bS $ with $ \frac{1}{1-\rho^{(t)}}\, (2BU+2KU)$.
Therefore, the total computational complexity in each iteration of FBS is $ 12BUK + 2U(B+K). $
Note that we ignore the complexity of the permutation problem for two reasons. 
First, this problem only needs to be solved when the large-scale fading components of the UEs change, which is at lower rate than the JED problem and mainly depends on UE locations---this observation is even more obvious for the position-based permutation problem.
Second, solving these permutation problems mainly requires additions and other simple operations. Since we measure complexity by counting the number of multiplications, it is challenging to relate the complexity of such operations in a fair manner.

Due to the nonconvexity of JED, better initialization methods improve the performance and require fewer iterations for FBS to converge.
Solving the LASSO problem \fref{eq:l1 chEst} indeed yields a good initializer, but is also computationally intensive.
We therefore propose to use  the initialization techniques proposed in \fref{sec: initialization} to reduce complexity while still enabling excellent performance. 
Figure~\ref{fig:QPSK_iterations} and \ref{fig:complexity_reduction_CHEST_MSE} show the required iterations for convergence and the initialized channel estimation MSE results for a $ B=128 $ antenna system with $ U=128 $ UEs transmitting with QPSK over $ K=128 $ time slots, where $ T = 32$ ($ 25 $\%) are used for training.
The stopping condition requires the ratio between the norm of the estimated gradient in the current iteration and the maximum of the norm of the estimated gradient throughout all the iterations to be smaller than the tolerance.
For the unpermuted channels (shown as ``None'' in \fref{fig:QPSK_iterations}), ETFs are used for pilots; LS and $ \ell_1 $-norm channel estimator are used to initialize $ \bH $;  the L-MMSE equalizer using the variance of noise is selected to initialize $ \bS_D $.
For the permuted channels (shown as CSI and PHY in \fref{fig:QPSK_iterations}), MUBs are used as pilots; the block-wise James-Stein estimator aided with the MAD technique is used to initialize $ \widetilde{\bH} $ (permuted channel matrix); the L-MMSE equalizer is used to initialize $ \widetilde\bS_D $ as shown in \fref{sec: init Sd}.
Analogously, we use the $ \ell_1 $-norm channel estimator as the benchmark in both plots.

In \fref{fig:QPSK_iterations}, our proposed initialization techniques enable $ 90 $\% of the UEs to converge after $ 1600 $ iterations which achieves the same convergence speed as the $ \ell_1 $-norm benchmark.
In stark contrast, we see that the poor initialization generated by least square (LS) channel estimation 
requires more than $ 3500 $ iterations for $ 90\% $ of UEs to converge.
Specifically, after clustering the large entries into blocks on the diagonal, both of CSI and PHY permutation methods are able to halve the required iterations to $ 1600 $.
Besides, the block-diagonal channel matrix also enables distributed processing with JED in each block in future work, which could be the key to significantly reduce interconnect data rates and algorithm complexity. 
Clearly, the development of new methods that further reduce the complexity of JED are necessary to enable a successful deployment in practice.

In \fref{fig:complexity_reduction_CHEST_MSE}, we show the channel estimation MSE for different initialization methods
instead of the MSE of the JED algorithm.
Analogously, we use the $ \ell_1 $-norm-based method in unpermuted channels as the benchmark. 
We see that the $\ell_1$-norm method provides the lowest MSE for channel estimation and thus has the fastest convergence speed in the complexity comparison. 
In the same channel, the least square (LS) channel estimation in a fully-loaded system with only $ 25\% $ nonorthogonal pilots provides the worst channel estimation MSE and yields the lowest convergence speed.
The permuted channel matrices have four virtual cells on the diagonal which form the block-wise structure.
Such a structure and the usage of mutually unbiased bases (MUBs) enable local orthogonality in each virtual cell with which we perform LS and our proposed initialization techniques in each virtual cell to get $ 3 $dB gain for MSE.
Lower initialized channel estimation MSE and better shaped channel matrix together make the convergence of FBS comparable with the $ \ell_1 $-norm-based benchmark.

We emphasize that the computational complexity of our JED algorithm, even when reduced by the proposed initialization methods, remains to be the main bottleneck in practice. 
One of the goals of our paper is to demonstrate that densely populated scenarios benefit significantly from more sophisticated data-detection algorithms (cf.~\fref{fig:BPSK_64x128x128} and \ref{fig:QPSK_128x128x128}). However, even in conventional cell-free massive MU-MIMO scenarios with more AP than UE antennas, our JED algorithm significantly outperforms linear data detectors (cf.~\fref{fig:16QAM_256x128x128}).
As it can be seen in \fref{sec:MI results}, if we try to serve more UEs, even the centralized L-MMSE data detector fails at providing satisfying performance---alternative methods that enable decentralized data detection in cell free systems~\cite{bjornsonScalable2020} would struggle even more in such scenarios.
In short, JED buys performance advantages at higher complexity and issues with scalability to more UEs, but realizes a clear advantage in such densely-populated scenarios.
On the bright side, our JED algorithm requires essentially only matrix-vector multiplications, which is key to enabling efficient and parallel hardware implementations.
%

\subsection{Pilot Overhead vs. MI Trade-off}
Figure~\ref{fig:CFJED_Training_TradOff} shows the trade-off between the per-UE MI achieved by $ 90 $\% of UEs and the amount of used pilots (as a fraction of orthogonal training). 
We show the per-UE MI for the $ B = 64 $ system with BPSK, $B=128$ system with QPSK, and the $B=256$ system with 16-QAM; all the systems have $ U = 128 $ UEs transmitting over $K=128$ time slots. 
The three dashed lines correspond to upper bounds, assuming that a rate loss is only caused by pilot transmission. 
For example, the point $(50,2) $ in the blue dashed line shows that the maximum transmission rate with $ 16 $-QAM modulation is  $ 2$\,bits/symbol/UE with $ 50 $\% pilot overhead.
We see that our JED algorithm can asymptotically achieve a per-UE MI close to the upper bound for all the scenarios.
For $ 50\% $ pilot overhead, all three systems reach the upper bound.
The highest per-UE MI is achieved by with $ 28\% $, $ 18\% $, and $ 11\% $ for BPSK, QPSK, and 16-QAM, respectively. These peaks are lower than the upper limit by $ 0.175 $\,bits/symbol/UE, $ 0.255 $\,bits/symbol/UE, and $ 0.65 $\,bits/symbol/UE, respectively.
We also see that as the number of antennas increases, the minimum requirement of pilots decreases.
Conversely, conventional L-MMSE (shown as dotted lines) data detection with $\ell_1$-norm channel estimator is not able to achieve a satisfactory per-UE MI.
Concretely, L-MMSE fails entirely in the overloaded BPSK system and is lower than the peak of QPSK and $ 16 $-QAM curves by $ 1.16 $\,bits/symbol/UE and $ 2.2 $\,bits/symbol/UE, respectively.

From the RMSSE and BER results in \fref{fig:BPSK_64x128x128} to Figs.~\ref{fig:16QAM_256x128x128}, we see that centralized L-MMSE data detection gradually approaches our JED algorithm in these metrics by increasing the number of pilots. This trend is, however, not only because the presence of more pilots but also due to the number of AP and UE antennas.
Figure \ref{fig:CFJED_Training_TradOff} further illustrates that JED significantly reduces the required pilot overhead compared to the L-MMSE data detector in densely-populated scenarios.
For example, consider the trade-off realized by QPSK modulation in a densely-populated scenario ($B=U=128$; red curves).
We observe that JED reaches the highest per-UE MI with only $18\%$ pilot overhead.
In contrast, we see that the L-MMSE data detector performs way worse than JED  and approaches the upper limit (the dashed curve) only at around $50\%$ pilot overhead. 
For the overloaded scenario with BPSK modulation ($B=64$ and $U=128$; pink curves), JED achieves a peak per-UE MI at around $25\%$ pilot overhead---the L-MMSE data detector achieves zero per-UE MI across the board. 
In short, deploying nonorthogonal pilots in densely-populated scenarios strongly affects data detectors that separate channel estimation from data detection.
The reason is that such methods require accurate channel estimates, which results in high pilot overhead, whereas JED leverages payload data for channel estimation and the pilots mainly serve to resolve the uniqueness issue (cf.~\fref{sec:ambiguity}).
The trade-off analysis for conventional massive MU-MIMO systems \cite[Fig.~5]{ostman2021URLLC} shows a similar result---pilot contamination caused by the lack of available pilots will lead to severe performance degradation.
\section{Conclusions}\label{sec:conclusion}
We have proposed a novel joint estimation and detection (JED) algorithm for densely populated cell-free massive MU-MIMO systems.
We have formulated a suitable MAP-JED problem and have developed an algorithm that builds upon forward-backward splitting (FBS). In addition, we have shown that such JED algorithms can be initialized by clustering the user equipments (UEs) and access points (APs). 
By combining both techniques, we have shown that if the number of UEs approaches or even exceeds the number of APs, then reliable transmission is possible with $ 50\% $ or much fewer pilots that would be necessary for orthogonal training. 

We see many open research problems. First and foremost is the design of techniques that further reduce the complexity of our JED algorithm---a promising direction is our recent work in~\cite{songSJEDfuture} that solves a related JED problem in only $10$ iterations with the aid of a neural network.
Furthermore, methods as in \cite{li17d,li2018feedforward,bjornsonScalable2020,bjornsonMaking2020,interdonato2019Scalability} that decentralize data detection algorithms, so that complexity can be off-loaded to the APs and interconnect data rates can be  reduced, will be key for a practical deployment of JED---here, the formation of overlapping UE clusters might be beneficial.
Moreover, establishing stronger optimality results for our FBS algorithm is an interesting but extremely challenging topic.
In addition, our results show that JED performs well in systems where the effective number of transmitting UEs is lower than the number of AP antennas $B$, which indicates  that JED might be particularly useful for nonortghogonal multiple access (NOMA) with appropriate modifications.
Finally, a hardware prototype of our JED algorithm that is able to support the throughputs of future cell-free massive MU-MIMO systems would pave the way for a practical deployment of JED.


\appendices

\section{Proof of \fref{lem:biconvexity}}
\label{app:biconvexity}

We start by proving convexity in \bH with $ \bS_D $ fixed. In this case, the matrix $ \left[ \bS_T, \bS_D \right]  $ and the concave regularizer $-\gamma\|\bS_D\|_F^2$ are constants. Hence, \fref{eq:opt3} reduces to a quadratic problem that is obviously convex.
To prove the convexity in $ \bS_D $ with $\bH$ fixed, we rewrite the objective function in $ \bS_D $ as 
\aln{\label{eq:biconvex proof}
	& \hat f(\bS_D) 
		= \Tr \!\left(  \bY_D^H\bY_D + \bS_D^H {\bA} \bS_D - 2 \mathfrak{R}\! \left\lbrace \bY_D^H\bH\bS_D \right\rbrace  \right)\!, 
}
with $ {\bA} =  \left( \bH^H\bH - \gamma \bI_U \right) $. We now provide conditions for which the Hessian of $ \hat{f}(\bS_D) $ is positive semidefinite. 
Note that $ \hat{f}(\bS_D) $ is a real-valued function of complex-valued variables. 
As discussed in \fref{sec:FASTA}, $ \hat f (\bS_D)$ depends on two matrices $ \bS_D $ and $ \bS_D^* $, hence there exist four different complex-valued Hessian matrices for $ \hat{f} (\bS_D)$ with respect to all the ordered combinations of $ \bS_D $ and $ \bS_D^* $~\cite{hjorungnes11a}.
Since the second-order derivatives of the constant term and affine terms in \fref{eq:biconvex proof} are zero, we refer to the results in~\cite[Ex.~5.1, 5.4]{hjorungnes11a} and get the Hessian with $\widetilde{\bA}$ and $\widetilde{\bA}^T$ on the diagonal and $0$ otherwise.
Here, $ \widetilde{\bA} = \bA\kron \bI_U $. Since the eigenvalues of $\mathcal{H}(\hat{f}) $ are the eigenvalues of $ \bA $, $ \mathcal{H}(\hat{f}) $ is positive semidefinite as long as the smallest eigenvalue of $ \bA $ is non-negative---this holds if $\lambda_\text{min} - \gamma \geq 0,$ where $ \lambda_\text{min}$ is the smallest eigenvalue of $ \bH^H\bH $ and $ \gamma $ is the regularizer of the concave term in \fref{eq:opt3}.  

\section{Proof of \fref{lem:phasepermutationambiguity}}
\label{app:phase-permutation-ambiguity}

If a phase-permutation ambiguity exists, then at least two columns of \bF will be identical after a phase change, i.e.,  $	\bmf_b = e^{j\phi}\bmf_{b'}$,  $ b\neq b' $. 
By inserting this into $ \normtwo{\bmf_b}^2 =  \nu$, we have
\aln{\label{eq: amb2}
	\nu =\abs{\bmf_b^H\bmf_b} =  \abs{e^{j\phi}\ \bmf_b^H\bmf_{b'}}\!.
}
Hence, if $ \kappa <\nu $, then \fref{eq: amb2} cannot not hold, which means that no pair of  columns of $ \bF $ are the same after a phase change.

\section{Proof of \fref{thm:monotone}}\label{app:fbsproof}
We start by rewriting the FBS update as follows:
\aln{
	\bZ\kp 
	&= \argmin_\bZ g(\bZ)  \textstyle
	+ \frac{1}{2\taut} \normfro{\bZ- \bZ\ko +\taut \nbfZt}^2    \label{eq:proof1}
	\\
	& = \argmin_\bZ  \Big\{  g(\bZ) + f(\bZ\ko)+ \frac{1}{2\taut}\normfro{\bZ - \bZ\ko}^2  \notag \\
	& \,\quad \qquad \qquad +  \innR{\bZ - \bZ\ko, \nbfZt}  \Big\}  \label{eq:proof2} .
}
Note that \fref{eq:proof1} and \fref{eq:proof2} differ by only additive constants, so both formulations attain the same minimizer.    

We begin by proving monotonicity of the algorithm. Since the choice $\bZ=\bZ\kp$ minimizes the expression in \fref{eq:proof2}, choosing instead $\bZ=\bZ\ko$ will result in a larger (or equal) value for this expression.  
More formally, 
\aln{g(\bZ\kp) + f(\bZ\ko) +  \innR{\bZ\kp - \bZ\ko, \nbfZt } + \notag
\\
 \frac{1}{2\taut}\normfro{\bZ\kp - \bZ\ko}^2  \le  g(\bZ\ko) + f(\bZ\ko). }

Combining this inequality with the line search condition~\fref{eq:linesearch} results in  
\aln{	\label{eq:monotonicity}
	g(\bZ\kp) + f(\bZ\kp) \le  g(\bZ\ko) + f(\bZ\ko).
}
{The minimizer $ \bZ\kp $ is always achieved after calculating the proximal in each iteration of FBS, indicating that \fref{eq:monotonicity} holds for every $ t $.} Therefore, we see that the method is monotonic {throughout the iterates.}

To prove convergence of the iterates, 
{we start by defining the subgradient of a real-valued function $ g: \complexset^{B\times (U+K)} \rightarrow \reals$.
A matrix $ \bG_{\bZ\ko} $ is called a subgradient of $ g $ at $ \bZ\ko$ if for any $ \bZ \in \complexset^{B\times (U+K)}$ we have 
\aln{\label{eq:subgradient def}
	g(\bZ) - g(\bZ\ko) \geq \innR{  \bG_{\bZ\ko}, \bZ-\bZ\ko  }.
}
}
We now consider the function 
\aln{\hat g(\bZ) = g(\bZ) + \frac{\gamma}{2}\normfro{\bZ-\bZ\ko}^2.}
This function is convex, and thus lies above its linear approximation.  We therefore have
\aln{ 
	\hat g(\bZ\ko) \; \ge \;  &\hat g(\bZ\kp) +  \innR{ \bZ\ko- \bZ\kp,\widehat \bG_{\bZ\kp}  }
	\nonumber \\
	 =\; & g(\bZ\kp) + \frac{\gamma}{2}\normfro{\bZ\kp-\bZ\ko}^2  \notag \\
&	  +  \innR{ \bZ\ko- \bZ\kp, \bG_{\bZ\kp}  }  
  -\gamma \normfro{\bZ\kp-\bZ\ko}^2  \nonumber 
	\\ 
	=\; & g(\bZ\kp)  +  \innR{ \bZ\ko- \bZ\kp, \bG_{\bZ\kp}  } \notag \\
&   -\frac{\gamma}{2}\normfro{\bZ\kp-\bZ\ko}^2, \label{proof3} 
}
where {$\widehat \bG_{\bZ\kp}= \bG _{\bZ\kp}+ \gamma(\bZ\kp-\bZ\ko)$} is a subgradient of $\hat g$ at $\bZ\kp,$ and $\bG_\bZ $ is a subgradient of $g$ at $\bZ\kp.$

Adding \eqref{proof3} with \eqref{eq:linesearch}, and noting that $\hat g(\bZ\ko)= g(\bZ\ko)$, we get
\aln{ 
	h(\bZ\kp) 
	&=  f(\bZ\kp)+g(\bZ\kp)  
	\notag \\
	& \le  f(\bZ\ko) + g(\bZ\ko) +\frac{1+\taut\gamma}{2\taut}\normfro{\bZ\kp-\bZ\ko}^2
	\notag \\
	& \quad+ \innR{\bZ\kp- \bZ\ko, \nbfZt+\bG_{\bZ\ko} } . \label{eq:convergence cond1}
}
{To simplify the inner product, we obtain from the optimality condition for \fref{eq:proof2} at $ \bZ\kp $} that 
\aln{
	\bG_{\bZ\kp} + \nbfZt + \frac{1}{\taut}(\bZ\kp - \bZ\ko) = \zeroes.  \label{eq:convergence cond2}
}
Combining \fref{eq:convergence cond1} with \fref{eq:convergence cond2}, we have
\aln{ 
	h(\bZ\kp) \le \,&  h(\bZ\ko) +\frac{1+\taut\gamma}{2\taut}\normfro{\bZ\kp-\bZ\ko}^2  \notag \\
	 & - \frac{1}{\taut}  \normfro{\bZ\kp- \bZ\ko}^2 \notag
	\\
	 =\,&  h(\bZ\ko) - \frac{1-\taut\gamma }{2\taut}\normfro{\bZ\kp-\bZ\ko}^2. \label{eq: proof4}
}
Summing this result over the first $T-1$ iterations, we obtain
\aln{h(\bZ^{(0)}) - h(\bZ^{(T)}) \ge \sum_{{ t}=0}^{T-1}  \frac{1-\taut\gamma }{2\taut}\normfro{\bZ\kp-\bZ\ko}^2.}
Because we have assumed $\taut<1/\gamma$ and the stepsizes are uniformly bounded away from zero, this guarantees that  $\{\bZ^T\}$ approaches some limit point $\bZ^{\text{opt}}$ as $t \rightarrow \infty.$ 

\balance
\bibliographystyle{IEEEtran}
\bibliography{bibs/VIPabbrv,bibs/publishers,bibs/confs-jrnls,bibs/vipbib_hc}

\begin{thebibliography}{10}
\providecommand{\url}[1]{#1}
\csname url@samestyle\endcsname
\providecommand{\newblock}{\relax}
\providecommand{\bibinfo}[2]{#2}
\providecommand{\BIBentrySTDinterwordspacing}{\spaceskip=0pt\relax}
\providecommand{\BIBentryALTinterwordstretchfactor}{4}
\providecommand{\BIBentryALTinterwordspacing}{\spaceskip=\fontdimen2\font plus
\BIBentryALTinterwordstretchfactor\fontdimen3\font minus
  \fontdimen4\font\relax}
\providecommand{\BIBforeignlanguage}[2]{{%
\expandafter\ifx\csname l@#1\endcsname\relax
\typeout{** WARNING: IEEEtran.bst: No hyphenation pattern has been}%
\typeout{** loaded for the language `#1'. Using the pattern for}%
\typeout{** the default language instead.}%
\else
\language=\csname l@#1\endcsname
\fi
#2}}
\providecommand{\BIBdecl}{\relax}
\BIBdecl

\bibitem{songMinimizing2020}
H.~Song, X.~You, C.~Zhang, O.~Tirkkonen, and C.~Studer, ``Minimizing pilot
  overhead in cell-free massive {{MIMO}} systems via joint estimation and
  detection,'' in \emph{Proc. IEEE Int. Workshop Signal Process. Advances
  Wireless Commun. (SPAWC)}, May 2020, pp. 1--5.

\bibitem{ngoCellfree2017}
H.~Q. Ngo, A.~Ashikhmin, H.~Yang, E.~G. Larsson, and T.~L. Marzetta,
  ``Cell-free massive {MIMO} versus small cells,'' \emph{{IEEE} Trans. Wireless
  Commun.}, vol.~16, no.~3, pp. 1834--1850, Mar. 2017.

\bibitem{buzzi2017cell}
S.~Buzzi and C.~D'Andrea, ``Cell-free massive {MIMO}: User-centric approach,''
  \emph{{IEEE} Wireless Commun. Lett.}, vol.~6, no.~6, pp. 706--709, Dec. 2017.

\bibitem{hoang2018cell}
T.~M. Hoang, H.~Q. Ngo, T.~Q. Duong, H.~D. Tuan, and A.~Marshall, ``Cell-free
  massive {MIMO} networks: Optimal power control against active
  eavesdropping,'' \emph{{IEEE} Trans. Commun.}, vol.~66, no.~10, pp.
  4724--4737, Oct. 2018.

\bibitem{zhang2020Prospective}
J.~Zhang, E.~Bj{\"o}rnson, M.~Matthaiou, D.~W.~K. Ng, H.~Yang, and D.~J. Love,
  ``Prospective multiple antenna technologies for beyond {{5G}},'' \emph{{IEEE}
  J. Sel. Areas Commun.}, vol.~38, no.~8, pp. 1637--1660, Aug. 2020.

\bibitem{basharUplink2019}
M.~Bashar, K.~Cumanan, A.~G. Burr, M.~Debbah, and H.~Q. Ngo, ``On the uplink
  max-min {SINR} of cell-free massive {MIMO} systems,'' \emph{{IEEE} Trans.
  Wireless Commun.}, vol.~18, no.~4, pp. 2021--2036, Apr. 2019.

\bibitem{basharMax2019}
M.~Bashar, K.~Cumanan, A.~G. Burr, H.~Q. Ngo, M.~Debbah, and P.~Xiao, ``Max-min
  rate of cell-free massive {MIMO} uplink with optimal uniform quantization,''
  \emph{{IEEE} Trans. Commun.}, vol.~67, no.~10, pp. 6796--6815, Oct. 2019.

\bibitem{maiPilot2018}
T.~C. Mai, H.~Q. Ngo, M.~Egan, and T.~Q. Duong, ``Pilot power control for
  cell-free massive {MIMO},'' \emph{{IEEE} Trans. Veh. Technol.}, vol.~67,
  no.~11, pp. 11\,264--11\,268, Nov. 2018.

\bibitem{sabbaghPilot2018}
R.~Sabbagh, C.~Pan, and J.~Wang, ``Pilot allocation and sum-rate analysis in
  cell-free massive {{MIMO}} systems,'' in \emph{Proc. IEEE Int. Conf. Commun.
  (ICC)}, May 2018, pp. 1--6.

\bibitem{doanPerformance2017}
T.~X. Doan, H.~Q. Ngo, T.~Q. Duong, and K.~Tourki, ``On the performance of
  multigroup multicast cell-free massive {MIMO},'' \emph{{IEEE} Commun. Lett.},
  vol.~21, no.~12, pp. 2642--2645, 2017.

\bibitem{parkOptimizing2018}
S.-H. Park, O.~Simeone, Y.~C. Eldar, and E.~Erkip, ``Optimizing pilots and
  analog processing for channel estimation in cell-free massive {MIMO} with
  one-bit {ADC}s,'' in \emph{Proc. IEEE Int. Workshop Signal Process. Advances
  Wireless Commun. (SPAWC)}, Jun. 2018, pp. 1--5.

\bibitem{liu2020Graph}
H.~Liu, J.~Zhang, S.~Jin, and B.~Ai, ``Graph coloring based pilot assignment
  for cell-free massive {{MIMO}} systems,'' \emph{{IEEE} Trans. Veh. Technol.},
  vol.~69, no.~8, pp. 9180--9184, Aug. 2020.

\bibitem{attarifarModified2019}
M.~Attarifar, A.~Abbasfar, and A.~Lozano, ``Modified conjugate beamforming for
  cell-free massive {{MIMO}},'' \emph{{IEEE} Wireless Commun. Lett.}, vol.~8,
  no.~2, pp. 616--619, Apr. 2019.

\bibitem{atzeni2021Distributed}
I.~Atzeni, B.~Gouda, and A.~T{\"o}lli, ``Distributed precoding design via
  over-the-air signaling for cell-free massive {{MIMO}},'' \emph{{IEEE} Trans.
  Wireless Commun.}, vol.~20, no.~2, pp. 1201--1216, Feb. 2021.

\bibitem{basharEnhanced2018}
M.~Bashar, K.~Cumanan, A.~G. Burr, M.~Debbah, and H.~Q. Ngo, ``Enhanced max-min
  sinr for uplink cell-free massive {MIMO} systems,'' in \emph{Proc. IEEE Int.
  Conf. Commun. (ICC)}, May 2018, pp. 1--6.

\bibitem{maryopiUplink2019}
D.~Maryopi, M.~Bashar, and A.~Burr, ``On the uplink throughput of zero forcing
  in cell-free massive {MIMO} with coarse quantization,'' \emph{{IEEE} Trans.
  Veh. Technol.}, vol.~68, no.~7, pp. 7220--7224, Jul. 2019.

\bibitem{attarifar2018random}
M.~Attarifar, A.~Abbasfar, and A.~Lozano, ``Random vs structured pilot
  assignment in cell-free massive {MIMO} wireless networks,'' in \emph{Proc.
  IEEE Int. Conf. Commun. (ICC)}, May 2018, pp. 1--6.

\bibitem{basharCellFree2018}
M.~Bashar, K.~Cumanan, A.~G. Burr, H.~Q. Ngo, and M.~Debbah, ``Cell-free
  massive {MIMO} with limited backhaul,'' in \emph{Proc. IEEE Int. Conf.
  Commun. (ICC)}, May 2018, pp. 1--7.

\bibitem{basharPerformance2018}
M.~Bashar, H.~Q. Ngo, A.~G. Burr, D.~Maryopi, K.~Cumanan, and E.~G. Larsson,
  ``On the performance of backhaul constrained cell-free massive {MIMO} with
  linear receivers,'' in \emph{Proc. {IEEE} Conf. Rec. Asilomar Conf. Signals,
  Sys., and Comp.}, Oct. 2018, pp. 624--628.

\bibitem{interdonato2019Scalability}
G.~Interdonato, P.~Frenger, and E.~G. Larsson, ``Scalability aspects of
  cell-free massive {{MIMO}},'' in \emph{Proc. IEEE Int. Conf. Commun. (ICC)},
  May 2019, pp. 1--6.

\bibitem{bjornsonScalable2020}
E.~Bj{\"o}rnson and L.~Sanguinetti, ``Scalable cell-free massive {{MIMO}}
  systems,'' \emph{{IEEE} Trans. Commun.}, pp. 4247--4261, 2020.

\bibitem{bjornsonMaking2020}
------, ``Making cell-free massive {MIMO} competitive with {MMSE} processing
  and centralized implementation,'' \emph{{IEEE} Trans. Wireless Commun.},
  vol.~19, no.~1, pp. 77--90, Jan. 2020.

\bibitem{alshamary2015optimal}
H.~A.~J. Alshamary, T.~Al-Naffouri, A.~Zaib, and W.~Xu, ``Optimal non-coherent
  data detection for massive {SIMO} wireless systems: A polynomial complexity
  solution,'' in \emph{Proc. IEEE Signal Process. Signal Process. Edu.
  Workshop}, Aug. 2015, pp. 172--177.

\bibitem{pham2009joint}
T.-H. Pham, Y.-C. Liang, and A.~Nallanathan, ``A joint channel estimation and
  data detection receiver for multiuser {MIMO} {IFDMA} systems,'' \emph{{IEEE}
  Trans. Commun.}, vol.~57, no.~6, pp. 1857--1865, June 2009.

\bibitem{prasad2015joint}
R.~Prasad, C.~R. Murthy, and B.~D. Rao, ``Joint channel estimation and data
  detection in {MIMO-OFDM} systems: A sparse {Bayesian} learning approach,''
  \emph{{IEEE} Trans. Signal Process.}, vol.~63, no.~20, pp. 5369--5382, Oct.
  2015.

\bibitem{kofidis2017joint}
E.~Kofidis, C.~Chatzichristos, and A.~L. de~Almeida, ``Joint channel
  estimation/data detection in {MIMO-FBMC/OQAM} systems---a tensor-based
  approach,'' in \emph{Proc. {IEEE} European Signal Process. Conf. (EUSIPCO)},
  Sept. 2017, pp. 420--424.

\bibitem{wen15b}
C.-K. Wen, C.-J. Wang, S.~Jin, K.-K. Wong, and P.~Ting, ``{B}ayes-optimal joint
  channel-and-data estimation for massive {MIMO} with low-precision {ADC}s,''
  \emph{{IEEE} Trans. Signal Process.}, vol.~64, no.~10, pp. 2541--2556, Jul.
  2015.

\bibitem{castaneda2017vlsi}
O.~Casta{\~n}eda, T.~Goldstein, and C.~Studer, ``{VLSI} designs for joint
  channel estimation and data detection in large {SIMO} wireless systems,''
  \emph{{IEEE} Trans. Circuits Syst. {I}}, vol.~65, no.~3, pp. 1120--1132, Mar.
  2017.

\bibitem{jiangJoint2020}
\BIBentryALTinterwordspacing
S.~Jiang, X.~Yuan, X.~Wang, C.~Xu, and W.~Yu, ``Joint user identification,
  channel estimation, and signal detection for grant-free noma,''
  \emph{arXiv:2001.03930}, Jul. 2020. [Online]. Available:
  \url{https://arxiv.org/abs/2001.03930}
\BIBentrySTDinterwordspacing

\bibitem{Yilmaz19a}
B.~Yilmaz and A.~Erdogan, ``Channel estimation for massive {MIMO}: A semiblind
  algorithm exploiting {QAM} structure,'' in \emph{Proc. Asilomar Conf.
  Signals, Syst., Comput.}, Nov. 2019.

\bibitem{feng2017noncoherent}
J.~Feng, H.~Gao, T.~Wang, T.~Lv, and W.~Guo, ``A noncoherent differential
  transmission scheme for multiuser massive {MIMO} systems,'' in \emph{Proc.
  IEEE Wireless Commun. Netw. Conf. (WCNC)}, Mar. 2017, pp. 1--6.

\bibitem{zhangBlind2018}
J.~Zhang, X.~Yuan, and Y.-J.~A. Zhang, ``Blind signal detection in massive
  {MIMO}: Exploiting the channel sparsity,'' \emph{{IEEE} Trans. Commun.},
  vol.~66, no.~2, pp. 700--712, Feb. 2018.

\bibitem{liuSuperresolution2019}
H.~Liu, X.~Yuan, and Y.~J. Zhang, ``Super-resolution blind channel-and-signal
  estimation for massive {{MIMO}} with one-dimensional antenna array,''
  \emph{{IEEE} Trans. Signal Process.}, vol.~67, no.~17, pp. 4433--4448, Sep.
  2019.

\bibitem{dingSparsity2019}
T.~Ding, X.~Yuan, and S.~C. Liew, ``Sparsity learning-based multiuser detection
  in grant-free massive-device multiple access,'' \emph{{IEEE} Trans. Wireless
  Commun.}, vol.~18, no.~7, pp. 3569--3582, Jul. 2019.

\bibitem{xueBlind2020}
\BIBentryALTinterwordspacing
Y.~Xue, Y.~Shen, V.~Lau, J.~Zhang, and K.~B. Letaief, ``Blind data detection in
  massive {MIMO} via $\ell_3$-norm maximization over the {Stiefel} manifold,''
  Apr. 2020. [Online]. Available: \url{https://arxiv.org/abs/2004.12301}
\BIBentrySTDinterwordspacing

\bibitem{binbindaiSparse2013}
B.~Dai and W.~Yu, ``Sparse beamforming for limited-backhaul network {MIMO}
  system via reweighted power minimization,'' in \emph{Proc. IEEE Global
  Commun. Conf. (GLOBECOM)}, Dec. 2013, pp. 1962--1967.

\bibitem{hanSparse2019}
\BIBentryALTinterwordspacing
D.~Han, J.~Park, and N.~Lee, ``Sparse joint transmission for cell-free massive
  {MIMO}: A sparse {PCA} approach,'' Dec. 2019. [Online]. Available:
  \url{https://arxiv.org/abs/1912.05231}
\BIBentrySTDinterwordspacing

\bibitem{shiGroup2014}
Y.~Shi, J.~Zhang, and K.~B. Letaief, ``Group sparse beamforming for green
  cloud-{RAN},'' \emph{{IEEE} Trans. Wireless Commun.}, vol.~13, no.~5, pp.
  2809--2823, May 2014.

\bibitem{vanchien2020Joint}
T.~Van~Chien, E.~Bj{\"o}rnson, and E.~G. Larsson, ``Joint power allocation and
  load balancing optimization for energy-efficient cell-free massive {{MIMO}}
  networks,'' \emph{{IEEE} Trans. Wireless Commun.}, vol.~19, no.~10, pp.
  6798--6812, Oct. 2020.

\bibitem{guoDistributed2019}
M.~Guo and M.~C. Gursoy, ``Distributed sparse activity detection in cell-free
  massive {MIMO} systems,'' in \emph{Proc. {IEEE} Global Conf. Signal Inf.
  Process. (GLOBALSIP)}, Nov. 2019, pp. 1--5.

\bibitem{shiLargescale2015}
Y.~Shi, J.~Zhang, K.~B. Letaief, B.~Bai, and W.~Chen, ``Large-scale convex
  optimization for ultra-dense cloud-{RAN},'' \emph{{IEEE} Wireless Commun.
  Mag.}, vol.~22, no.~3, pp. 84--91, Jun. 2015.

\bibitem{jinChannel2019a}
Y.~Jin, J.~Zhang, S.~Jin, and B.~Ai, ``Channel estimation for cell-free
  {mmWave} massive {MIMO} through deep learning,'' \emph{{IEEE} Trans. Veh.
  Technol.}, vol.~68, no.~10, pp. 10\,325--10\,329, Oct. 2019.

\bibitem{mirfarshbafanBeamspace2019}
S.~H. Mirfarshbafan, A.~{Gallyas-Sanhueza}, R.~Ghods, and C.~Studer,
  ``Beamspace channel estimation for massive {{MIMO mmWave}} systems: Algorithm
  and {{VLSI}} design,'' \emph{arXiv:1910.00756 [cs, eess, math]}, Oct. 2019.

\bibitem{interdonato2019Downlink}
G.~Interdonato, H.~Q. Ngo, P.~Frenger, and E.~G. Larsson, ``Downlink training
  in cell-free massive {{MIMO}}: {{A}} blessing in disguise,'' \emph{{IEEE}
  Trans. Wireless Commun.}, vol.~18, no.~11, pp. 5153--5169, Nov. 2019.

\bibitem{huCellfree2019}
X.~Hu, C.~Zhong, X.~Chen, W.~Xu, H.~Lin, and Z.~Zhang, ``Cell-free massive
  {MIMO} systems with low resolution {ADCs},'' \emph{{IEEE} Trans. Commun.},
  vol.~67, no.~10, pp. 6844--6857, Oct. 2019.

\bibitem{riera-palouClustered2018}
F.~{Riera-Palou}, G.~Femenias, A.~G. Armada, and A.~{P{\'e}rez-Neira},
  ``Clustered cell-free massive {MIMO},'' in \emph{Proc. IEEE Global Commun.
  Conf. (GLOBECOM)}, Dec. 2018, pp. 1--6.

\bibitem{liuTabuSearchBased2020}
H.~Liu, J.~Zhang, X.~Zhang, A.~Kurniawan, T.~Juhana, and B.~Ai,
  ``Tabu-search-based pilot assignment for cell-free massive {MIMO} systems,''
  \emph{{IEEE} Trans. Veh. Technol.}, vol.~69, no.~2, pp. 2286--2290, Feb.
  2020.

\bibitem{buzziUserCentric2018}
S.~Buzzi, C.~D'Andrea, and C.~D'Elia, ``User-centric cell-free massive {MIMO}
  with interference cancellation and local {ZF} downlink precoding,'' in
  \emph{Proc. IEEE Int. Symp. Wirel. Comm. Syst. (ISWCS)}, Aug. 2018, pp. 1--5.

\bibitem{interdonatoUbiquitous2019}
G.~Interdonato, E.~Bj{\"o}rnson, H.~Quoc~Ngo, P.~Frenger, and E.~G. Larsson,
  ``Ubiquitous cell-free massive {MIMO} communications,'' \emph{EURASIP J.
  Wireless Commun. Netw.}, vol. 2019, p. 197, Dec. 2019.

\bibitem{dandrea2020User}
C.~D'Andrea and E.~G. Larsson, ``User association in scalable cell-free massive
  {{MIMO}} systems,'' in \emph{Proc. {IEEE} Conf. Rec. Asilomar Conf. Signals,
  Sys., and Comp.}, Nov. 2020, pp. 826--830.

\bibitem{huangEfficient2020}
X.~Huang, X.~Zhu, Y.~Jiang, and Y.~Liu, ``Efficient enhanced k-means clustering
  for semi-blind channel estimation of cell-free massive {{MIMO}},'' in
  \emph{Proc. IEEE Int. Conf. Commun. (ICC)}, Jun. 2020, pp. 1--6.

\bibitem{durt2010mutually}
T.~Durt, B.-G. Englert, I.~Bengtsson, and K.~{\.Z}yczkowski, ``On mutually
  unbiased bases,'' \emph{Intl. J. of Quantum Info.}, vol.~8, no.~04, pp.
  535--640, June 2010.

\bibitem{gesbert2003theory}
D.~Gesbert, M.~Shafi, D.-S. Shiu, P.~J. Smith, and A.~Naguib, ``From theory to
  practice: An overview of {MIMO} space--time coded wireless systems,''
  \emph{{IEEE} J. Sel. Areas Commun.}, vol.~21, no.~3, pp. 281--302, Apr. 2003.

\bibitem{goldsteinField2016}
\BIBentryALTinterwordspacing
T.~Goldstein, C.~Studer, and R.~G. Baraniuk, ``A field guide to
  forward-backward splitting with a {FASTA} implementation,'' Nov. 2014.
  [Online]. Available: \url{https://arxiv.org/abs/1411.3406}
\BIBentrySTDinterwordspacing

\bibitem{pedersenSparse2015a}
N.~L. Pedersen, C.~Navarro~Manch{\'o}n, M.-A. Badiu, D.~Shutin, and B.~H.
  Fleury, ``Sparse estimation using {{Bayesian}} hierarchical prior modeling
  for real and complex linear models,'' \emph{EURASIP J. Signal Process.}, vol.
  115, pp. 94--109, Oct. 2015.

\bibitem{eldar2010Blocksparse}
Y.~C. Eldar, P.~Kuppinger, and H.~Bolcskei, ``Block-sparse signals: Uncertainty
  relations and efficient recovery,'' \emph{{IEEE} Trans. Signal Process.},
  vol.~58, no.~6, pp. 3042--3054, Jun. 2010.

\bibitem{jeon16a}
C.~Jeon, A.~Maleki, and C.~Studer, ``On the performance of mismatched data
  detection in large {MIMO} systems,'' in \emph{Proc. IEEE Int. Symp. Inf.
  Theory (ISIT)}, May 2016.

\bibitem{abbasi2019performance}
E.~Abbasi, F.~Salehi, and B.~Hassibi, ``Performance analysis of convex data
  detection in {MIMO},'' in \emph{Proc. IEEE Int. Conf. Acoust., Speech, Signal
  Process. (ICASSP)}, May 2019, pp. 4554--4558.

\bibitem{shahabuddin2017admm}
S.~Shahabuddin, M.~Juntti, and C.~Studer, ``{ADMM}-based infinity norm
  detection for large {MU-MIMO}: Algorithm and {VLSI} architecture,'' in
  \emph{Proc. IEEE Int. Symp. Circuits and Syst. (ISCAS)}, May 2017, pp. 1--4.

\bibitem{shah2016biconvex}
S.~Shah, A.~K. Yadav, C.~D. Castillo, D.~W. Jacobs, C.~Studer, and
  T.~Goldstein, ``Biconvex relaxation for semidefinite programming in computer
  vision,'' in \emph{Eur. Conf. Comput. Vision}, Sep. 2016, pp. 717--735.

\bibitem{ngoEVDbased2012}
H.~Q. Ngo and E.~G. Larsson, ``{EVD}-based channel estimation in multicell
  multiuser {MIMO} systems with very large antenna arrays,'' in \emph{Proc.
  IEEE Int. Conf. Acoust., Speech, Signal Process. (ICASSP)}, Mar. 2012, pp.
  3249--3252.

\bibitem{hjorungnes11a}
A.~Hj{\o}rungnes, \emph{Complex-Valued Matrix Derivatives: With Applications in
  Signal Processing and Communications}.\hskip 1em plus 0.5em minus 0.4em\relax
  Cambridge Univ. Press, 2011.

\bibitem{troppDesigning2005}
J.~Tropp, I.~Dhillon, R.~Heath, and T.~Strohmer, ``Designing structured tight
  frames via an alternating projection method,'' \emph{{IEEE} Trans. Inf.
  Theory}, vol.~51, no.~1, pp. 188--209, Jan. 2005.

\bibitem{welchLower1974}
L.~Welch, ``Lower bounds on the maximum cross correlation of signals,''
  \emph{{IEEE} Trans. Inf. Theory}, vol.~20, no.~3, pp. 397--399, May 1974.

\bibitem{troppJust2006}
J.~Tropp, ``Just relax: Convex programming methods for identifying sparse
  signals in noise,'' \emph{{IEEE} Trans. Inf. Theory}, vol.~52, no.~3, pp.
  1030--1051, Mar. 2006.

\bibitem{jamesEstimation1992a}
W.~James and C.~Stein, ``\BIBforeignlanguage{en}{Estimation with quadratic
  loss},'' in \emph{\BIBforeignlanguage{en}{Breakthroughs in {{Statistics}}}},
  S.~Kotz and N.~L. Johnson, Eds.\hskip 1em plus 0.5em minus 0.4em\relax {New
  York, NY}: Springer, 1992, pp. 443--460.

\bibitem{gallyas-sanhueza2021Blind}
A.~{Gallyas-Sanhueza} and C.~Studer, ``Blind {{SNR}} estimation and
  nonparametric channel denoising in multi-antenna {{mmWave}} systems,'' in
  \emph{Proc. IEEE Int. Conf. Commun. (ICC)}, Jun. 2021, pp. 1--7.

\bibitem{fogelConvex2015}
F.~Fogel, R.~Jenatton, F.~Bach, and A.~{d'Aspremont}, ``Convex {{Relaxations}}
  for {{Permutation Problems}},'' \emph{SIAM J. Matrix Anal. Appl.}, vol.~36,
  no.~4, pp. 1465--1488, Jan. 2015.

\bibitem{tangOptimizing2019}
W.~Tang, Y.~Yang, L.~Zeng, and Y.~Zhan, ``Optimizing {MSE} for clustering with
  balanced size constraints,'' \emph{Symmetry}, vol.~11, no.~3, p. 338, 2019.

\bibitem{tang2001mobile}
A.~Tang, J.~Sun, and K.~Gong, ``Mobile propagation loss with a low base station
  antenna for {NLOS} street microcells in urban area,'' in \emph{Proc. IEEE
  Veh. Technol. Conf. Spring (VTC-Spring)}, vol.~1, May 2001, pp. 333--336.

\bibitem{shannon48a}
C.~E. Shannon, ``A mathematical theory of communication,'' \emph{Bell Sys.
  Tech. J.}, vol.~27, no.~3, pp. 379--423, Jul. 1948.

\bibitem{zhang2006non}
J.~Zhang, ``Non-asymptotic capacity lower bound for non-coherent {SIMO}
  channels with memory,'' in \emph{Proc. IEEE Int. Symp. Inf. Theory (ISIT)},
  July 2006, pp. 1272--1276.

\bibitem{3gpp138-101}
3GPP, ``{5G}; {NR}; user equipment ({UE}) radio transmission and reception,''
  Oct. 2018, {TS} 38.101 version 15.3.0 Rel.~15.

\bibitem{castanedaFiniteAlphabet2020}
O.~Casta\~neda, S.~Jacobsson, G.~Durisi, T.~Goldstein, and C.~Studer,
  ``Finite-alphabet {MMSE} equalization for all-digital massive {MU-MIMO
  {mmWave}} communication,'' \emph{{IEEE} J. Sel. Areas Commun.}, 2020.

\bibitem{ostman2021URLLC}
J.~{\"O}stman, A.~Lancho, G.~Durisi, and L.~Sanguinetti, ``{{URLLC}} with
  massive {{MIMO}}: Analysis and design at finite blocklength,'' \emph{{IEEE}
  Trans. Wireless Commun.}, pp. 1--1, 2021.

\bibitem{songSJEDfuture}
H.~Song, X.~You, C.~Zhang, and C.~Studer, ``Soft-output joint channel
  estimation and data detection using deep unfolding,'' in \emph{IEEE Inf.
  Theory Workshop (ITW)}, 2021, invited paper.

\bibitem{li17d}
K.~Li, R.~R. Sharan, Y.~Chen, T.~Goldstein, J.~R. Cavallaro, and C.~Studer,
  ``Decentralized baseband processing for massive {MU}-{MIMO} systems,''
  \emph{IEEE J. Emerging Sel. Topics Circuits Syst.}, vol.~7, no.~4, pp.
  491--507, Dec. 2017.

\bibitem{li2018feedforward}
K.~Li, C.~Jeon, J.~R. Cavallaro, and C.~Studer, ``Feedforward architectures for
  decentralized precoding in massive {MU-MIMO} systems,'' in \emph{Proc.
  Asilomar Conf. Signals, Syst., Comput.}, Pacific Grove, CA, USA, Oct. 2018,
  pp. 1659--1665.

\end{thebibliography}
\balance

\end{document}